\documentclass[preprint,3p,authoryear,times]{elsarticle}

\journal{Transportation Research Part C}
\usepackage[fleqn]{amsmath}
\usepackage[]{amsfonts}
\usepackage[]{amsthm}
\usepackage[]{amssymb}
\usepackage{empheq}
\usepackage{lscape} 

\usepackage[]{algorithm2e}
\usepackage[english]{babel}
\usepackage[T1]{fontenc}

\usepackage[colorlinks=true,linkcolor=blue]{hyperref} 
\usepackage{graphicx}
\usepackage{caption,subcaption}
\usepackage{tikz}
\usepackage{pstricks}
\usepackage{xstring}
\usepackage{multirow}
\usepackage{hhline}
\usepackage{url}

\definecolor{darkblue}{rgb}{0,0,0.5}

\hypersetup{colorlinks=true,linkcolor=black,citecolor=darkblue,urlcolor=darkblue} 

\newcommand{\newref}[2]{\hyperref[#2]{#1~\ref*{#2}}} 

\usepackage{longtable}
\usepackage{pdflscape}

\usepackage[toc,page]{appendix}

\usepackage{bbding}
\usepackage{tabularray}

\newcommand{\an}{\mathrm{a}}
\newcommand{\st}{\mathrm{s}}

\begin{document}

\renewcommand{\sectionautorefname}{Sect.}
\renewcommand{\subsectionautorefname}{Sect.}
\renewcommand{\subsubsectionautorefname}{Sect.}
\renewcommand{\figureautorefname}{Fig.}
\renewcommand{\tableautorefname}{Table}
\renewcommand{\equationautorefname}{Eq.}

\graphicspath{{./Figures/}}

\begin{frontmatter}

\title{On the Role of Non-Localities in Fundamental Diagram Estimation}

\author[1,2]{Jing Liu}
\author[1]{Fangfang Zheng\corref{cor1}}
\ead{fzheng@swjtu.edu.cn}
\author[1]{Boxi Yu}
\author[2]{Saif Jabari}

\address[1]{School of Transportation and Logistics, National Engineering Laboratory of Integrated Transportation Big Data Application Technology, National United Engineering Laboratory of Integrated and Intelligent Transportation, Southwest Jiaotong University Western Hi-tech Zone Chengdu, Sichuan 611756, P.R. China}

\address[2]{Division of Engineering, New York University Abu Dhabi, Saadiyat Island PO Box 129188 - Abu Dhabi, United Arab Emirates}

\cortext[cor1]{Corresponding author}

\begin{sloppypar} 
\begin{abstract}
We consider the role of non-localities in speed-density data used to fit fundamental diagrams from vehicle trajectories. We demonstrate that the use of anticipated densities results in a clear classification of speed-density data into stationary and non-stationary points, namely, acceleration and deceleration regimes and their separating boundary. The separating boundary represents a locus of stationary traffic states, i.e., the fundamental diagram. To fit fundamental diagrams, we develop an enhanced cross entropy minimization method that honors equilibrium traffic physics. We illustrate the effectiveness of our proposed approach by comparing it with the traditional approach that uses local speed-density states and least squares estimation. Our experiments show that the separating boundary in our approach is invariant to varying trajectory samples within the same spatio-temporal region, providing further evidence that the separating boundary is indeed a locus of stationary traffic states.
\end{abstract}


\end{sloppypar} 
\end{frontmatter}


\begin{sloppypar} 
\section*{Highlights} 
\begin{itemize}
\item {We approach the modeling of the FD as a binary classification problem, specifically focusing on distinguishing acceleration regimes.}
\item {Our proposed method introduces a set of empirical samples that consider non-localities, allowing us to model the FD while maintaining its inherent physical equilibrium.}
\item {We propose an enhanced cross-entropy loss function for fitting the FD model using these non-locality samples.}
\item {Our approach consistently generates an invariant FD across various datasets collected at the same location.}
\end{itemize}

\section{Introduction} \label{Sect1}
Fundamental diagrams (FDs) date back to \cite{greenshields1935study}, and they remain an area of continuing research interest, particularly, methods of fitting FDs \citep{9703273,bramich2023fitfun,liu2023linearizing}. The enduring appeal of FDs lies in their ability to summarize the functional relationship between macroscopic traffic flow variables in equilibrium. FDs serve as essential inputs for continuous traffic flow models \citep{MAKRIDIS2020102803} and find extensive applications in areas such as traffic control \citep{wang2014local,HEYDECKER2011206,DFREJO201915}, capacity analysis \citep{qin2023stabilizing}, traffic state estimation \citep{thodi2022incorporating,zheng2018traffic}, prediction \citep{liu2021dynamic,YANG2021102862}, and identification \citep{KALAIR2021103178}.

\begin{figure}[!ht]
  \centering
  \includegraphics[width=0.6\textwidth]{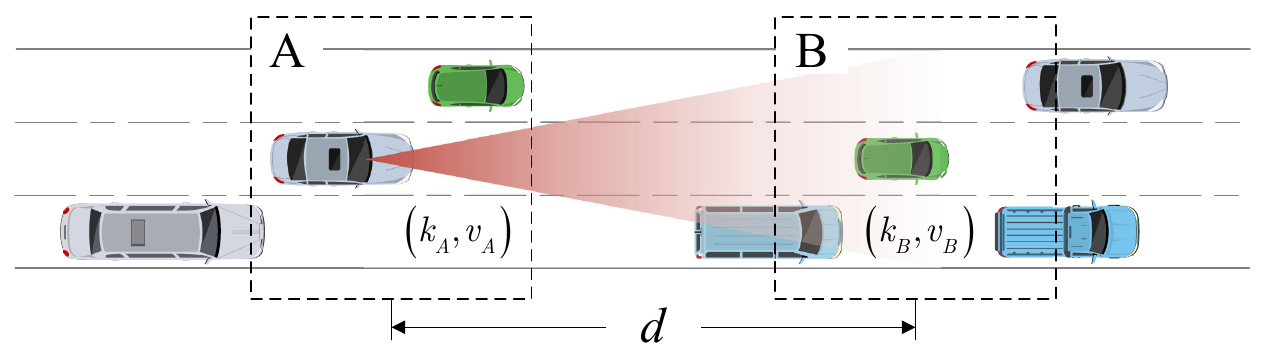}
  \caption{Speed adaptation in response to downstream traffic conditions}
  \label{Fig1}
\end{figure}

In the context of FD fitting methods, all approaches in the literature attempt to relate variables locally, i.e., they relate estimated densities and speeds in the same spatio-temporal location (e.g., ($k_{\mathrm{A}}$,$v_{\mathrm{A}}$) and ($k_{\mathrm{B}}$,$v_{\mathrm{B}}$) depicted in \autoref{Fig1}). We will refer to samples created this way as \textit{local density ($K$) speed ($V$)} samples, or LKV for short. The common fitting techniques use some variant of least square estimation (LSE) to determine the parameters of a specific FD model. The fitting methods implicitly assume that the dependent variable (i.e., speed) is independent across different densities and follows a Gaussian distribution. Additionally, they assume that the Gaussian distribution has a mean that aligns with the equilibrium speed as modeled by the FD, and a constant variance that is independent of density (although few papers have allowed for heteroskedasticity \citep{jabari2014probabilistic}, a constant variance remains a common implicit assumption). The estimation of model parameters is then performed using maximum likelihood estimation (MLE) \citep{9703273}.

The nature of LKV samples is incompatible with the implicit assumptions of LSE-based techniques summarized above. First, traffic flow cannot instantaneously transition from one equilibrium state to another. The non-stationary transition speed is influenced by the desired speed, which is determined by downstream traffic conditions (as depicted in \autoref{Fig1}, where the density in region B exceeds that in the upstream region A, causing upstream vehicles to adjust their speed in anticipation of traffic in region B rather than region A), thus violating the implicit independence assumption. Several researchers attempted to address this issue by excluding non-stationary data from FD fitting \citep{jabari2014probabilistic}. This results in removal of a large amount of data from the sample, which is impractical. As the LKV samples do not contain equilibrium information, other approaches consider external information (e.g., the arrival rate \citep{cassidy1998bivariate}) or make assumptions about the relationship between equilibrium state and the coefficient of speed variation \citep{seo2019fundamental}. Second, although the speed near a certain density can be modeled as a Gaussian distribution, the parameters (mean and variance) vary with density \citep{helbing1997fundamentals,jabari2013stochastic}. Therefore, despite the popularity of LSE methods applied to LKV data for FD modeling, these drawbacks cannot be ignored.

Estimating the desired speed of traffic flow under a specific density is challenging due to the hidden nature of this variable, which can only be observed in rare instances of ideal equilibrium conditions. In most situations, traffic conditions are dynamic, and it takes time for the traffic speed to align with the corresponding desired speed, resulting in hysteresis \citep{laval2011hysteresis}. However, the relationship between the desired speed and the current speed can be deduced from the acceleration. When the desired speed associated with the anticipated density is lower than the current speed, negative acceleration occurs in the traffic flow, whereas positive acceleration is observed when the desired speed exceeds the current speed. Hence, acceleration/deceleration behavior solely depends on current and desired speeds in each sample independently. Thus, adding acceleration/deceleration behavior overcomes the dependence drawback mentioned above. It also naturally delineates stationary and non-stationary data in the macroscopic relation, thereby allowing for fitting of true equilibrium relations (i.e., FDs) to the data. 

In this paper, we propose a novel way to create samples by substituting the current density with the anticipated density and utilize the acceleration property as a label, to model the FD. We will refer to these samples as \textit{non-local density (K) speed (V)} samples, or NLKV for short. By doing so, we can frame the equilibrium relationship as a binary classification problem, using the sign of the acceleration label. Additionally, we propose an enhanced cross-entropy (ECE) loss as the optimization objective to minimize and fit the FD model parameters using the NLKV samples.

The rest of this paper is organized as follows. \autoref{Sect2} provides a review of the relevant research on the estimation of empirical samples, FD models, and optimization methods employed to fit the models. In \autoref{Sect3}, we present the assumptions of this study and outline the structure of NLKV samples and the ECE loss. \autoref{Sect4} describes the method for creating the NLKV samples from trajectories, including the estimation of continuous speed, density, and acceleration fields of the trajectories, as well as the anticipated density for each spatio-temporal location. The NLKV samples fitted with the ECE loss (NLKV+ECE) can be regarded as a parallel approach to fitting the LKV samples using LSE (LKV+LSE). \autoref{Sect5} compares the two approaches, examining the properties of the samples themselves and the fitting results with FD models. Finally, \autoref{Sect6} concludes this paper and provides additional discussions.

\section{Related Research}\label{Sect2}
The development of FDs of real-world traffic involves the use of local samples created from field-collected data within small spatio-temporal locations, including estimated flow rates ($q$), densities ($k$) and macroscopic speeds ($v$). Each $(v,k,q)$ tuple represents a sample point in empirical $v-k$ (or $q-k$ and $q-v$) plots. When point sensors like loop detectors are used to collect the field data, the spatial extent is defined by the detectors themselves. One estimates flow rates from the traffic counts, densities from sensor occupancies \citep{PAPAGEORGIOU20081}, and speed from individual vehicle speeds \citep{treiber2013traffic} (or utilizing the relation $q = k v$). In the case of trajectory data, the $(v,k,q)$ values for each spatio-temporal location can be estimated using Edie’s method \citep{edie1963discussion}.

Macroscopic relations can be classified into two types: physical and empirical models. Physical models involve the development of theories of traffic dynamics, such as car-following behavior \citep{gazis1961nonlinear,newell1961nonlinear,del1995functional,ZHENG2023104276} and continuous fluid approximations \citep{greenberg1959analysis}, to establish functional forms for the macroscopic relations. This approach is advantageous as it allows traffic flow theories to be tested against empirical observations. Physical models also allow for the development of FD relationships (i.e., macroscopic relations of equilibrium traffic states). On the other hand, empirical models formulate analytical expressions for macroscopic relationships with free parameters that are adjusted to provide a best fit to the empirical samples. These free parameters are then interpreted in terms of traffic flow properties, such as free-flow speed, capacity, and jam density, to provide the analytical expression with phenomenological meaning (e.g., \cite{greenshields1935study,underwood1961speed,drake1965statistical}). A subclass of empirical models uses data-mining methods (e.g., \cite{sun2014data,hadiuzzaman2018adaptive,shi2021physics,bramich2023fitfun}), sometimes referred to as non-parametric models, where the free parameters generally lack interpretability in the context of traffic flow properties. However, when provided with sufficient empirical samples, these methods can demonstrate remarkable flexibility and fitting performance. In a study conducted by \cite{9703273}, it was observed that Sun's empirical-based model \citep{sun2014data} exhibited the best fitting performance among the 50 models using loop detector data from 25 cities. This finding underscores the potential and efficacy of empirical models.

Most techniques utilize a variant of LSE to fit macroscopic models to empirical samples. This approach involves determining the model parameters by minimizing the sum of squared estimation errors. The accuracy of the fitting result is influenced by the distribution of the empirical samples. \cite{qu2015fundamental} acknowledged the existence of different speed distributions at various densities and proposed a density-weighted optimization function. Traditionally, the error distribution is assumed to follow a Gaussian distribution. However, it is now recognized that this assumption is insufficient for modeling LKV samples due to the complex nature of the observed noise, which can be attributed to various factors such as driver behavior, traffic dynamics, and hysteresis effects \citep{9703273}.

\section{Preliminary and Methodology}\label{Sect3}

\subsection{The equilibrium state and NLKV samples}\label{Sect3.1} 
\begin{figure}[!ht]
  \begin{subfigure}{.48\textwidth}
    \centering
    \caption{deceleration}
    \includegraphics[width=.7\linewidth]{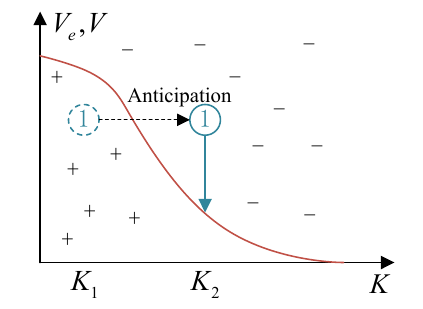}
    \label{Fig2a}
  \end{subfigure}
  \begin{subfigure}{.48\textwidth}
    \centering
    \caption{acceleration}
    \includegraphics[width=.7\linewidth]{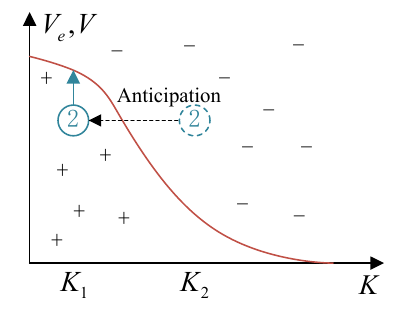}
    \label{Fig2b}
  \end{subfigure}
  \caption{Illustration of speed adaptation and anticipation: (a) deceleration; (b) acceleration}
  \label{Fig2}
\end{figure}

Equilibrium states in traffic are stationary states (i.e. when the vehicles are not accelerating or decelerating). Classically, one also assumes homogeneous drivers (e.g., identical reaction times, identical desired speeds, and identical safety distances). Our focus is on the former: stationary states. 
Naturally, the desired speed associated with an increasing density follows a monotonically decreasing and continuous function 
\citep{9703273,liu2023linearizing,van2015genealogy}. As drivers anticipate the traffic state they are about to enter, they assign an anticipated density to the current traffic state. Subsequently, this anticipated density determines the corresponding desired speed (modeled by an FD). As a result, drivers proactively adjust their current speed to approach the desired speed. This adjustment is manifested as follows: if the current speed exceeds the desired speed corresponding to the anticipated density, the driver decelerates (\autoref{Fig2a}). Conversely, they accelerate if the current speed falls below the desired speed (\autoref{Fig2b}). In cases where the current speed is close to the desired speed, slight acceleration or deceleration may occur due to random factors.

We propose a novel way to create samples (NLKV samples) from trajectory data, considering non-localities, to model the equilibrium FD relationship. The NLKV samples have anticipated density ($k^{\an}$) and current speed ($v$) as variables and their acceleration/deceleration state (0 if accelerating, 1 if decelerating) as labels. \autoref{table1} provides the comparison of the structural differences between the LKV and NLKV samples. 
\begin{table}[!ht]
\centering
\caption{Data form of LKV and NLKV samples}\label{table1}
\begin{tblr}{
  cells = {c},
  cell{1}{1} = {r=2}{},
  cell{1}{2} = {c=2}{},
  cell{1}{4} = {c=3}{},
  hline{1,3,6} = {-}{},
  hline{2} = {2-6}{},
}
Sample                          & LKV     &       & NLKV                &       &                            \\
                                & Density & Speed & Anticipated density & Speed & Acceleration sign (0 or 1) \\
{Spatio-temporal\\location 1}   & $k_1$   & $v_1$ & $k^{\an}_1$           & $v_1$ & $y_1$                      \\
$\vdots$                           & $\vdots$   & $\vdots$ & $\vdots$               & $\vdots$ & $\vdots$                      \\
{Spatio-temporal\\location~$n$} & $k_n$   & $v_n$ & $k^{\an}_n$           & $v_n$ & $y_n$                      
\end{tblr}
\end{table}

The following assumptions serve as guiding principles in our proposed model:
\begin{enumerate}[1)]
  \item The anticipated density $k^{\an}$ is determined by the current traffic state. Drivers accurately anticipate $k^{\an}$ based on their current speed $v$, downstream density, and a transition time $t^{\mathrm{m}}$.
  \item The desired speed $v^{\an}$ corresponding to $k^{\an}$ exists and is determined by the function $v^{\an}=f(k^{\an})$, where $f(\cdot)$ is the $v-k$ relationship to be determined.
  \item The desired speed $v^{\an}$ corresponding to the anticipated density $k^{\an}$ is adopted as the target for the current traffic to adjust its speed accordingly.
  \item The density variation from $k$ to $k^{\an}$ is monotone, either increasing or decreasing.
\end{enumerate}

Assumption 1) states that drivers possess accurate anticipation skills, enabling them to estimate $k^{\an}$ correctly. We divide space and time into discrete intervals. For a specific space interval $j$ and time interval $i$, $k^{\an}$ is estimated as the density of space interval $j+v t^{\mathrm{m}}$ and time interval $i+t^{\mathrm{m}}$. The parameter $t^{\mathrm{m}}$ is inspired by the definition of decision sight distance, which refers to the distance required to detect an unexpected source or hazard on a roadway, recognize its potential threat, select an appropriate speed and path, and safely and efficiently complete the required maneuver \citep{green2004policy}. Decision sight distance is estimated as $v_{\mathrm{des}} t^{\mathrm{m}}$, where $v_{\mathrm{des}}$ is the design speed \citep{mcgee1979decision}. It varies from 10.2 to 14.5 seconds depending on the type of road (i.e., rural, suburban, or urban roads) \citep{green2004policy}. 

Assumptions 2) and 3) reflect the definition of stationary states and the dynamical characteristics of traffic flow, respectively. With the monotone decreasing property of the $v-k$ relationship, assumption 4) ensures that the acceleration $a$ during the adjustment of traffic flow from the current speed $v$ to the anticipated desired speed $v^{\an}=f(k^{\an})$ maintains a consistent sign. If $v$ is perceived to be higher than $v^{\an}$ corresponding to $k^{\an}$, vehicles decelerate and continue to decelerate until they reach speed $v^{\an}$. If $v$ is perceived to be lower than $v^{\an}$, vehicles accelerate and continue to accelerate until they reach speed $v^{\an}$. During this process, no alternating acceleration and deceleration behaviors occur. This is summarized by the following condition:
%
\begin{equation}\label{Eq1}
\mathrm{sgn}~(v^{\an}-v) = \mathrm{sgn}~a,
\end{equation}
where $a$ represents the acceleration during the adjustment from $v$ to $v^{\an}$ and $\mathrm{sgn}$ is the sign function. 
According to the assumptions, the desired speed $v^{\an}$ under the anticipated density $k^{\an}$ satisfies
\begin{equation}\label{Eq2}
\left\{
\begin{aligned}
v^{\an} & < v \mbox{ if and only if }  a < 0 \\
v^{\an} & > v \mbox{ if and only if }  a > 0 \\
v^{\an} & = v \mbox{ if and only if }  a = 0 \\
\end{aligned}
\right.
.
\end{equation}
%
\autoref{Eq2} says that the desired speed of vehicles is higher than the current speed if they have positive acceleration. Conversely, if they are decelerating, the current speed is higher than the desired speed. At equilibrium, the acceleration is 0, signifying that there is no change in speed over time. In this state, the current speed matches the desired speed corresponding to the anticipated density.

By plotting the anticipated density against the current speed and labeling the data points with the corresponding sign of acceleration (referred to as the NLKV samples), a scatter plot can be generated to visually represent the relationship between anticipated density, current speed, and the sign of acceleration. The boundary in the scatter plot is determined by the anticipated density and its corresponding desired speed, which serves as a separator, dividing the samples into two distinct regions. The region above the boundary represents the samples where the current speed is higher than the desired speed (indicating negative acceleration), while the region below the boundary represents samples where the current speed is lower than the desired speed (indicating positive acceleration). By modeling the separation curve for samples with positive and negative accelerations, the desired speed corresponding to the anticipated density represented by the $v-k$ relationship can be obtained. The points on the curve indicate the scenario where acceleration is zero, representing a state of equilibrium.

However, it is important to note that the NLKV samples created in the field traffic may not exhibit ideal properties described above. In practical scenarios, the regions of acceleration and deceleration often overlap as the system approaches stationary. This overlap can be attributed to unobserved features (random features) and the heterogeneity of vehicles in the system.

Taking these factors into consideration, we further make the assumption that the probability of deceleration, denoted as $\pi$, is related to the difference between the current speed $v$ and the desired speed $v^{\an}$, following a logistic distribution. Specifically, we have
%
\begin{equation}\label{Eq3}
    p(y|z) = \pi^y (1-\pi)^{1-y} = \bigg( \frac{1}{1+e^{-z}} \bigg)^y \bigg( \frac{e^{-z}}{1+e^{-z}} \bigg)^{1-y},
\end{equation}
where $y$ represents the sign of acceleration, with $y=1$ indicating a negative acceleration in the current traffic flow, and $y=0$ representing a positive acceleration. The variable $z$ is the difference between the current speed and their desired speed under the anticipated density $k^{\an}$. \autoref{Eq3} implies that the deceleration probability $\pi<0.5$ if $z<0$, whereas $\pi>0.5$ if $z>0$. Additionally, if $v\ll v^{\an}$, then $\pi\to 0$. This indicates a very small probability of negative acceleration and a high probability of positive acceleration. 
Conversely, if $v\gg v^{\an}$, we have that $\pi\to 1$, i.e., an overwhelmingly high probability of deceleration. 

We now consider the $v-k$ relationship $v^{\an}=f(k^{\an};\vec{\theta})$, where $\vec{\theta}$, a vector, represents the parameters of $f$. The sign of acceleration given $z=v-v^{\an}$, a binary variable, naturally follows a binomial distribution with trial parameter 1 and success probability 
\begin{equation}
    \pi = \frac{1}{1+e^{-(v-f(k^{\an};\vec{\theta}))}}
\end{equation}
(i.e., a Bernoulli distribution): $\{y|f(k^{\an};\vec{\theta}),v\} \sim \mathrm{Binomial}(1,\pi)$. The probability density function of this distribution is given by
\begin{equation}\label{Eq4}
    p(y|f(k^{\an};\vec{\theta}),v) = \pi^y (1-\pi)^{1-y} = \bigg( \frac{1}{1+e^{-(v-f(k^{\an};\vec{\theta}))}} \bigg)^y \bigg( \frac{e^{-(v-f(k^{\an};\vec{\theta}))}}{1+e^{-(v-f(k^{\an};\vec{\theta}))}} \bigg)^{1-y}, ~ y \in \{0,1\}.
\end{equation}

\subsection{The loss function with NLKV samples}\label{Sect3.2}
In the last subsection, we have established the NLKV samples and determined the distribution of the sign of their acceleration $y$. The $v-k$ relationship can be modeled as a binary classification problem, with $y$ as the dependent variable. To estimate the parameters of a specific FD model, the likelihood function can be formulated assuming that $y$ follows a Bernoulli distribution with a deceleration probability $\pi$, that is $\{y|f(k^{\an};\vec{\theta}),v\}\sim \mathrm{Binomial}(1,\pi)$, based on the discussion above. From \autoref{Eq4}, the likelihood function, denoted as $L(\vec{\theta})$, is defined as the probability of observing the given sample (of size $m$) $\{k^{\an}_i,v_i,y_i\}_{i=1}^m$ conditioned on the model parameters $\vec{\theta}$ and the $v-k$ relationship $f(\cdot)$:
\begin{equation}
    L(\vec{\theta}) := \prod_{i=1}^m \bigg( \frac{1}{1+e^{-(v_i-f(k^{\an}_i;\vec{\theta}))}} \bigg)^{y_i} \bigg( \frac{e^{-(v_i-f(k^{\an}_i;\vec{\theta}))}}{1+e^{-(v_i-f(k^{\an}_i;\vec{\theta}))}} \bigg)^{1-y_i}.
\end{equation}

By maximizing the likelihood function with respect to the parameter $\vec{\theta}$, we can estimate the model parameters that best fit the observed samples and the FD model using maximum likelihood estimation (MLE). This is equivalent to minimizing the negative log-likelihood function, that is, solving the following optimization problem:
\begin{align}\label{Eq6}
    \underset{\vec{\theta}}{\mathrm{minimize}}~ - \log \big(L(\vec{\theta})\big) &= \sum_{i=1}^m \bigg( y_i \log \Big( 1+e^{-(v_i-f(k^{\an}_i;\vec{\theta}))} \Big) + (1-y_i) \big(v_i-f(k^{\an}_i;\vec{\theta})\big) + (1-y_i) \log \Big( 1+e^{-(v_i-f(k^{\an}_i;\vec{\theta}))} \Big) \bigg) \nonumber \\
    &= \sum_{i=1}^m \bigg( (1-y_i) \big(v_i-f(k^{\an}_i;\vec{\theta})\big) + \log \Big( 1+e^{-(v_i-f(k^{\an}_i;\vec{\theta}))} \Big) \bigg).
\end{align}

\autoref{Eq6} sums the cross-entropy loss of each sample and can, therefore, be influenced by bias in the sample of acceleration and deceleration data (as can be seen in the second line after canceling similar terms). To address this issue, we enhance the objective function by introducing sample rate weighting. Additionally, we aim to increase the convergence speed towards the minimum of \autoref{Eq6} and impose a greater penalty in case of a false classification when there is a significant absolute distance between the current speed and the anticipated desired speed. Therefore, we propose to minimize the following enhanced cross entropy (ECE) loss to estimate the parameters of the FD model: 
\begin{equation}\label{Eq7}
    \underset{\vec{\theta}}{\mathrm{minimize}}~ \frac{1}{m}\sum_{i=1}^m \bigg( \omega y_i \log \Big( 1+e^{-(v_i-f(k^{\an}_i;\vec{\theta}))} \Big) + (1-\omega) (1-y_i) \big(v_i-f(k^{\an}_i;\vec{\theta})\big) + (1-\omega) (1-y_i) \log \Big( 1+e^{-(v_i-f(k^{\an}_i;\vec{\theta}))} \Big) \bigg),
\end{equation}
%
where 
\begin{equation}
    \omega :=\frac{\# \left\{ y_i|y_i=0 \right\} }{\# \left\{ y_i \right\} } = \frac{\# \left\{ y_i|y_i=0 \right\} }{m }
\end{equation}
represents the fraction of positive accelerations in the sample. 
We also divided the loss by the number of samples $m$, which does not affect the optimization problem but allows us to interpret the loss function obtained in this way.


\section{NLKV Samples Estimated from Trajectories}\label{Sect4}

In order to estimate the NLKV samples from trajectories, it is necessary to first estimate the macroscopic fields for speed, density, and acceleration. These fields should exhibit a sufficient level of smoothness to accurately capture the overall patterns and trends in the traffic flow. Subsequently, the anticipated density for each spatio-temporal state can be obtained based on the previously estimated macroscopic fields. The anticipated density represents the density that vehicles expect to reach at each specific spatio-temporal location, considering their current speeds. Finally, the estimated anticipated density can be paired with the corresponding speed and sign of acceleration for each state of the traffic flow, enabling the formation of the NLKV samples.

\subsection{Estimation of the speed, density, and acceleration fields}\label{Sect4.1} 

\begin{figure}[!ht]
  \centering
  \includegraphics[width=0.45\textwidth]{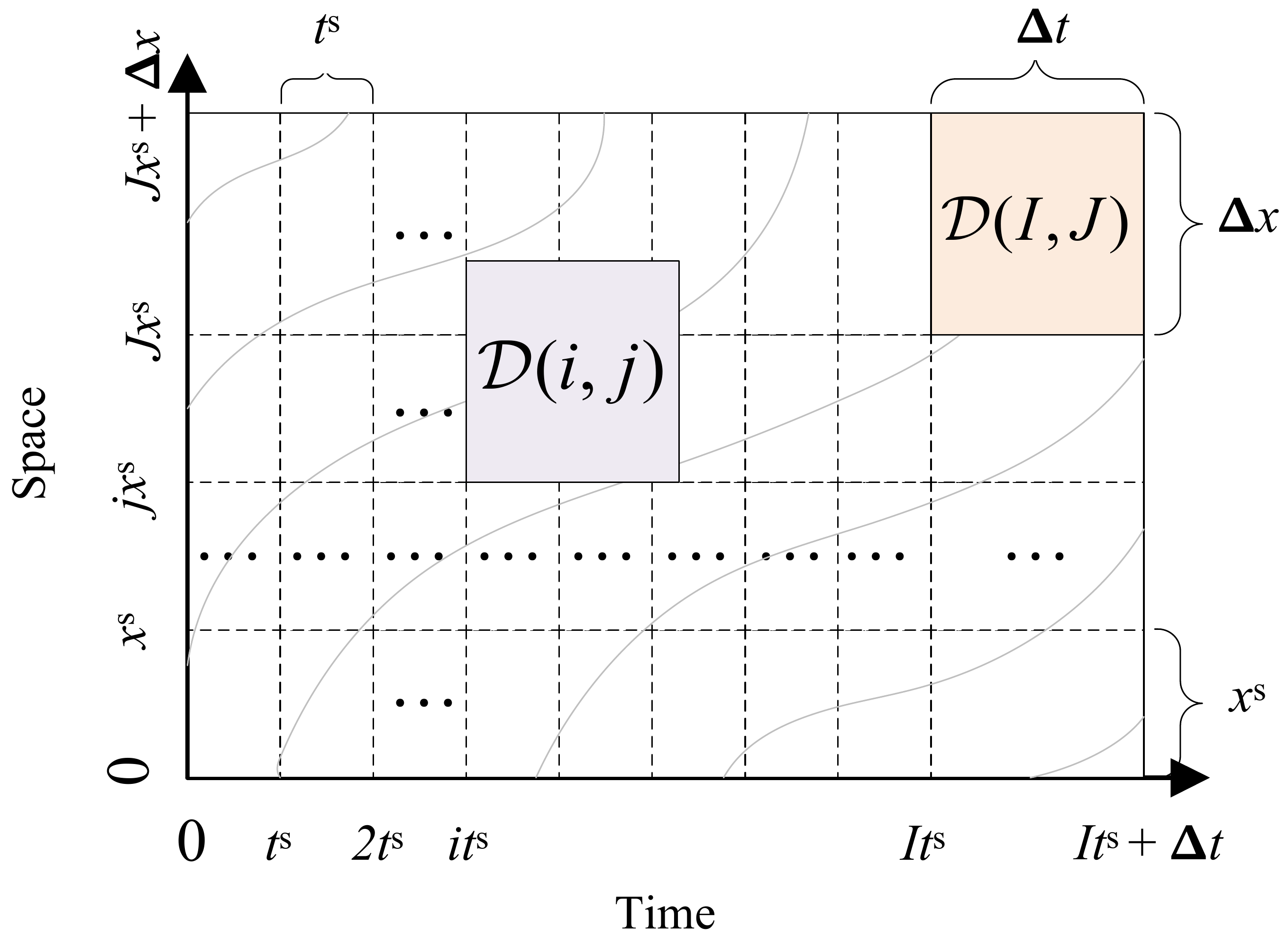}
  \caption{Discretization of the spatio-temporal domain}
  \label{Fig3}
\end{figure}

We denote the spatio-temporal domain by $\mathcal{D}=\mathcal{X}\times\mathcal{T}$ representing a road section $\mathcal{X}$ and time interval $\mathcal{T}$, which we divide into homogeneous spatio-temporal intervals $\mathcal{D}(i,j)$, where $i\in \{0,...,I\}$, $j\in \{0,...,J\}$, and $i$,$j$ are integers, as shown in \autoref{Fig3}. The intervals $\mathcal{D}(i,j)$ are $\Delta x$ long in the space dimension and $\Delta t$ long in the time dimension. We consider sliding step sizes to capture non-localities: we denote the \emph{sliding} space and time step sizes by 
$x^{\st} \ll \Delta x$ and $t^{\st} \ll \Delta t$, respectively. Thus, the spatio-temporal interval $\mathcal{D}(i,j)$ covers $i t^{\st}$ to $i t^{\st}+\Delta t$ in time and $j x^{\st}$ to $j x^{\st}+\Delta x$ in space. With the sliding intervals, we can discretize the entire spatio-temporal domain $\mathcal{D}$ into $I\times J$ subdomains with overlaps, where 
\begin{equation}
    I=\left\lfloor \frac{T-\Delta t}{t^{\st}} \right\rfloor, ~ J=\left\lfloor \frac{X-\Delta x}{x^{\st}} \right\rfloor, 
\end{equation}
and $T := |\mathcal{T}|$ and $X:=|\mathcal{X}|$. Due to the overlapping spatio-temporal subdomains, there are partially duplicated trajectories between $\mathcal{D}(i,j)$ and $\mathcal{D}(i+1,j)$ as well as between $\mathcal{D}(i,j)$ and $\mathcal{D}(i,j+1)$. This enhances the smoothness of macroscopic traffic flow parameters estimated from subdomains in both time and space. As $t^{\st}$ and $x^{\st}$ approach zero in the limit, a continuous field of traffic flow states can be obtained within the corresponding $\{\mathcal{D}(i,j)\}$.

According to the general definition proposed by \cite{edie1963discussion}, the macroscopic parameters $k_{i,j}$, and $v_{i,j}$ of each subdomain $\mathcal{D}(i,j)$ can be calculated as
\begin{equation}\label{Eq8}
k_{i,j}=\frac{\sum_{n=1}^{N_{i,j}}t_n}{A_{i,j}},
\end{equation}
and
\begin{equation}\label{Eq9}
v_{i,j}=\frac{\sum_{n=1}^{N_{i,j}}x_n}{\sum_{n=1}^{N_{i,j}}t_n},
\end{equation}
%
where $N_{i,j}$ is the total number of vehicles in $\mathcal{D}(i,j)$. $x_n$ and $t_n$ are the travel distance and travel time of vehicle $n$ in $\mathcal{D}(i,j)$, respectively. $A_{i,j} = \Delta x \Delta t$ is the area of $\mathcal{D}(i,j)$. 
The speed field $\mathcal{V}\in \mathbb{R}^{I\times J}$ and density field $\mathcal{K}\in \mathbb{R}^{I\times J}$ can be obtained by applying \autoref{Eq8} and \autoref{Eq9} to each subdomain: $\mathcal{V}(i,j):=v_{i,j}$, $\mathcal{K}(i,j):=k_{i,j}$.

The acceleration $a_{i,j}$ in $\mathcal{D}(i,j)$ can be approximated as
\begin{equation}\label{Eq10}
a_{i,j}\approx \frac{v_{i+1,j+b}-v_{i,j}}{t^{\st}},
\end{equation}
%
where 
\begin{equation}
    b:=\left\lfloor \frac{v_{i,j} t^{\st}}{x^{\st}} \right\rfloor
\end{equation}
is the estimated number of spatial intervals traversed at the current speed within $t^{\st}$ units of time. 
\autoref{Eq10} assumes that the speed change between two adjacent temporal steps is linear. The acceleration field obtained by traversing all $\mathcal{D}(i,j)$ is represented by the acceleration field $\mathcal{A}\in \mathbb{R}^{I\times J}$, which is given by
\begin{equation}\label{Eq11}
\mathcal{A}\left( i,j \right)=
\begin{cases}
    a_{i,j} & ~ i\le I-1, j\le J-b \\
    \emptyset & \mbox{otherwise}
\end{cases}.
\end{equation}
%
Finally, we denote by $y(i,j)$ the sign of acceleration of subarea $\mathcal{D}(i,j)$ i.e., 
\begin{equation}\label{Eq12}
y(i,j)=
\begin{cases}
    0 & \mathcal{A}(i,j) > 0 \\
    1 & \mathcal{A}(i,j) < 0 \\
    \emptyset & \mathcal{A}(i,j) = \emptyset
\end{cases}.
\end{equation}

\subsection{Estimation of the anticipated density}\label{Sect4.2} 

As per assumption 1), the anticipated density $k^{\an}$ of $\mathcal{D}(i,j)$ is estimated as $k_{i+t^{\mathrm{m}},j+v t^{\mathrm{m}}}$, i.e., we have a shift in indices given as $i \mapsto i+t^{\mathrm{m}}$ and $j \mapsto j+v t^{\mathrm{m}}$. To align with the discrete spatio-temporal location, the anticipated density is adjusted as
\begin{equation}\label{Eq13}
k^{\an}_{i,j}=k_{i+\left\lfloor \frac{t^{\mathrm{m}}}{t^{\st}} \right\rfloor t^{\st},j+\left\lfloor \frac{v_{i,j} t^{\mathrm{m}}}{x^{\st}} \right\rfloor x^{\st}}.
\end{equation}
That is, $k^{\an}_{i,j} = k_{i',j'}$, where $i' := i+\left\lfloor t^{\mathrm{m}} / t^{\st} \right\rfloor t^{\st}$ and  $j' := j+\left\lfloor v_{i,j} t^{\mathrm{m}} / x^{\st} \right\rfloor x^{\st}$. 
Consequently, the \emph{anticipated densities} field, $\mathcal{K}^{\an}\in \mathbb{R}^{I\times J}$, is 
\begin{equation}\label{Eq14}
\mathcal{K}^{\an}(i,j)=
\begin{cases}
    k^{\an}_{i,j} & i\le I-\left\lfloor \dfrac{t^{\mathrm{m}}}{t^{\st}} \right\rfloor t^{\st},\ j\le J-\left\lfloor \dfrac{v_{i,j} t^{\mathrm{m}}}{x^{\st}} \right\rfloor x^{\st} \\
    \emptyset  & \mbox{otherwise}
\end{cases}.
\end{equation}

With these definitions, we obtain a complete NLKV sample $\left\{ \mathcal{K}^{\an}(i,j),\mathcal{V}(i,j),y(i,j) \right\}$ for the entire spatio-temporal domain. 
We only samples from domain intervals $\mathcal{D}(i,j)$ with $i\le \min\left( I-1,I-\left\lfloor t^{\mathrm{m}} / t^{\st} \right\rfloor t^{\st} \right)$ and $j\le \min\left( J-b,J-\left\lfloor v t^{\mathrm{m}} / x^{\st} \right\rfloor x^{\st} \right)$ for FD modeling.


\section{Comparison between the Fitting Approaches of LKV+LSE and NLKV+ECE}\label{Sect5}
This section provides a systematic comparison between fitting FDs using (classical) LKV samples and the proposed NLKV samples. For the former, we use traditional LSE; for the latter, we employ the ECE minimization method developed in \autoref{Sect3.2}. 
We introduce the field trajectory datasets used to perform the comparisons in \autoref{Sect5.1}. Moving on to \autoref{Sect5.2}, we estimate the LKV and NLKV samples for different trajectory datasets. We then compare the distribution of LKV and NLKV features for a specific dataset. \autoref{Sect5.3} introduces our FD models and the LSE optimization model used for FD fitting. The fitting results of the LKV+LSE and NLKV+ECE approaches with the same FD model are presented in \autoref{Sect5.4}. Finally, in \autoref{Sect5.5}, we compare the fitting results of different FD models using the LKV+LSE and NLKV+ECE approaches.

\subsection{The field trajectory dataset}\label{Sect5.1}

This study utilizes four trajectory datasets obtained from different freeway road links. The datasets are as follows:

\begin{itemize}
    \item Dataset 1: This dataset is obtained from the Hanshin Expressway link \#11, which was collected in the Zen Traffic Data (ZTD) source \citep{ZTD}. In the ZTD dataset, link \#11 is labeled as L001. Dataset 1 specifically corresponds to the trajectory dataset L001F001, representing one hour of data collection for this particular link.
    \item Dataset 2: This dataset is obtained from the Hanshin Expressway link \#4, also collected from the ZTD source. In the ZTD dataset, link \#4 is labeled as L002. Dataset 2 corresponds to the trajectory dataset L002F001, representing one hour of data collection for this particular link.
    \item Dataset 3: This dataset is obtained from the Next Generation Simulation Program (NGSIM, \cite{NGSIM}), specifically collected from the US101 link.
    \item Dataset 4: Trajectories from the NGSIM, specifically collected from the I80 link.
\end{itemize}

\begin{figure}[!ht]
  \begin{subfigure}{.33\textwidth}
    \centering
    \caption{Dataset 1: $\mathcal{V}$}
    \includegraphics[width=1\linewidth]{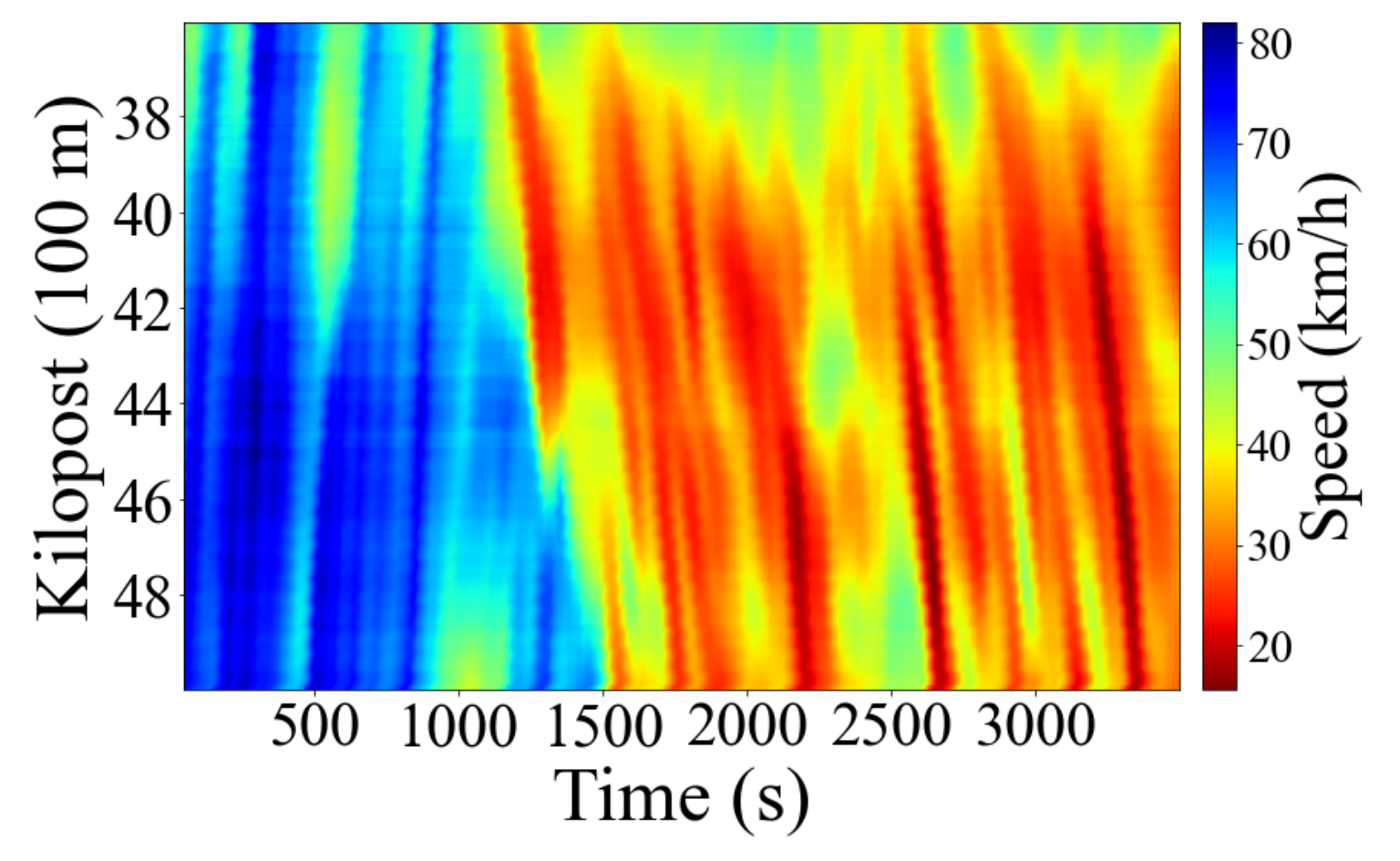}
    \label{Fig4a}
  \end{subfigure}
  \begin{subfigure}{.33\textwidth}
    \centering
    \caption{Dataset 1: $\mathcal{K}$}
    \includegraphics[width=1\linewidth]{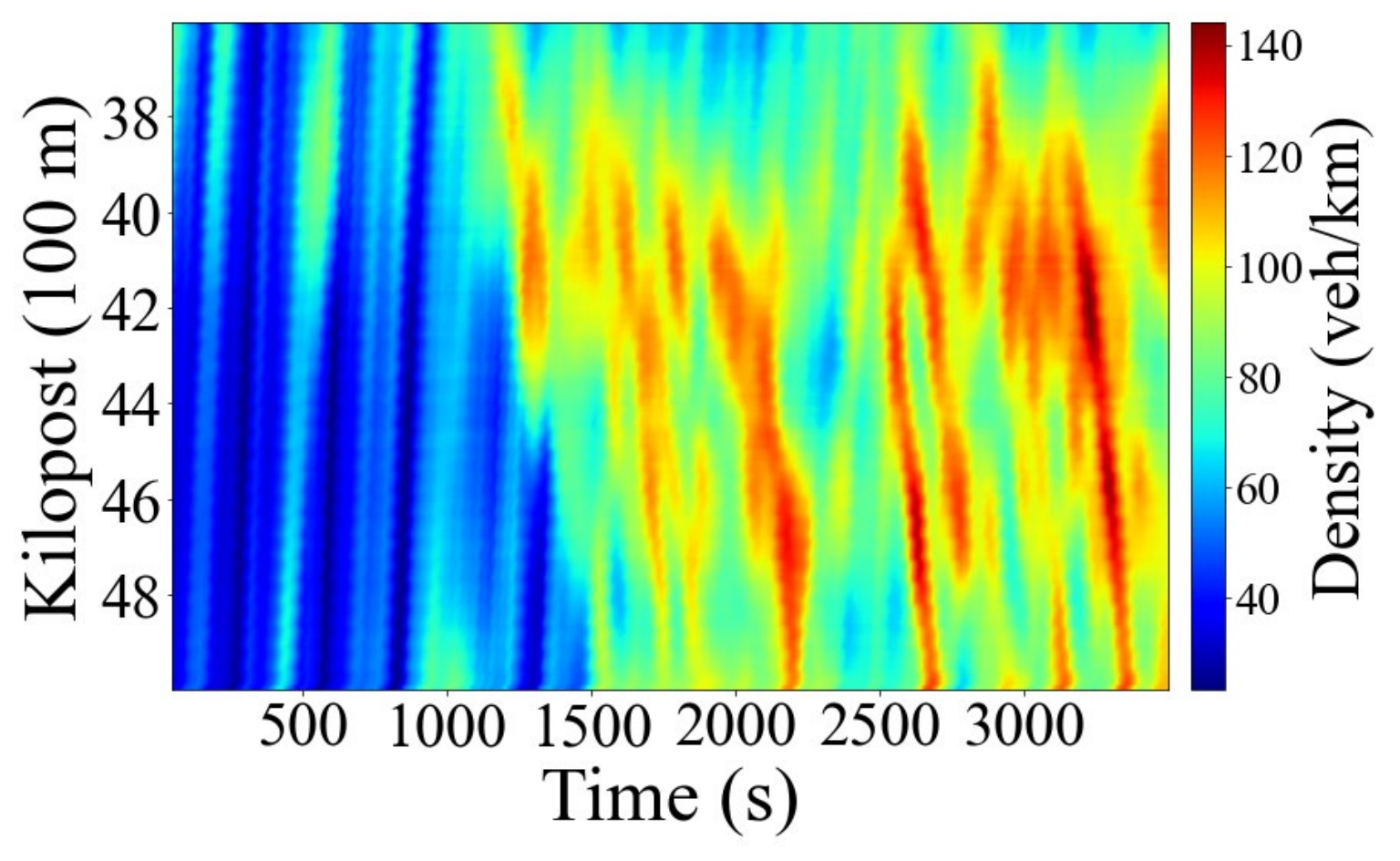}
    \label{Fig4b}
  \end{subfigure}
  \begin{subfigure}{.33\textwidth}
    \centering
    \caption{Dataset 1: $\mathcal{A}$}
    \includegraphics[width=1\linewidth]{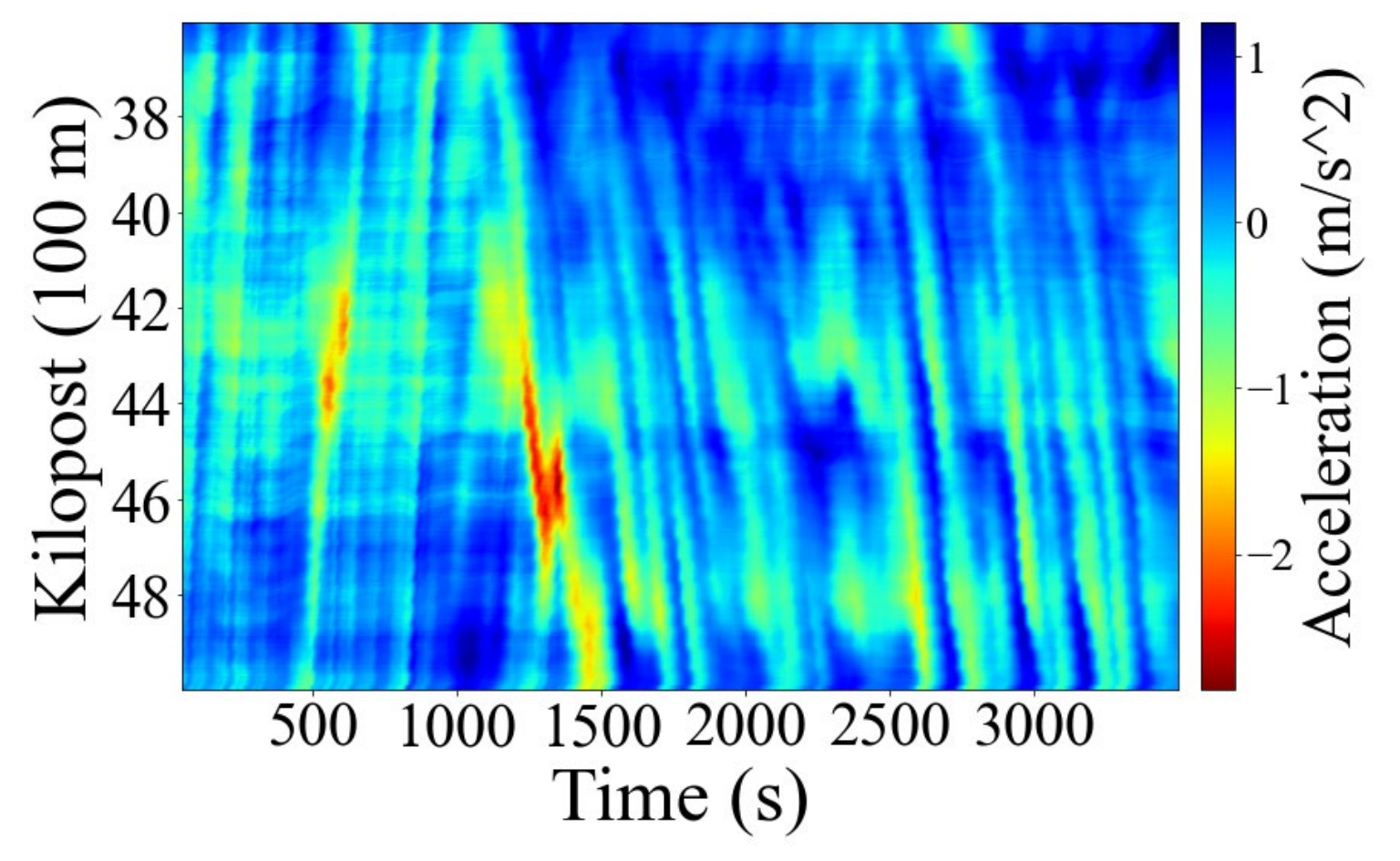}
    \label{Fig4c}
  \end{subfigure}
  \begin{subfigure}{.33\textwidth}
    \centering
    \caption{Dataset 2: $\mathcal{V}$}
    \includegraphics[width=1\linewidth]{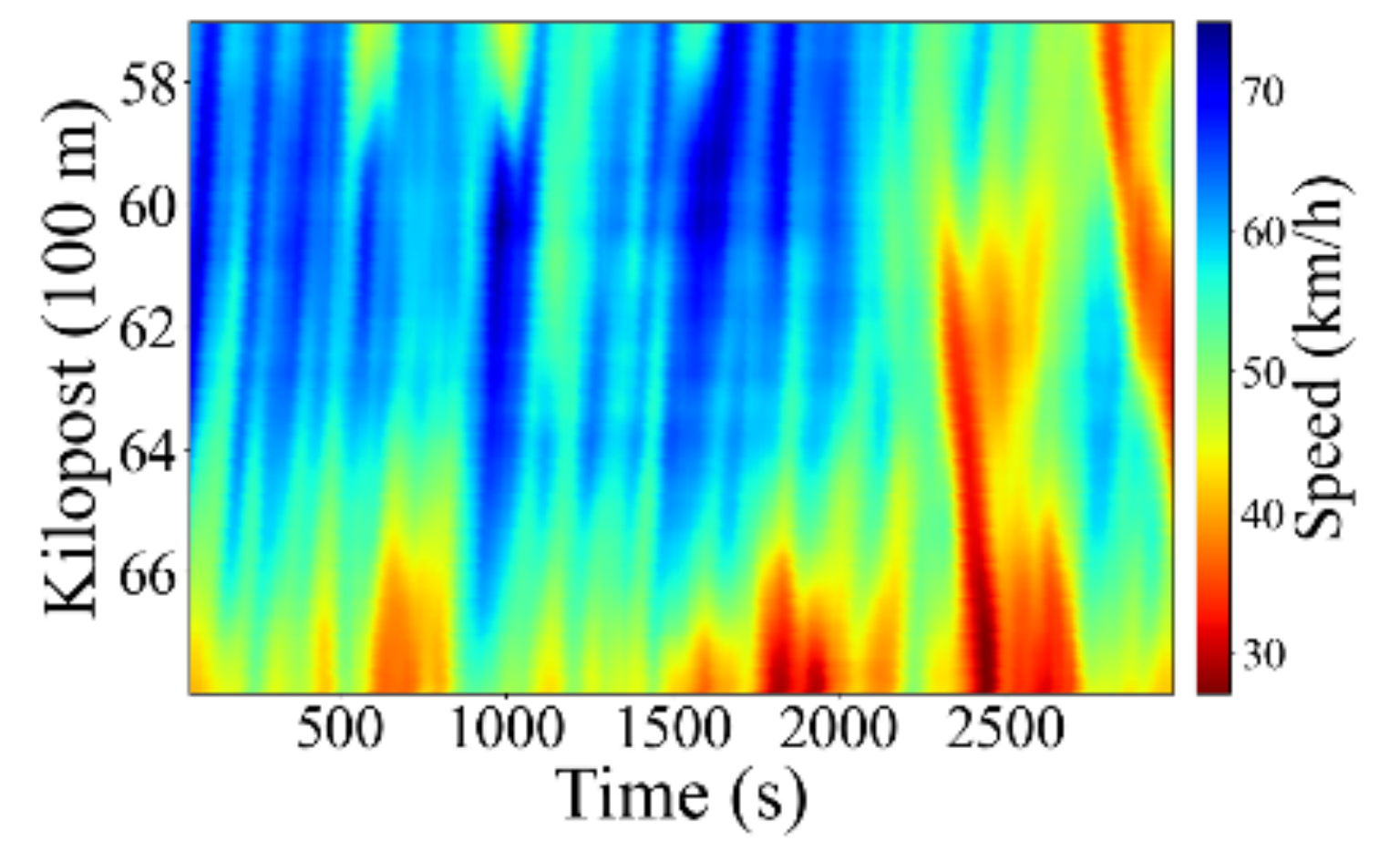}
    \label{Fig4d}
  \end{subfigure}
  \begin{subfigure}{.33\textwidth}
    \centering
    \caption{Dataset 2: $\mathcal{K}$}
    \includegraphics[width=1\linewidth]{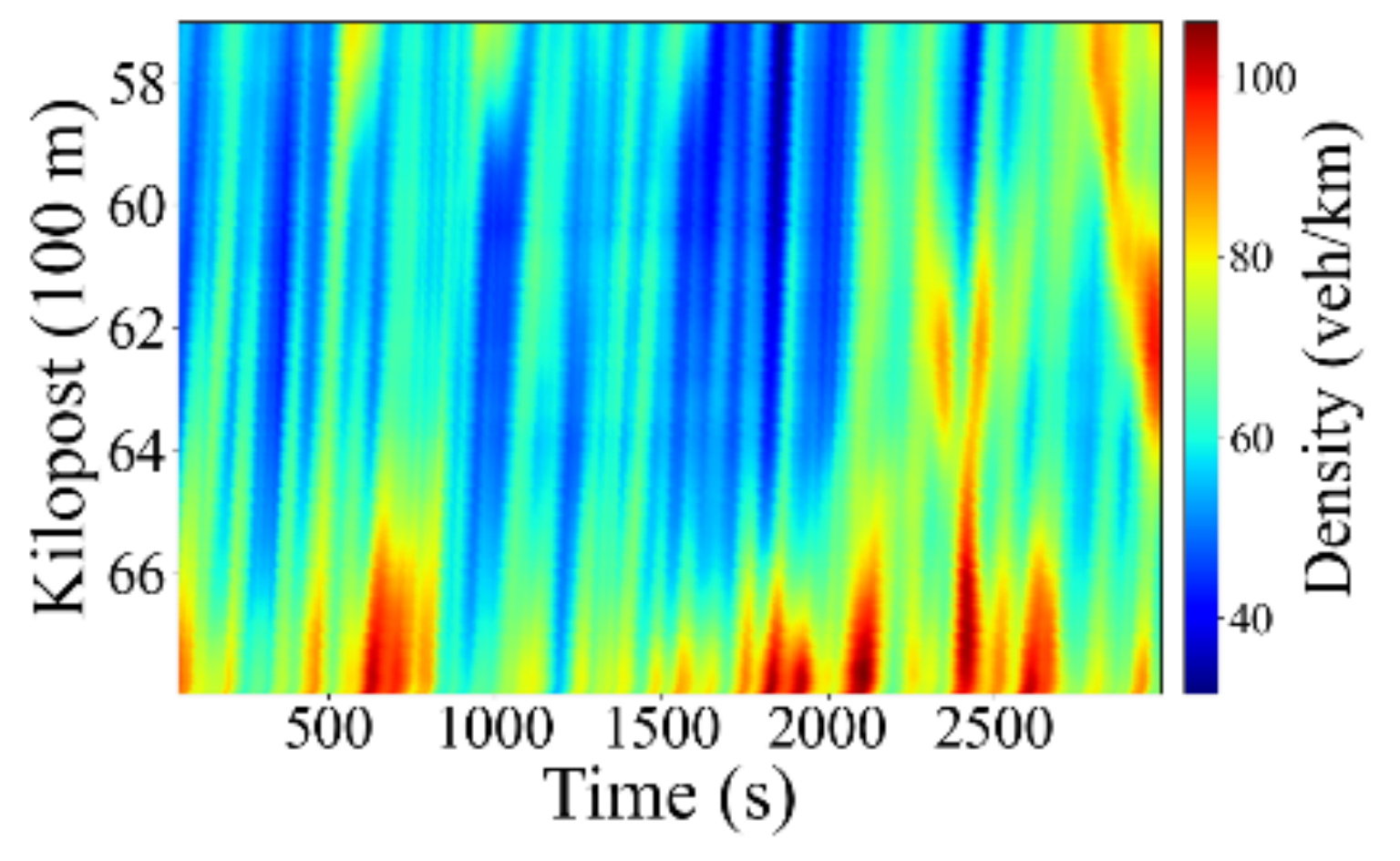}
    \label{Fig4e}
  \end{subfigure}
  \begin{subfigure}{.33\textwidth}
    \centering
    \caption{Dataset 2: $\mathcal{A}$}
    \includegraphics[width=1\linewidth]{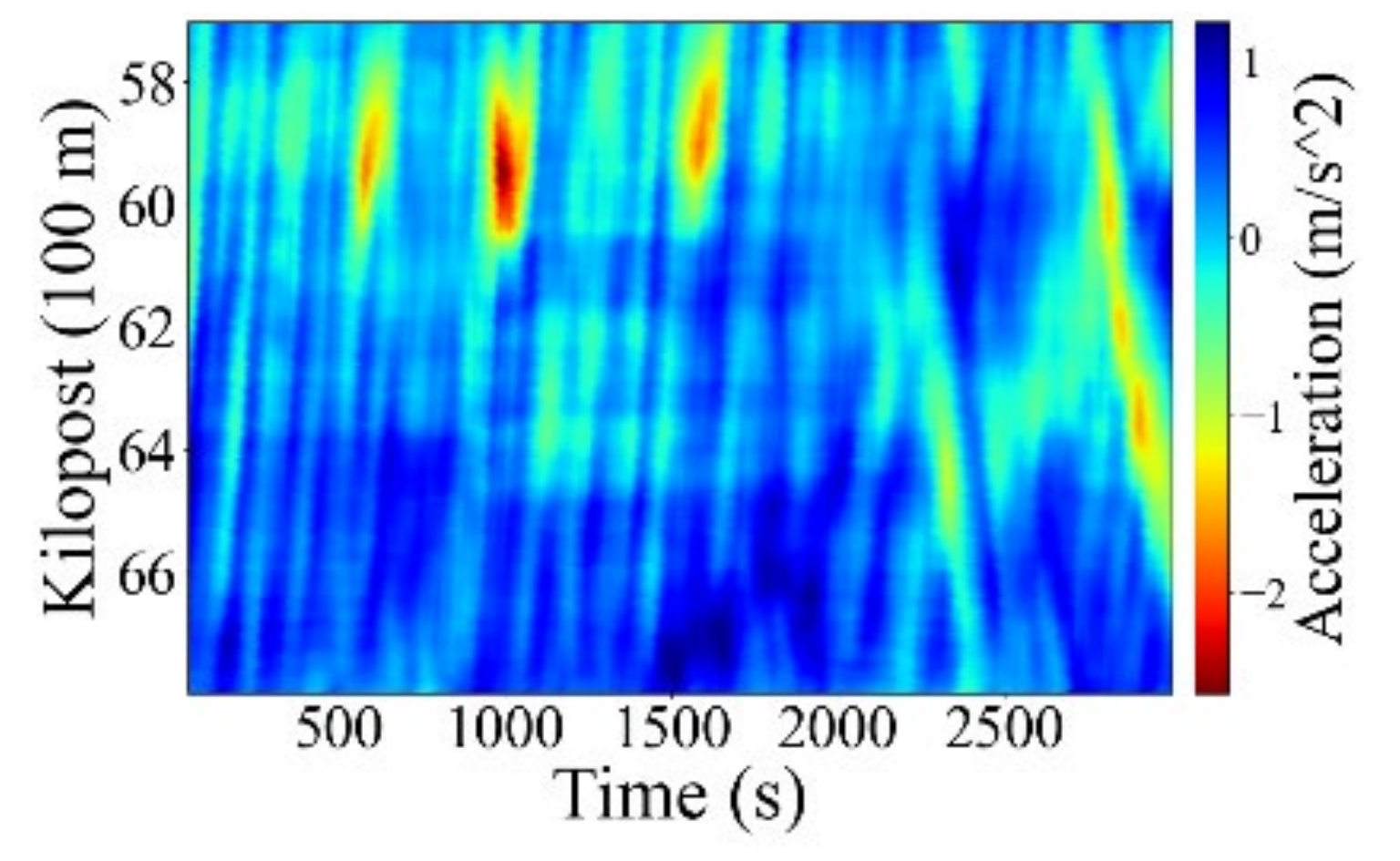}
    \label{Fig4f}
  \end{subfigure}
  \begin{subfigure}{.33\textwidth}
    \centering
    \caption{Dataset 3: $\mathcal{V}$}
    \includegraphics[width=1\linewidth]{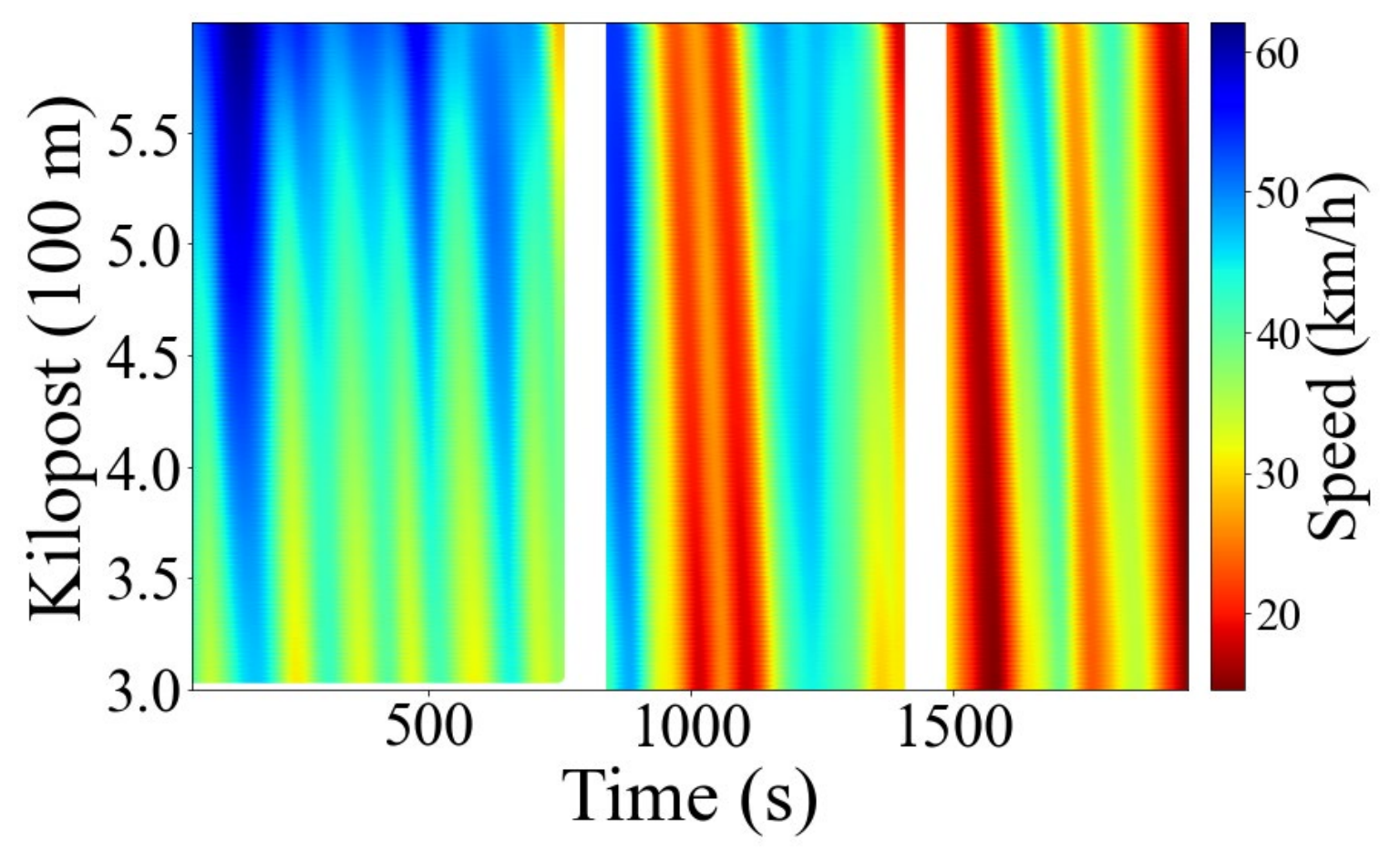}
    \label{Fig4g}
  \end{subfigure}
  \begin{subfigure}{.33\textwidth}
    \centering
    \caption{Dataset 3: $\mathcal{K}$}
    \includegraphics[width=1\linewidth]{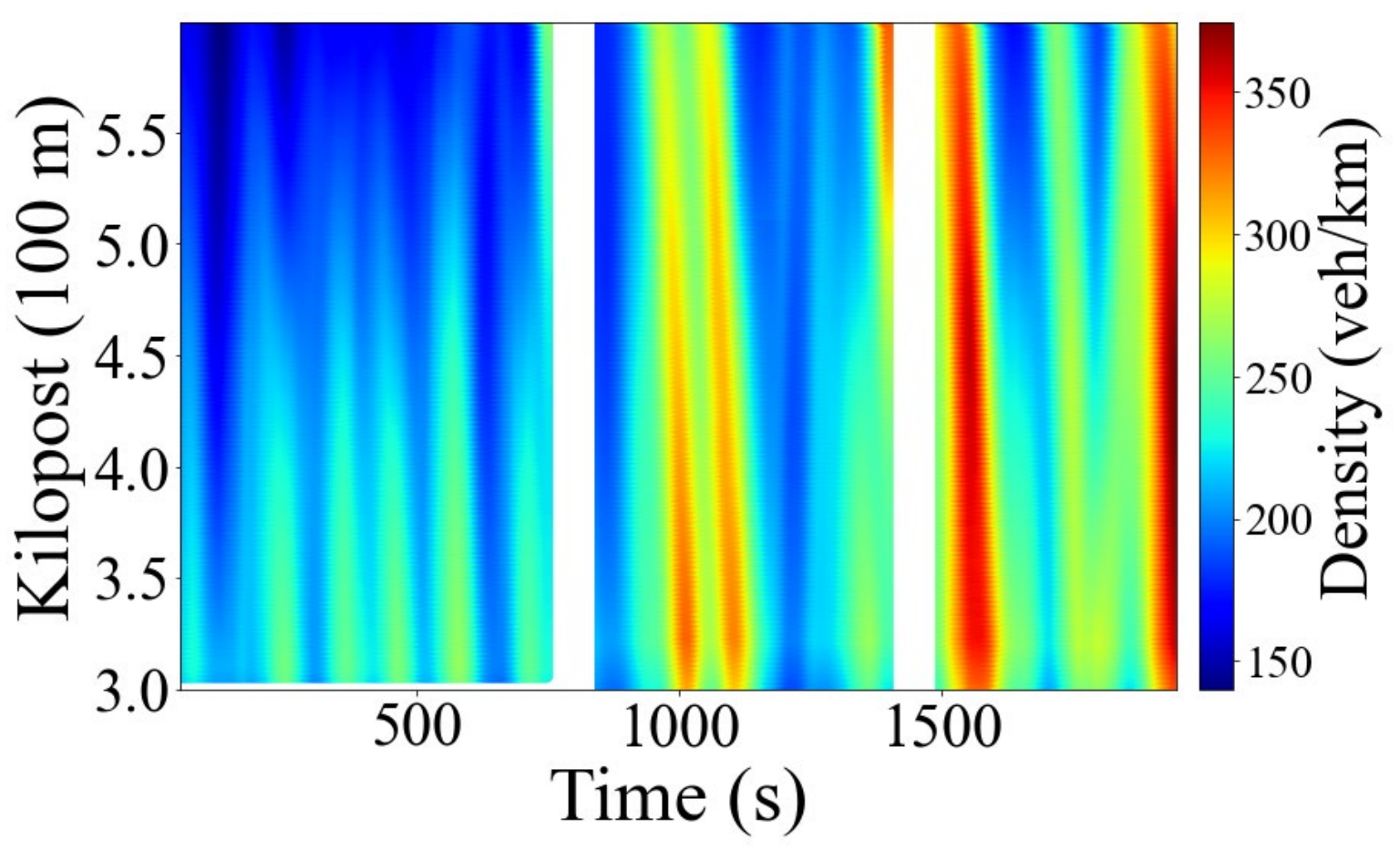}
    \label{Fig4h}
  \end{subfigure}
  \begin{subfigure}{.33\textwidth}
    \centering
    \caption{Dataset 3: $\mathcal{A}$}
    \includegraphics[width=1\linewidth]{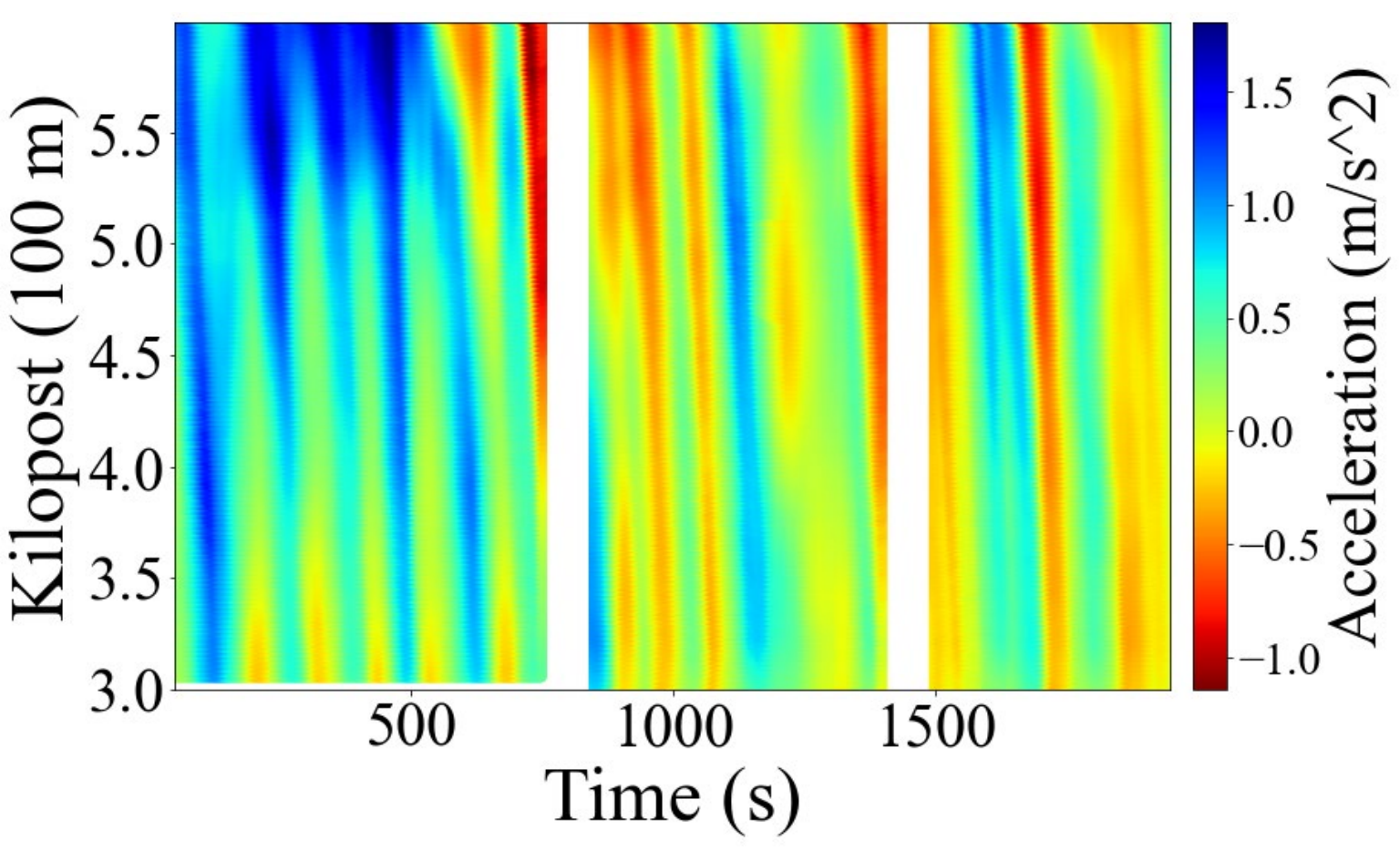}
    \label{Fig4i}
  \end{subfigure}
  \begin{subfigure}{.33\textwidth}
    \centering
    \caption{Dataset 4: $\mathcal{V}$}
    \includegraphics[width=1\linewidth]{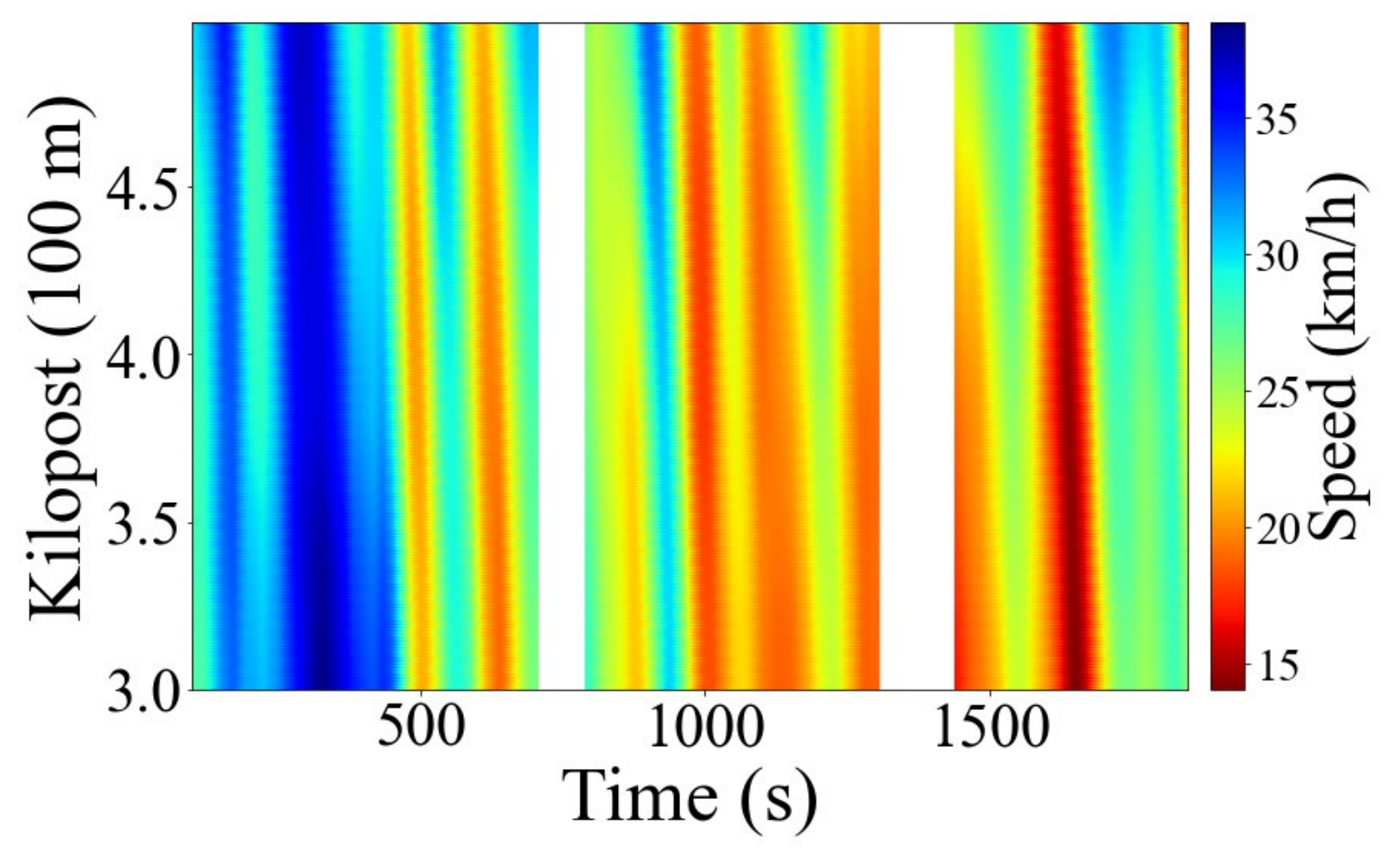}
    \label{Fig4j}
  \end{subfigure}
  \begin{subfigure}{.33\textwidth}
    \centering
    \caption{Dataset 4: $\mathcal{K}$}
    \includegraphics[width=1\linewidth]{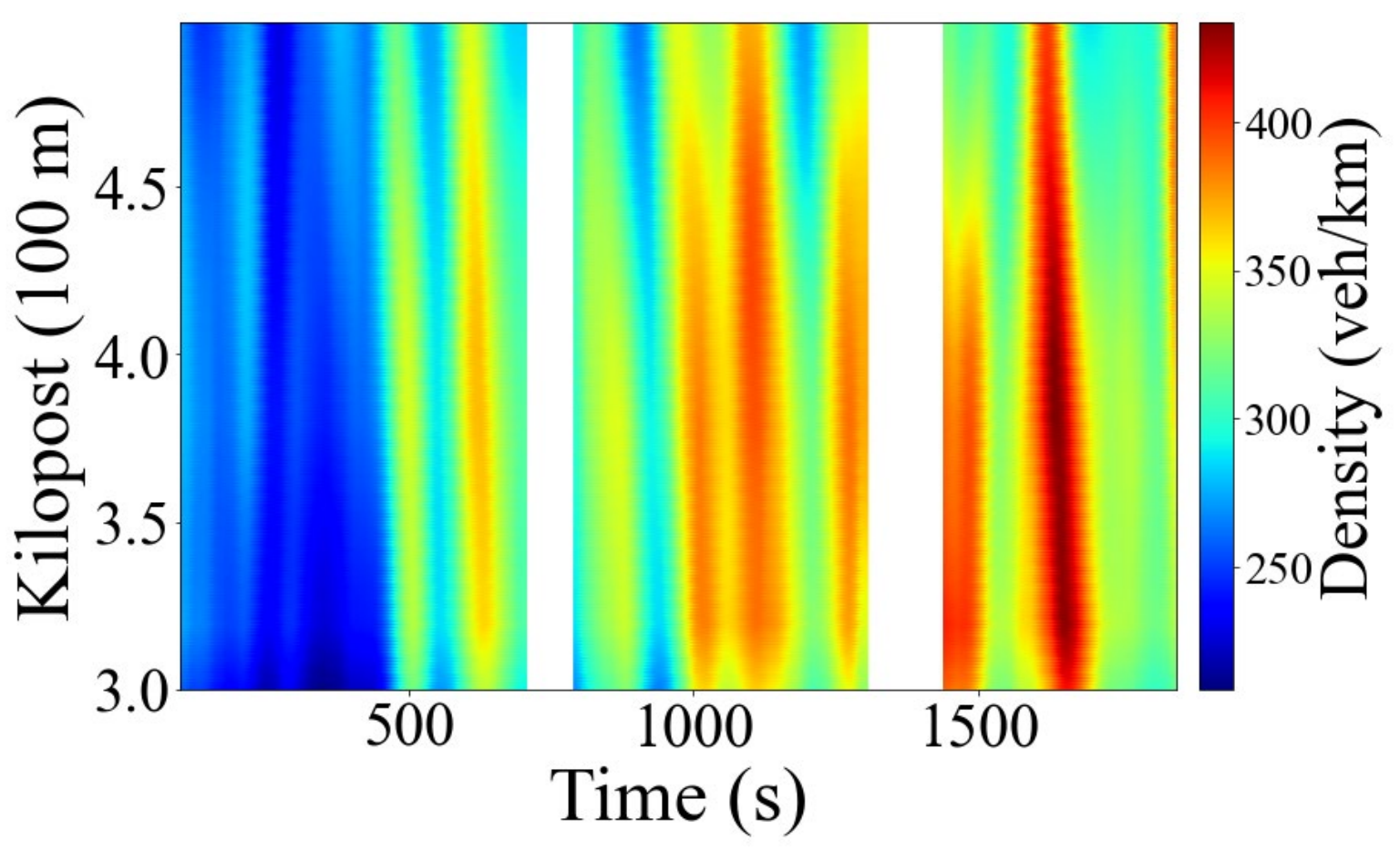}
    \label{Fig4k}
  \end{subfigure}
  \begin{subfigure}{.33\textwidth}
    \centering
    \caption{Dataset 4: $\mathcal{A}$}
    \includegraphics[width=1\linewidth]{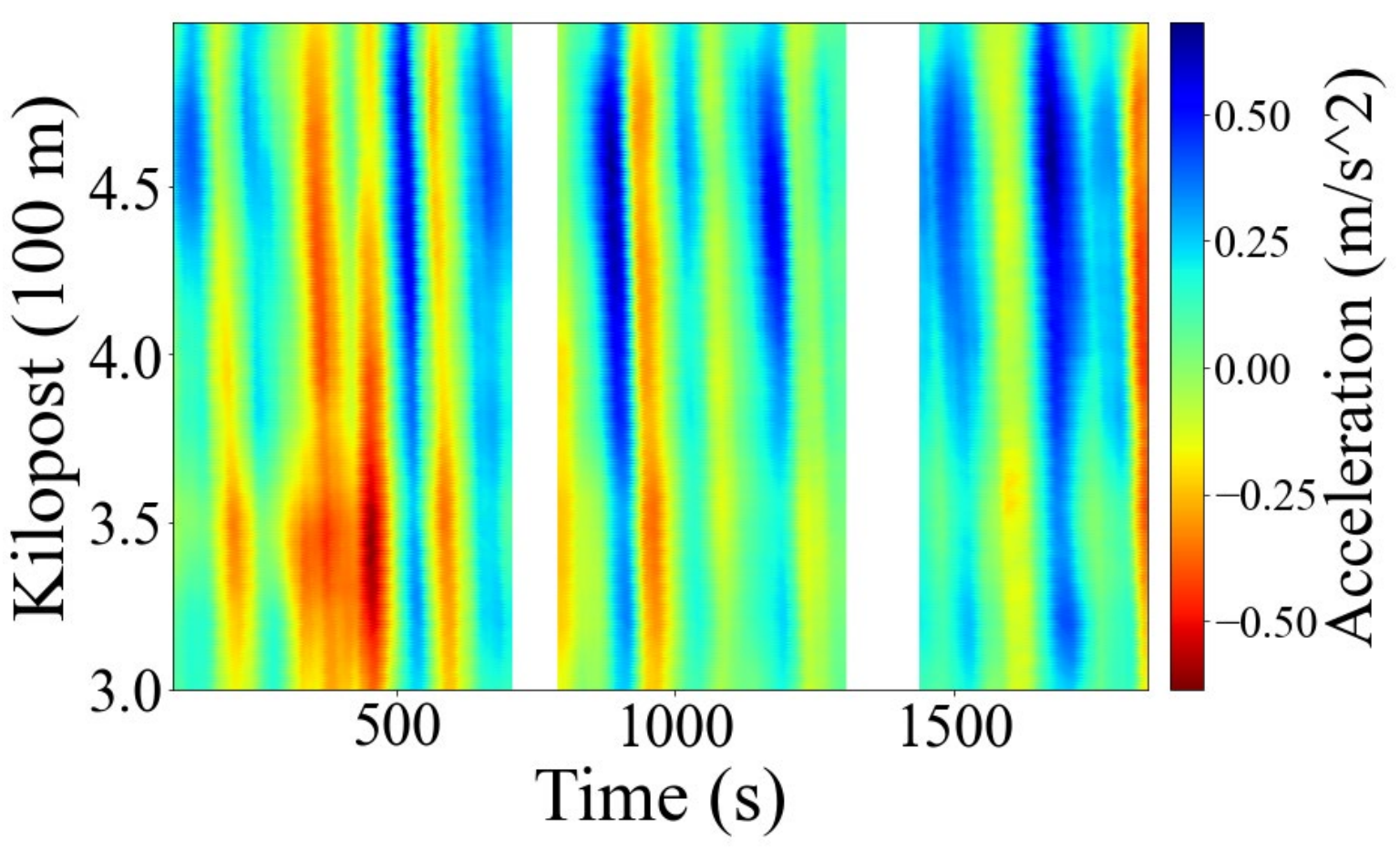}
    \label{Fig4l}
  \end{subfigure}
  \caption{The speed, density, and acceleration fields of the different trajectory datasets: (a)-(c) the fields of dataset 1; (d)-(f) the fields of dataset 2; (g)-(i) the fields of dataset 3; (j)-(l) the fields of dataset 4.}
  \label{Fig4}
\end{figure}

We begin by estimating the speed, density, and acceleration fields for each of the four trajectory datasets. The spatio-temporal domain is divided into subdomains with a time interval of $\Delta t=50s$ and a spatial interval of $\Delta x=300m$. This division ensures sufficient coverage for estimating the macroscopic traffic flow parameters within each subdomain. To balance continuity in the parameter fields and computational efficiency, smaller sliding steps in space and time are chosen, with $t^{\st}=2s$ and $x^{\st}=3m$. For the anticipated density estimation, the transition time $t^{\mathrm{m}}$ of 12 seconds is selected, which falls within the recommended range \citep{green2004policy}. 
The speed ($\mathcal{V}$), density ($\mathcal{K}$), and acceleration ($\mathcal{A}$) fields for the four datasets, developed using \autoref{Eq8}-\autoref{Eq11}, 
are presented in \autoref{Fig4}.

We note that the blanks present in (g)-(l) of \autoref{Fig4} are due to the gaps in the NGSIM data. The NGSIM dataset provides trajectories in separate 15-minute intervals. Therefore, the fields are estimated individually for each 15-minute subset of trajectories and then concatenated together. The blank areas in \autoref{Fig4} represent the boundaries between each 15-minute subset of trajectories resulting from the subdomain of the space-time field estimation. The blank areas do not affect our tests. 

\subsection{LKV and NLKV samples of different trajectory datasets}\label{Sect5.2}

\begin{figure}[!ht]
  \begin{subfigure}{.33\textwidth}
    \centering
    \caption{Dataset 1: LKV sample pairs}
    \includegraphics[width=1\linewidth]{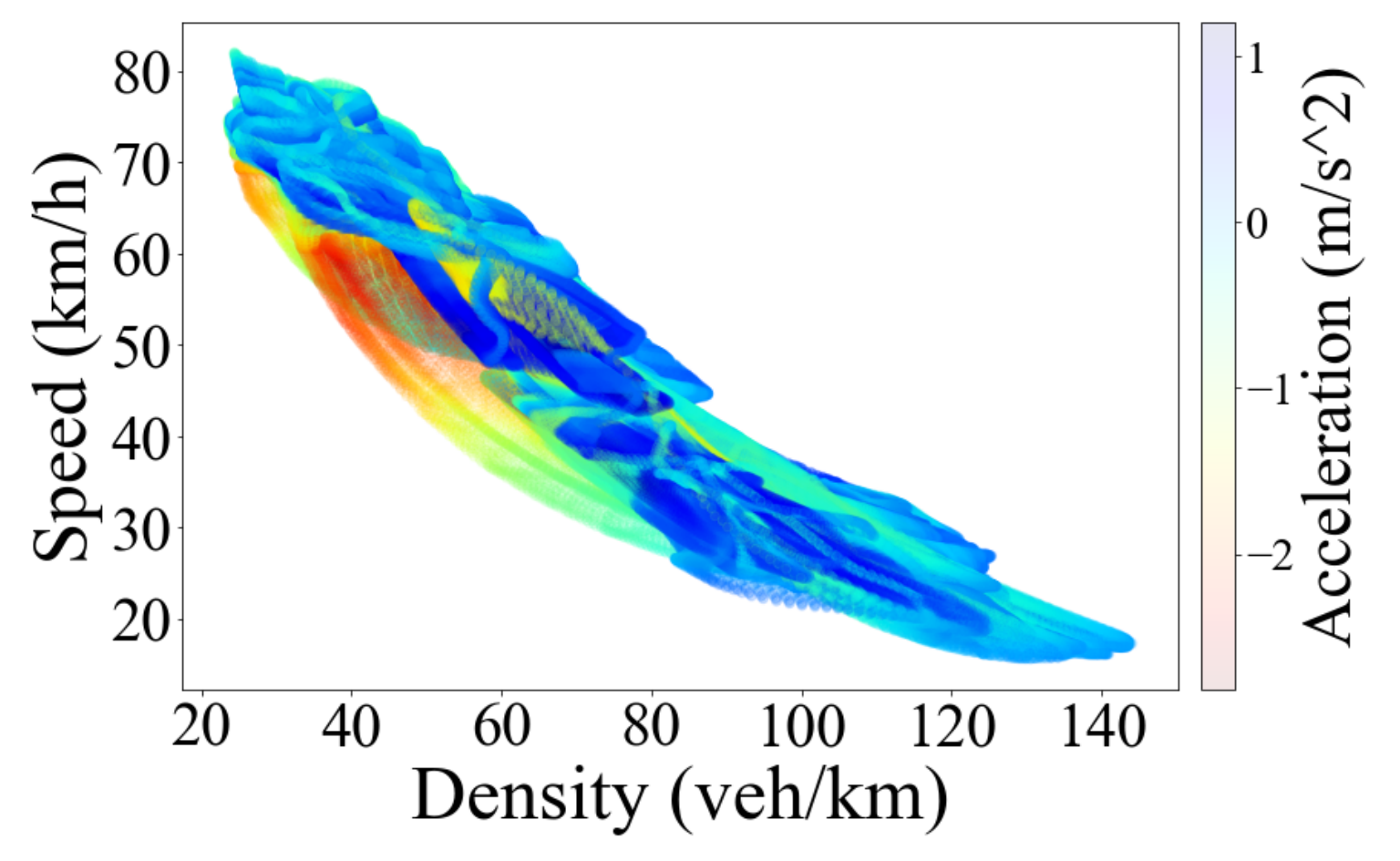}
    \label{Fig5a}
  \end{subfigure}
  \begin{subfigure}{.33\textwidth}
    \centering
    \caption{Dataset 1: NLKV sample pairs + $\mathcal{A}(i,j)$}
    \includegraphics[width=1\linewidth]{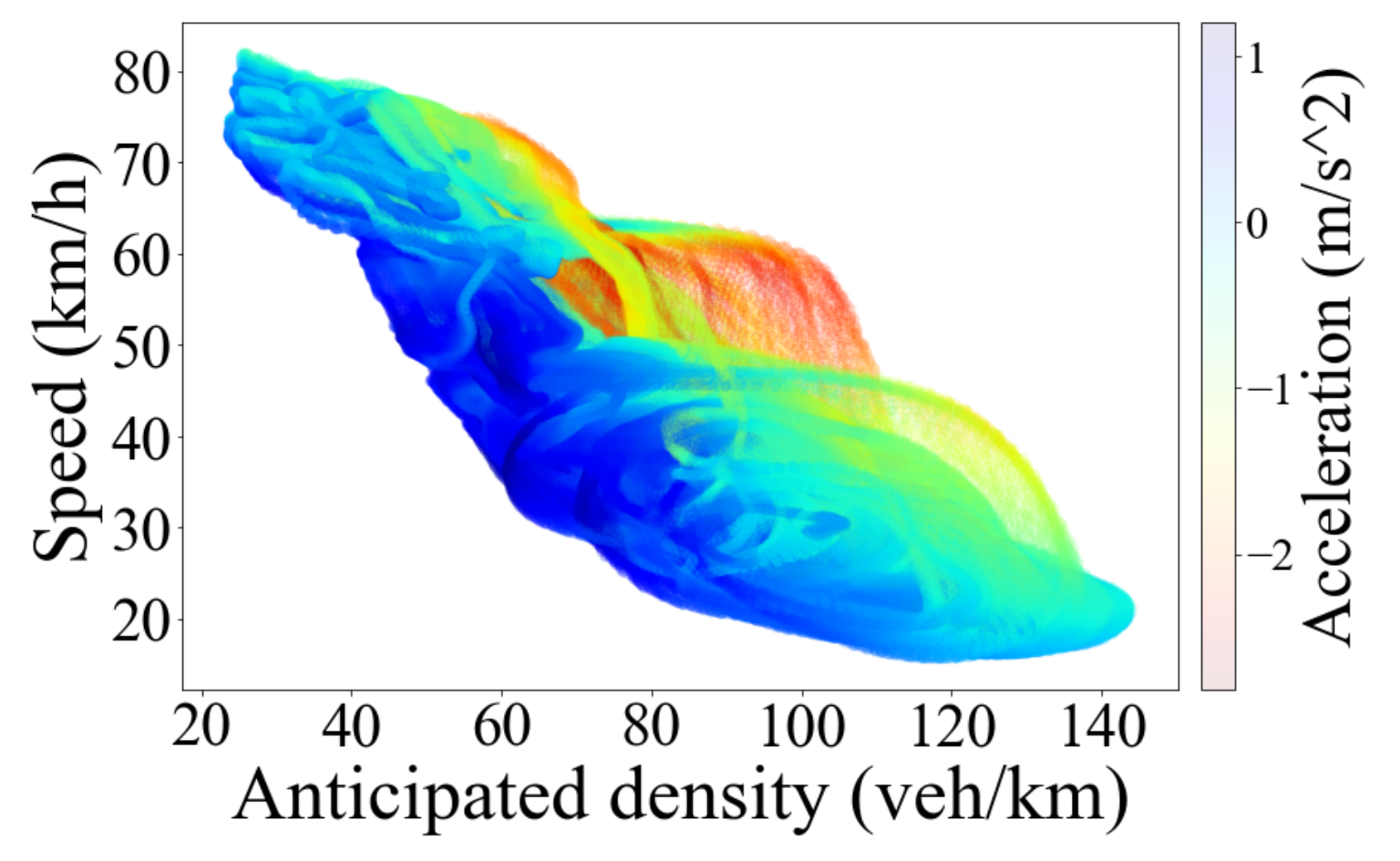}
    \label{Fig5b}
  \end{subfigure}
  \begin{subfigure}{.33\textwidth}
    \centering
    \caption{Dataset 1: NLKV sample pairs}
    \includegraphics[width=1\linewidth]{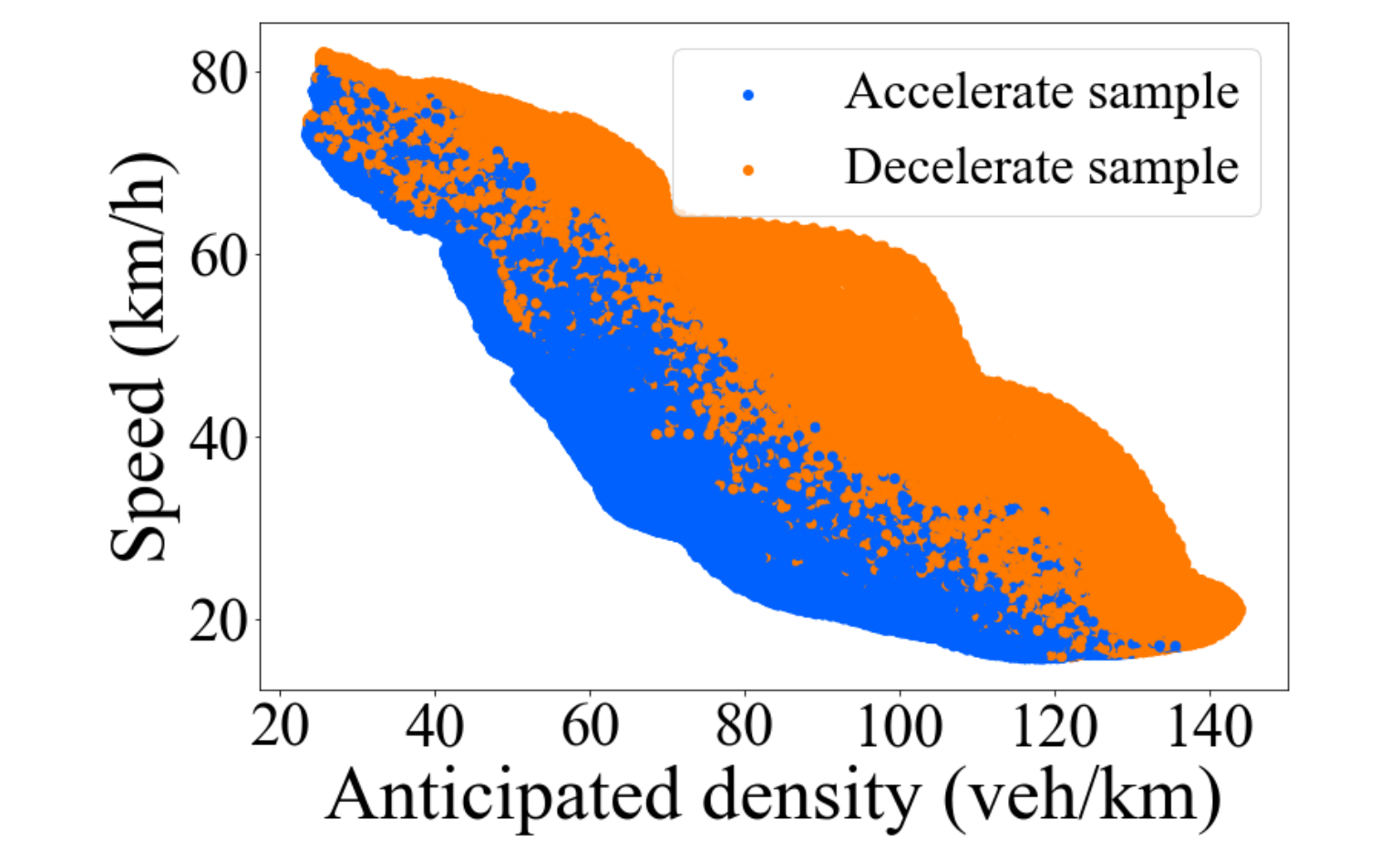}
    \label{Fig5c}
  \end{subfigure}
  \begin{subfigure}{.33\textwidth}
    \centering
    \caption{Dataset 2: LKV sample pairs}
    \includegraphics[width=1\linewidth]{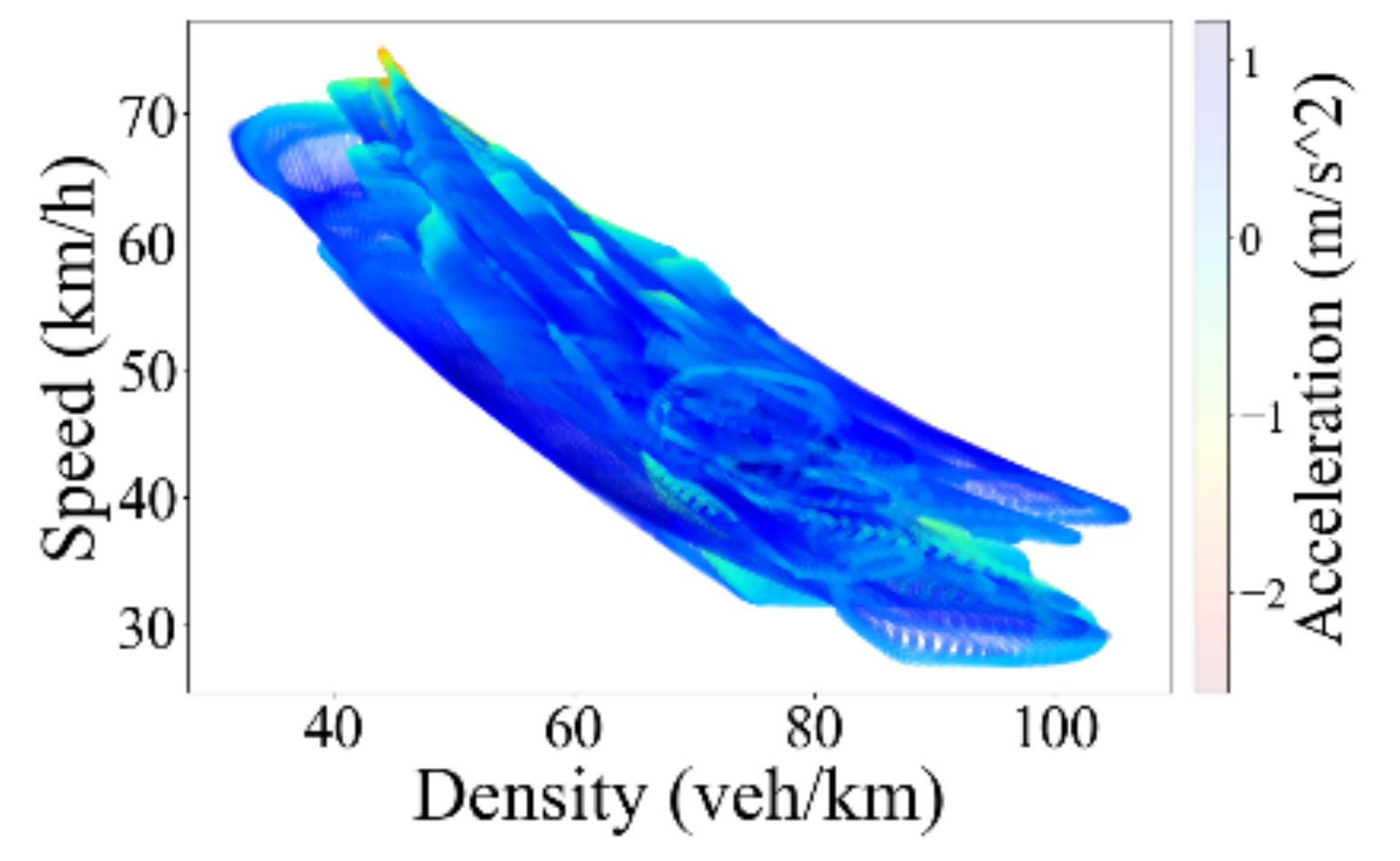}
    \label{Fig5d}
  \end{subfigure}
  \begin{subfigure}{.33\textwidth}
    \centering
    \caption{Dataset 2: NLKV sample pairs + $\mathcal{A}(i,j)$}
    \includegraphics[width=1\linewidth]{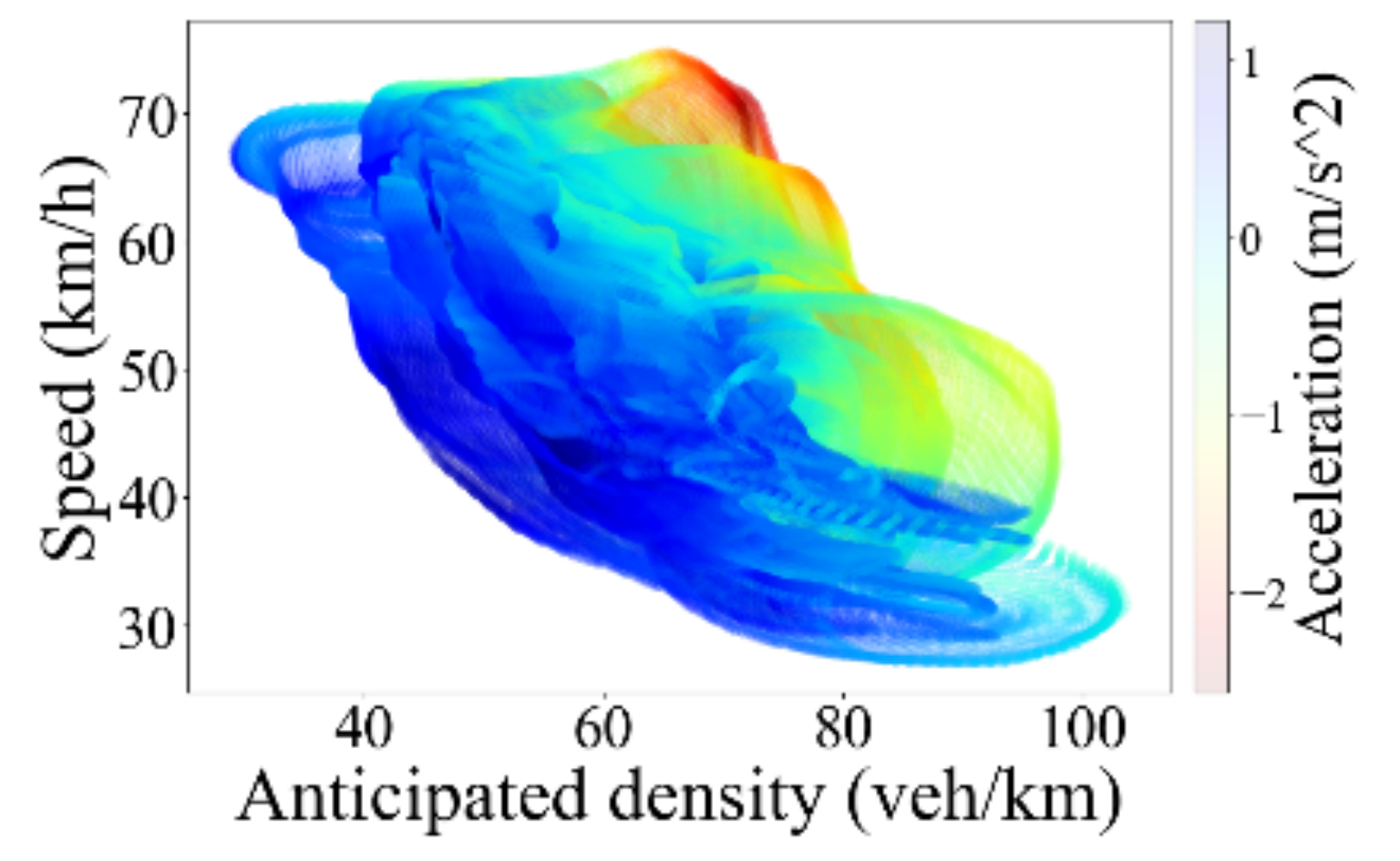}
    \label{Fig5e}
  \end{subfigure}
  \begin{subfigure}{.33\textwidth}
    \centering
    \caption{Dataset 2: NLKV sample pairs}
    \includegraphics[width=1\linewidth]{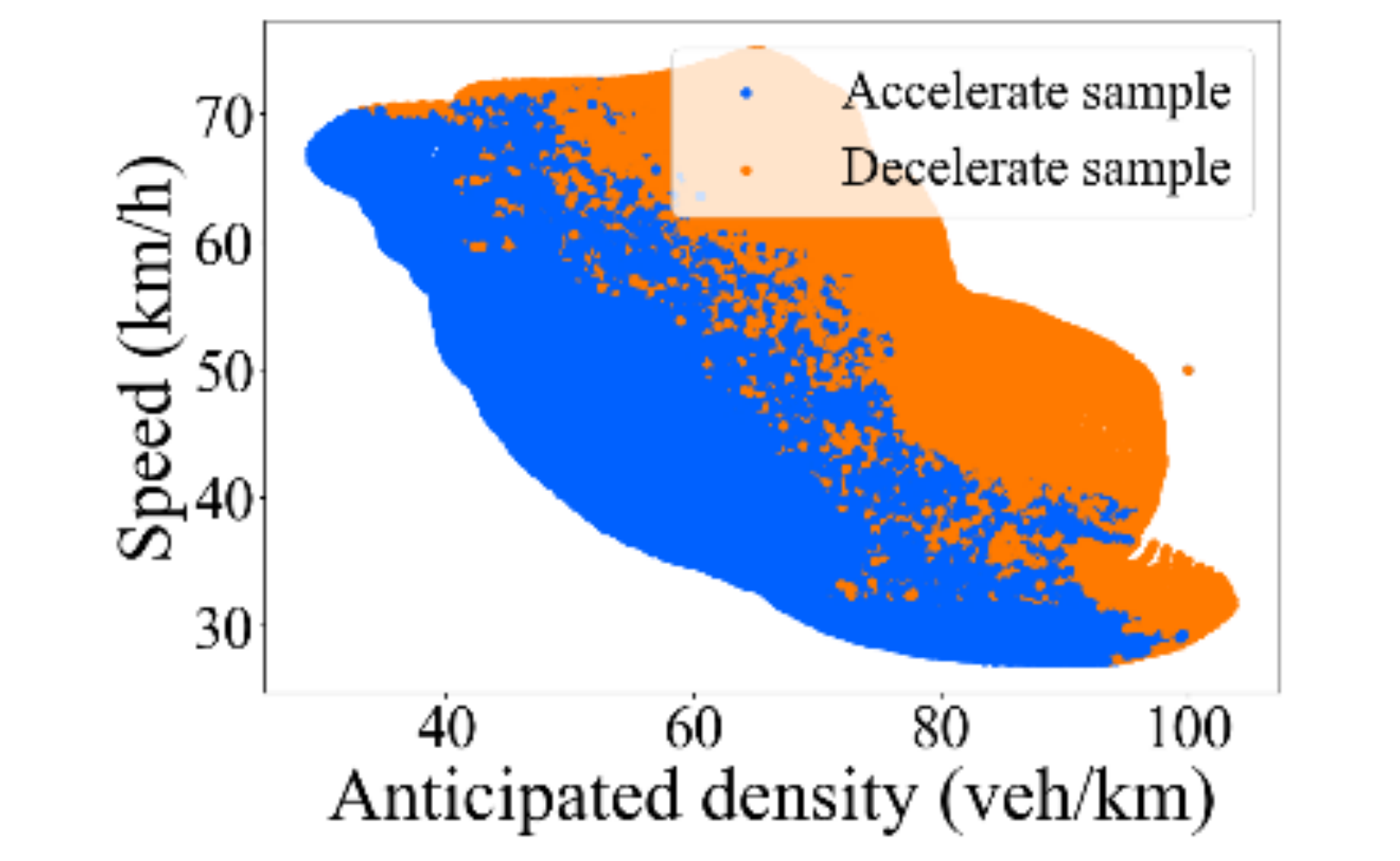}
    \label{Fig5f}
  \end{subfigure}
  \begin{subfigure}{.33\textwidth}
    \centering
    \caption{Dataset 3: LKV sample pairs}
    \includegraphics[width=1\linewidth]{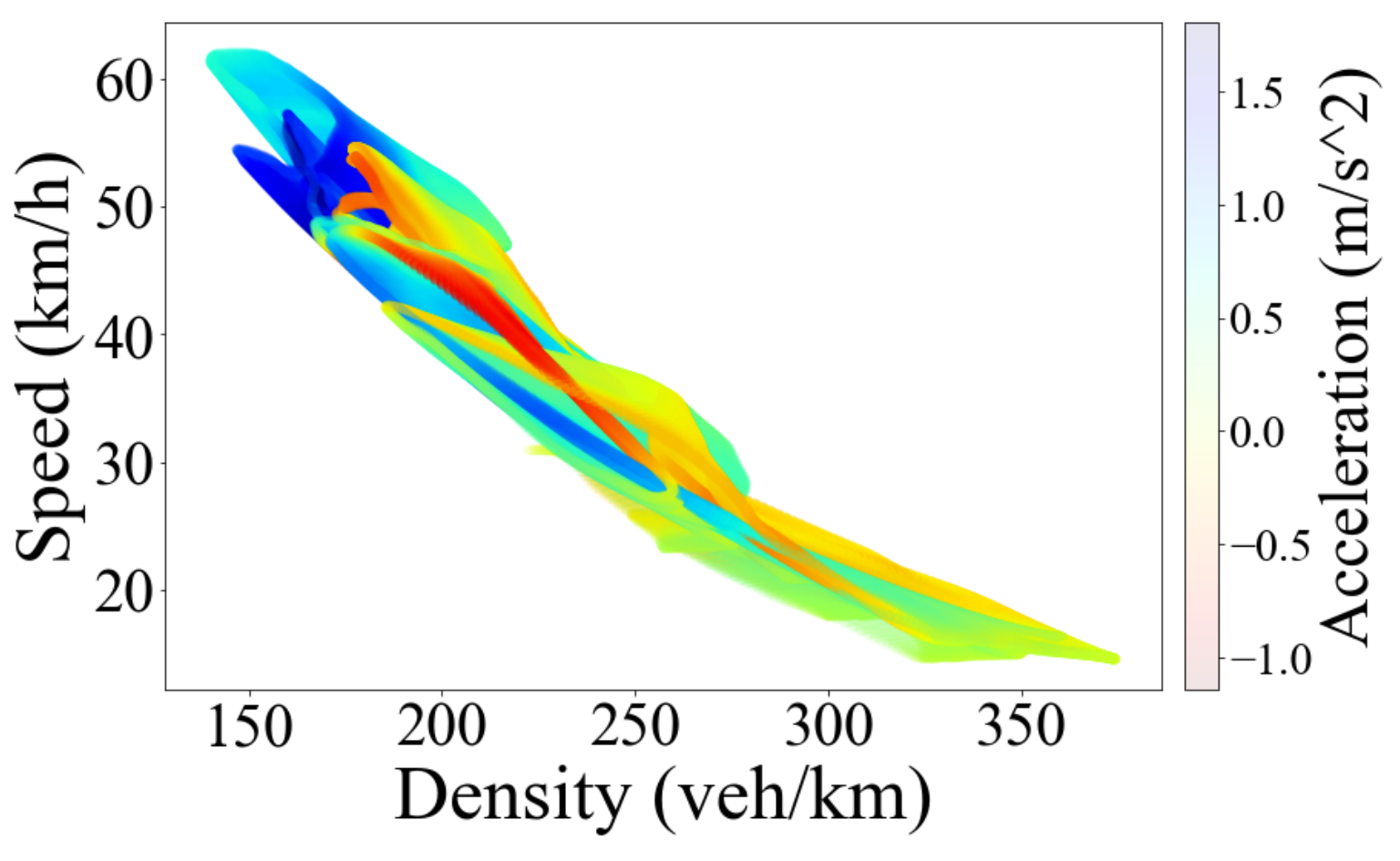}
    \label{Fig5g}
  \end{subfigure}
  \begin{subfigure}{.33\textwidth}
    \centering
    \caption{Dataset 3: NLKV sample pairs + $\mathcal{A}(i,j)$}
    \includegraphics[width=1\linewidth]{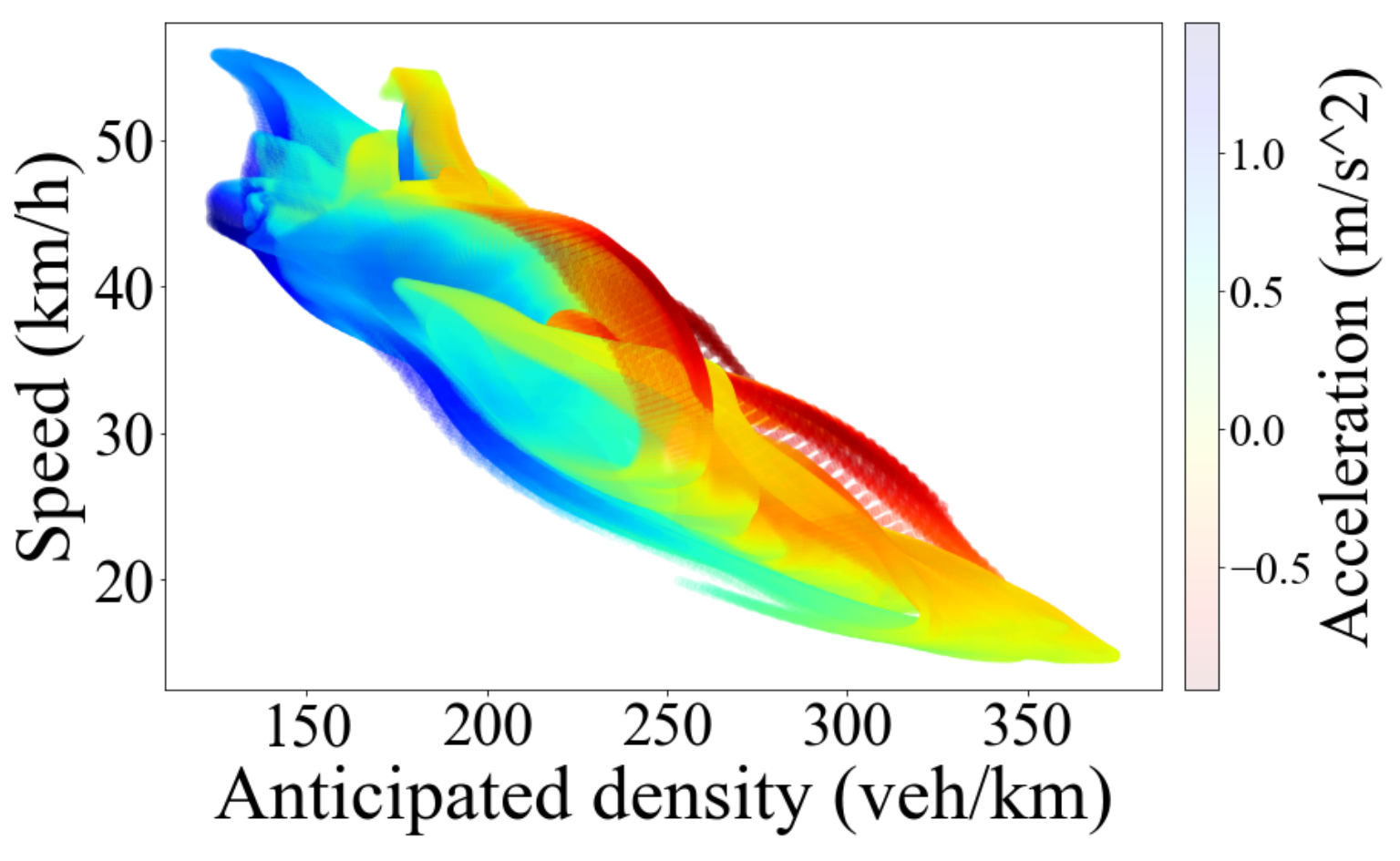}
    \label{Fig5h}
  \end{subfigure}
  \begin{subfigure}{.33\textwidth}
    \centering
    \caption{Dataset 3: NLKV sample pairs}
    \includegraphics[width=1\linewidth]{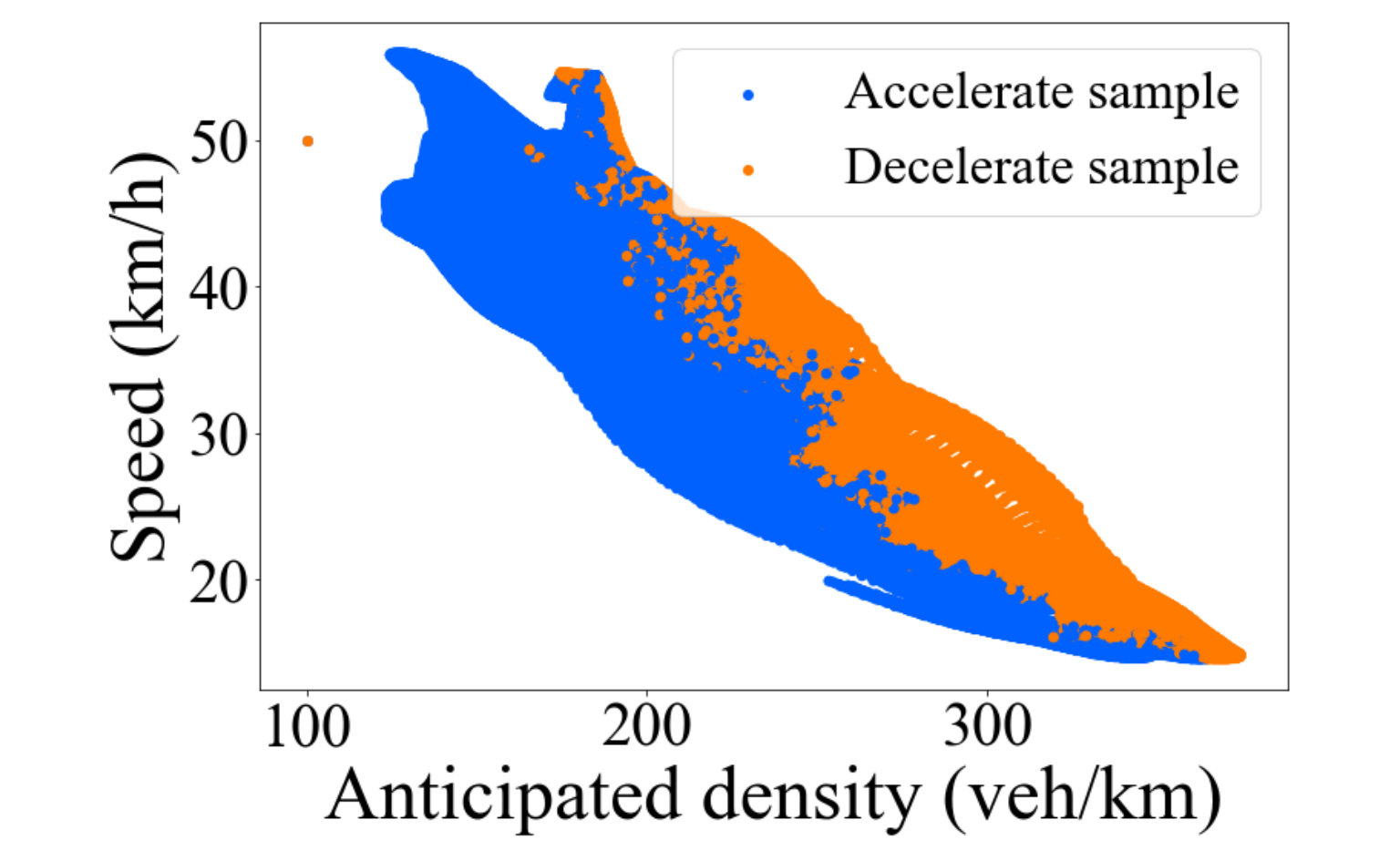}
    \label{Fig5i}
  \end{subfigure}
  \begin{subfigure}{.33\textwidth}
    \centering
    \caption{Dataset 4: LKV sample pairs}
    \includegraphics[width=1\linewidth]{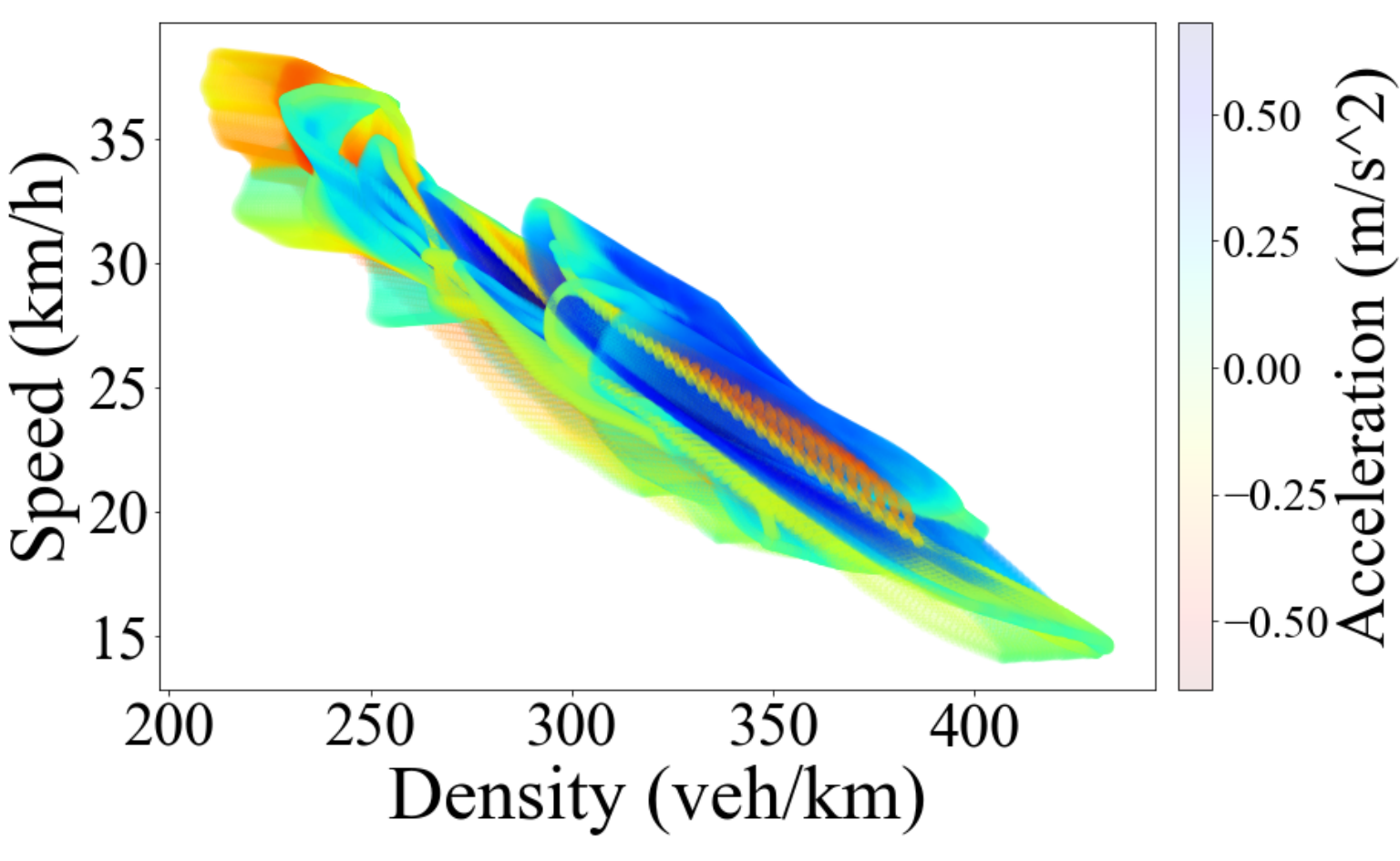}
    \label{Fig5j}
  \end{subfigure}
  \begin{subfigure}{.33\textwidth}
    \centering
    \caption{Dataset 4: NLKV sample pairs + $\mathcal{A}(i,j)$}
    \includegraphics[width=1\linewidth]{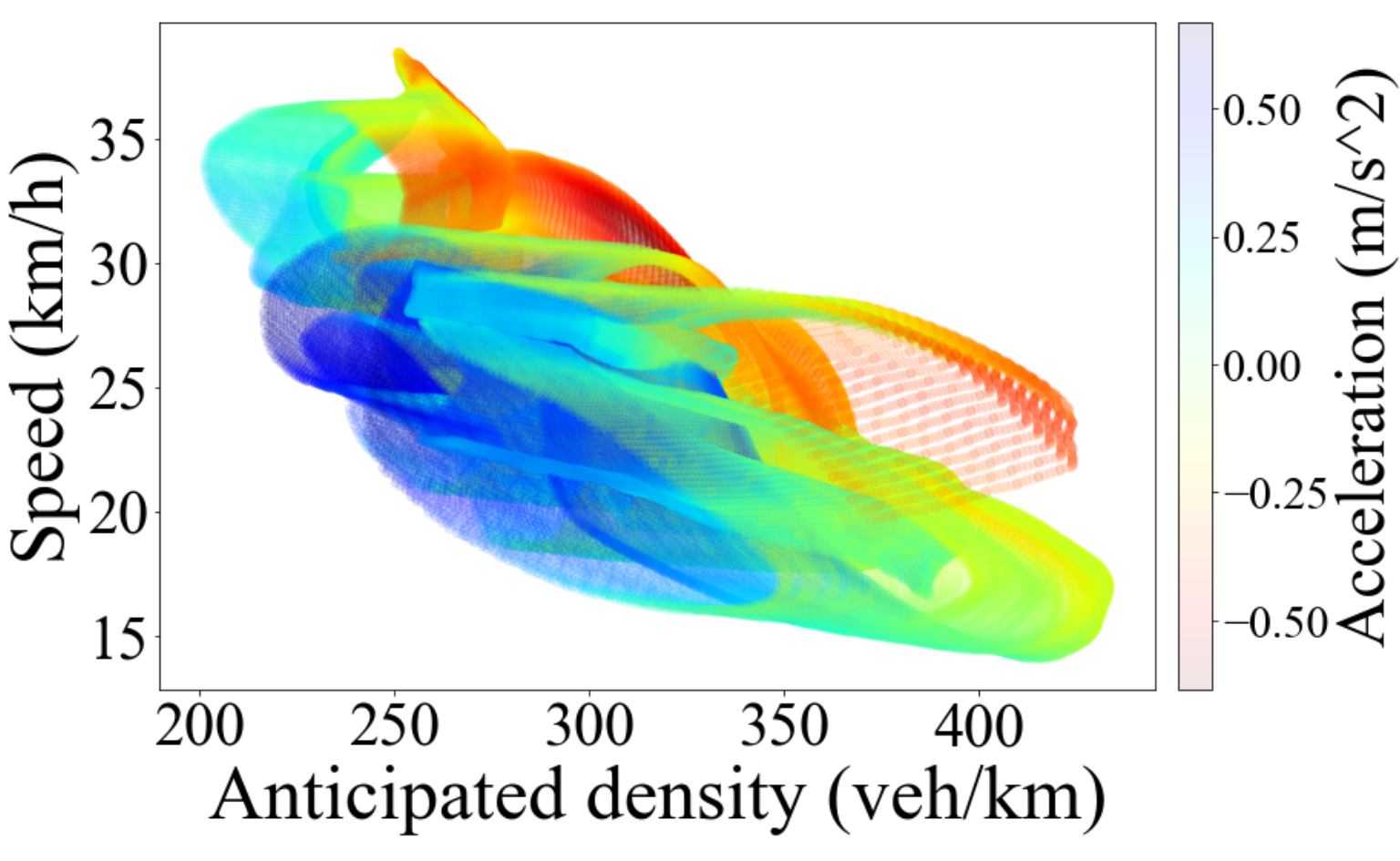}
    \label{Fig5k}
  \end{subfigure}
  \begin{subfigure}{.33\textwidth}
    \centering
    \caption{Dataset 4: NLKV sample pairs}
    \includegraphics[width=1\linewidth]{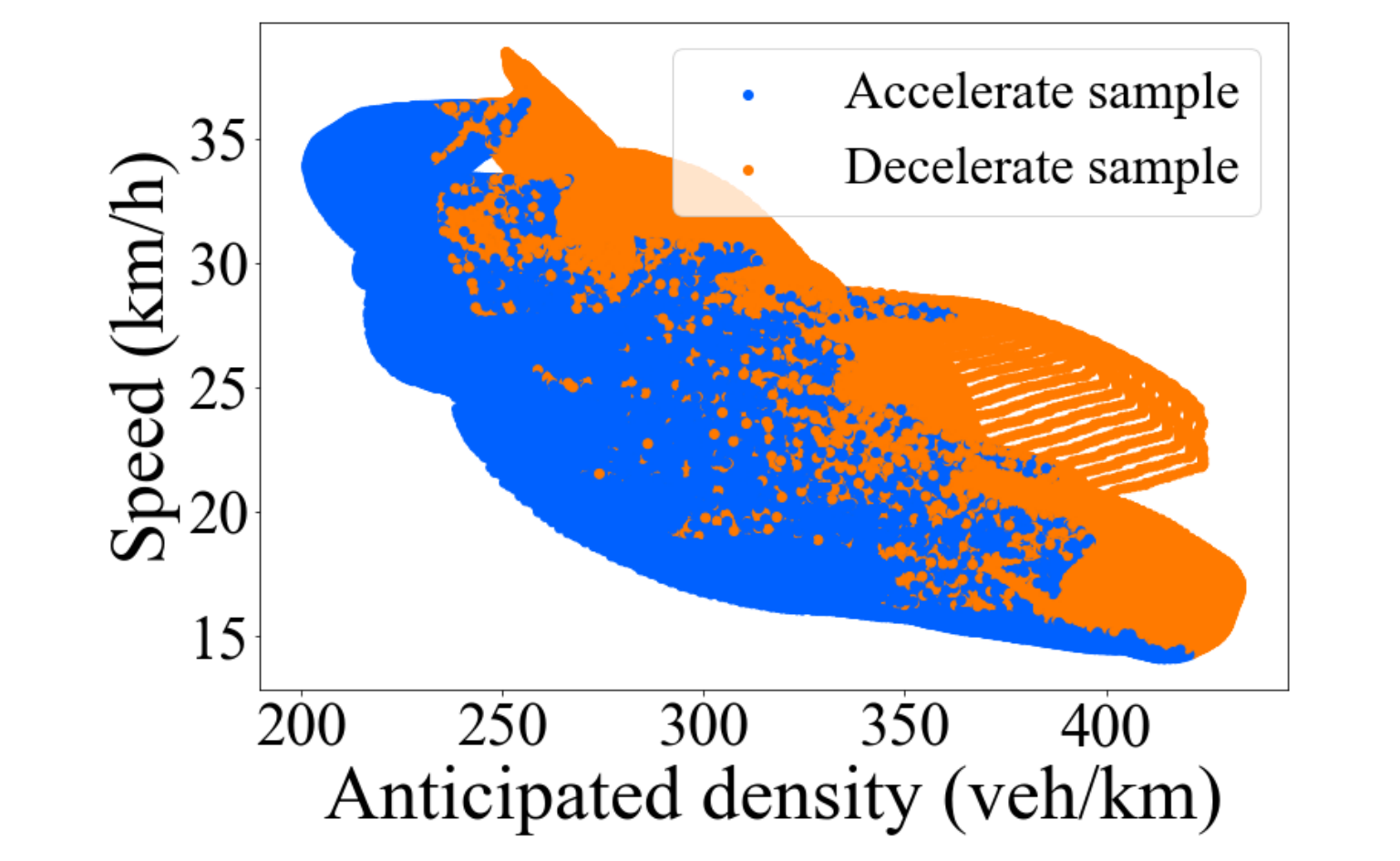}
    \label{Fig5l}
  \end{subfigure}
  \caption{The speed-density relationship derived using LKV samples and NLKV samples: (a), (d), (g), and (j) the LKV samples of datasets1- 4 colored by acceleration; (b), (e), (h), and (k) the NLKV samples of datasets 1- 4 colored by acceleration; (c), (f), (i), and (l) the NLKV samples of datasets 1- 4 indicated with acceleration and deceleration.}
  \label{Fig5}
\end{figure}

We begin by comparing the properties of LKV and NLKV samples across datasets 1 to 4, as shown in \autoref{Fig5}. The figures in the first column of \autoref{Fig5} depict the scattered LKV samples of each subdomain $\mathcal{D}(i,j)$, representing $\mathcal{K}(i,j)$--$\mathcal{V}(i,j)$ pairs, with colors corresponding to their respective acceleration $\mathcal{A}(i,j)$. The presence of traffic hysteresis can be clearly observed in (a), (d), (g), and (j) of \autoref{Fig5} by the absence of separation between acceleration regimes. This indicates that the speeds are transient, which is not suitable for fitting equilibrium relations. 

The second column of \autoref{Fig5} depicts scatter-plots of $\mathcal{K}^{\an}(i,j)$--$\mathcal{V}(i,j)$ pairs for each subdomain $\mathcal{D}(i,j)$, with colors representing their corresponding acceleration $\mathcal{A}(i,j)$. It is evident from these figures that with the non-locality, a distinct separation between acceleration, deceleration, and nearly steady state traffic flow area can be observed. The color gradient in this column of figures correlates with the following driver behaviors: 1) Negative (positive) acceleration occurs when a higher (lower) density than current is anticipated. 2) A higher absolute value of acceleration occurs when significant traffic state changes are expected.

The last column of figures in \autoref{Fig5} displays the NLKV samples, which represent scatter-plots of $\mathcal{K}^{\an}(i,j)$--$\mathcal{V}(i,j)$ pairs for each subdomain $\mathcal{D}(i,j)$, with colors indicating \emph{the sign} of their acceleration, denoted as $y(i,j)$. By classifying the empirical samples into two distinct sets of acceleration and deceleration, the separation becomes more pronounced. It is evident from \autoref{Fig5} that NLKV samples capture equilibrium states uninfluenced by hysteresis, making them more suitable for FD modeling.

\begin{figure}[!ht]
\centering
  \begin{subfigure}{.8\textwidth}
    \centering
    \caption{Observed heteroskedasticity in LKV data}
    \includegraphics[width=.5\linewidth]{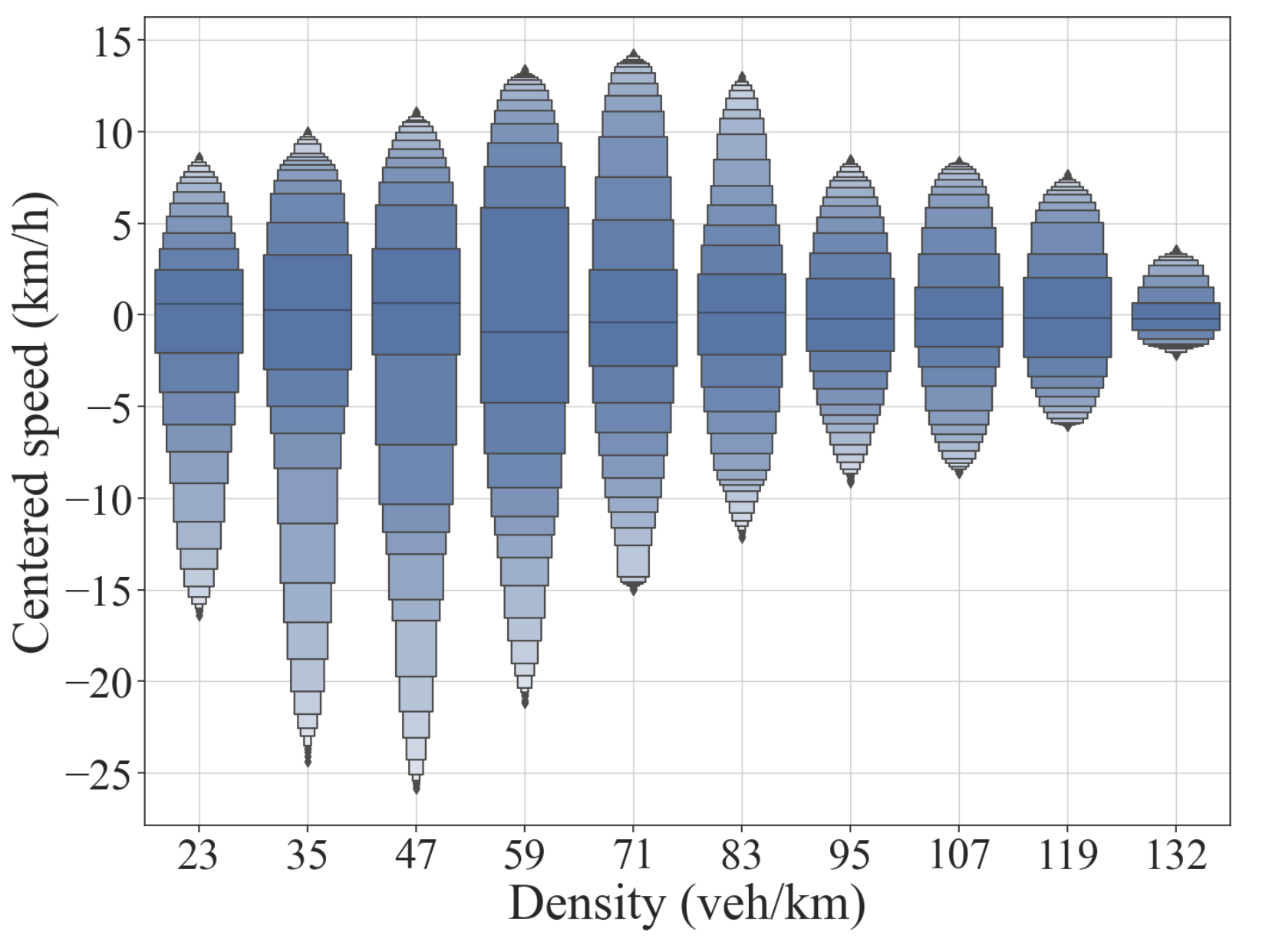}
    \label{Fig6a}
  \end{subfigure} \\
  \begin{subfigure}{.4\textwidth}
    \centering
    \caption{Probabilities of deceleration in NLKV data}
    \includegraphics[width=1\linewidth]{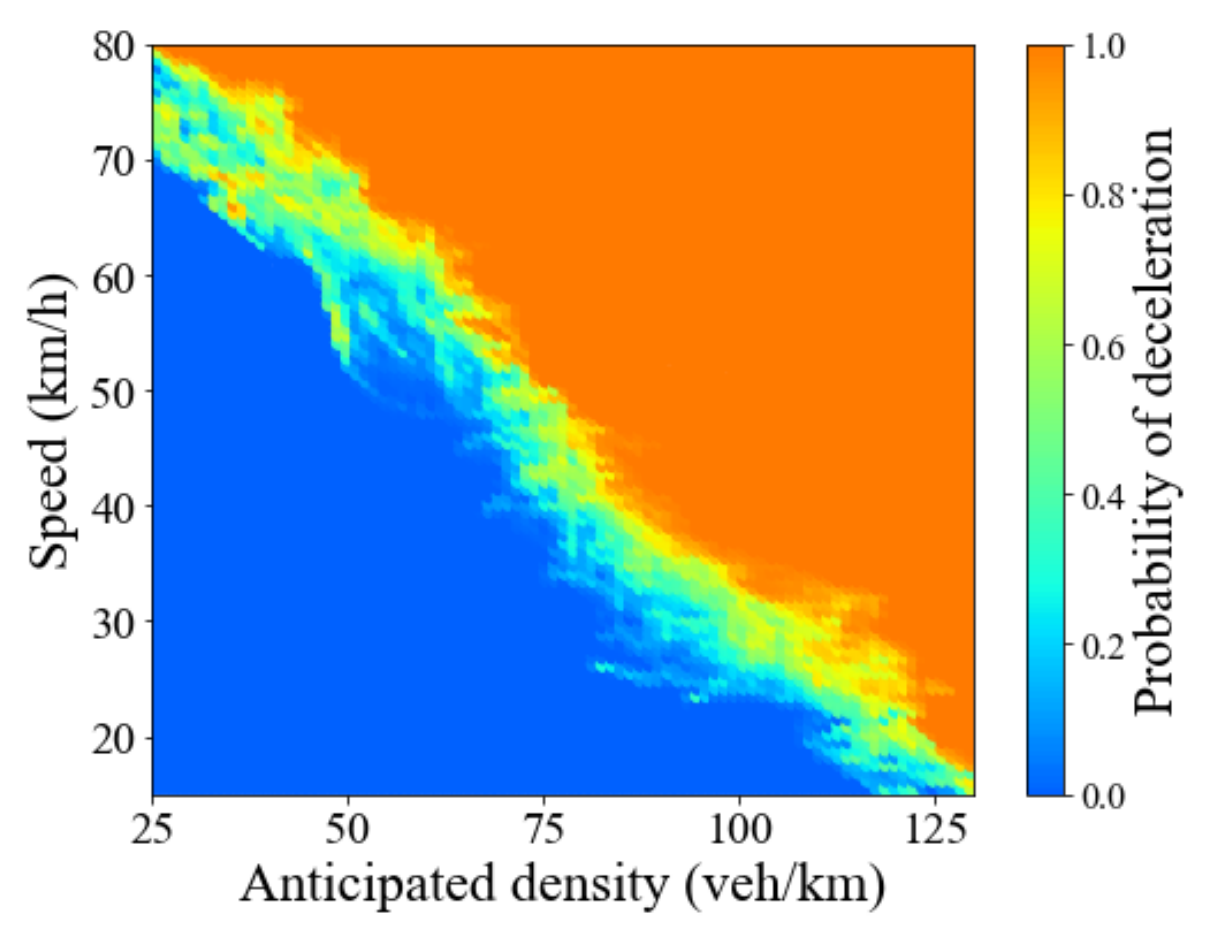}
    \label{Fig6b}
  \end{subfigure}
  \ \ \ \ 
  \begin{subfigure}{.4\textwidth}
    \centering
    \caption{Probabilities of deceleration in NLKV data with center speeds}
    \includegraphics[width=1\linewidth]{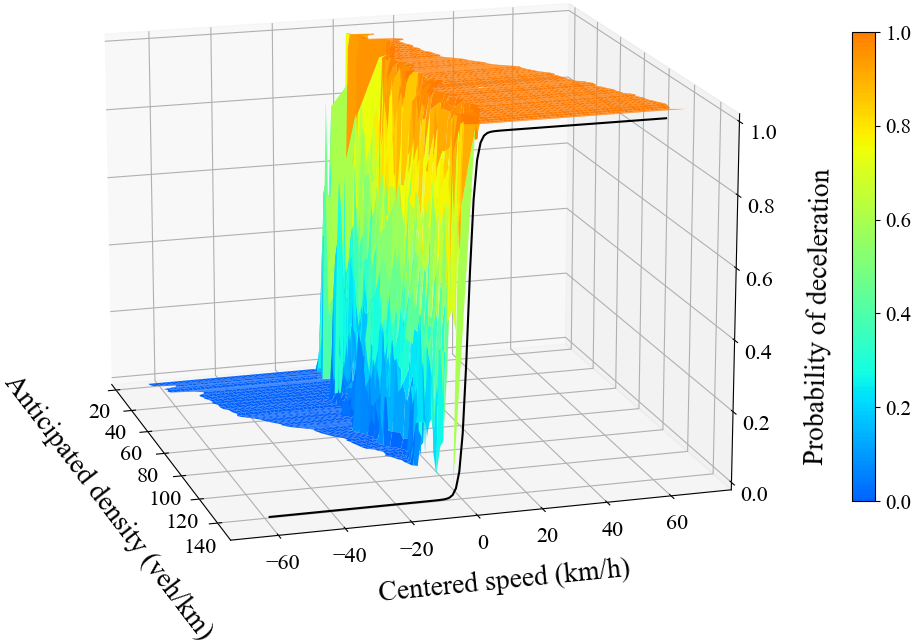}
    \label{Fig6c}
  \end{subfigure}
  \caption{The distribution of LKV and NLKV samples: (a) centered speeds vs. densities in LKV data; (b) probabilities of deceleration for varying anticipated density--speed pairs in NLKV data; (c) probabilities of deceleration with speeds centered around the $p=0.5$ points.}
  \label{Fig6}
\end{figure}

We further analyze the distribution of LKV and NLKV samples in dataset 1, and illustrate the results in \autoref{Fig6}. As mentioned in \autoref{Sect1}, the traditional fitting approach (LKV+LSE) implicitly assumes uniform variance in the sample. \autoref{Fig6a} illustrates that speed variance depends on traffic density and varies substantially from one density to another (speeds in the figure are centered around the mean). 
As a result, standard LSE-based fitting techniques are not suitable for LKV data.

On the other hand, the underlying statistical assumption in our ECE minimization method is that the signs of acceleration, given anticipated density and speed, follow an identical Bernoulli distribution. To test this assumption, we calculate deceleration probabilities using $p=n_{\mathrm{d}}/n$, where $n_{\mathrm{d}}$ represents the number of deceleration samples in the subarea, and $n$ is the total number of samples in the same subarea. These probabilities are depicted in \autoref{Fig6b} for varying anticipated density--speed pairs. We observe a similar pattern across different anticipated density--speed pairs (the width of the separating band across the diagonal does not change). Further, upon plotting the deceleration probabilities after centering the speeds in \autoref{Fig6c}, we observe logistic distributions that appear invariant to anticipated density, which corroborates the implicit statistical assumption underlying our approach.

\subsection{FD models and the LSE for comparison}\label{Sect5.3}

To compare the performance of the LKV+LSE and the proposed NLKV+ECE approaches, we select three FD models: Greenberg’s model \citep{greenberg1959analysis}, Smulders’ model \citep{smulders1990control}, and the Franklin-Newell model \citep{newell1961nonlinear,franklin1961structure}. These are classical parametric models of the FD. We note that Greenberg's model does not respect boundedness axioms in \citep{del1995functional}; the other two models do, as well as the concavity axiom. This is immaterial in this study as our purpose is to illustrate the value of non-localities in speed-density data and not to promote a particular parametric model. The functional forms of these models are given below:
\begin{equation}\label{Eq15}
\mbox{Greenberg's FD: }   ~\hat{v}(k;v_0,k_{\mathrm{jam}})=v_0 \ln \left( \frac{k_{\mathrm{jam}}}{k}\right),
\end{equation}
\begin{equation}\label{Eq16}
\mbox{Smulders'  FD: } ~ \hat{v}(k; v_0, k_{\mathrm{crit}}, k_{\mathrm{jam}})=
\left\{ 
  \begin{array}{cl}
    v_0 \left( 1-\dfrac{k}{k_{\mathrm{jam}}}\right)& k<k_{\mathrm{crit}}\\
    v_0 k_{\mathrm{crit}} \left( \dfrac{1}{k}-\dfrac{1}{k_{\mathrm{jam}}}\right)& k\geq k_{\mathrm{crit}}\\
  \end{array}
  ,
\right.
\end{equation}

\begin{equation}\label{Eq17}
\mbox{Franklin-Newell FD: } ~ \hat{v}(k;v_0,\lambda,k_{\mathrm{jam}})=v_0 \left( 1-e^{-\frac{\lambda}{v_0}\left( \frac{1}{k}-\frac{1}{k_{\mathrm{jam}}}\right)}\right),
\end{equation}
where $\hat{v}$ is a speed-density FD and $v_0$, $k_{\mathrm{jam}}$, $k_{\mathrm{crit}}$ and $\lambda$ are interpretable parameters of FD models. 
For the LKV+LSE approach, the fitting problem is given as
\begin{equation}\label{Eq18}
\underset{\vec{\theta}}{\mathrm{minimize}} ~  \frac{1}{m}\sum_{i=1}^{m}\Big( v_i-f\left( k_i;\vec{\theta} \right) \Big)^2.
\end{equation}

\textbf{We emphasize that \autoref{Eq18} is representative of the state of the art and not just one FD fitting method.} Our tests use parametric models for $f$, but $f$ can be any non-parametric model as well. In either case, what we see is that \autoref{Eq18} broadly overlooks equilibrium characteristics in the traffic. While our comparisons are focused on parametric FDs, one can easily argue that our findings generalize to any type of model $f$.

\subsection{Fitting results of LKV+LSE and NLKV+ECE approaches}\label{Sect5.4}

\subsubsection{Results of the same trajectory dataset and same FD model}\label{Sect5.4.1}

\begin{figure}[!ht]
\centering
  \begin{subfigure}{.45\textwidth}
    \centering
    \caption{Fitted FDs with the LKV samples}
    \includegraphics[width=.8\linewidth]{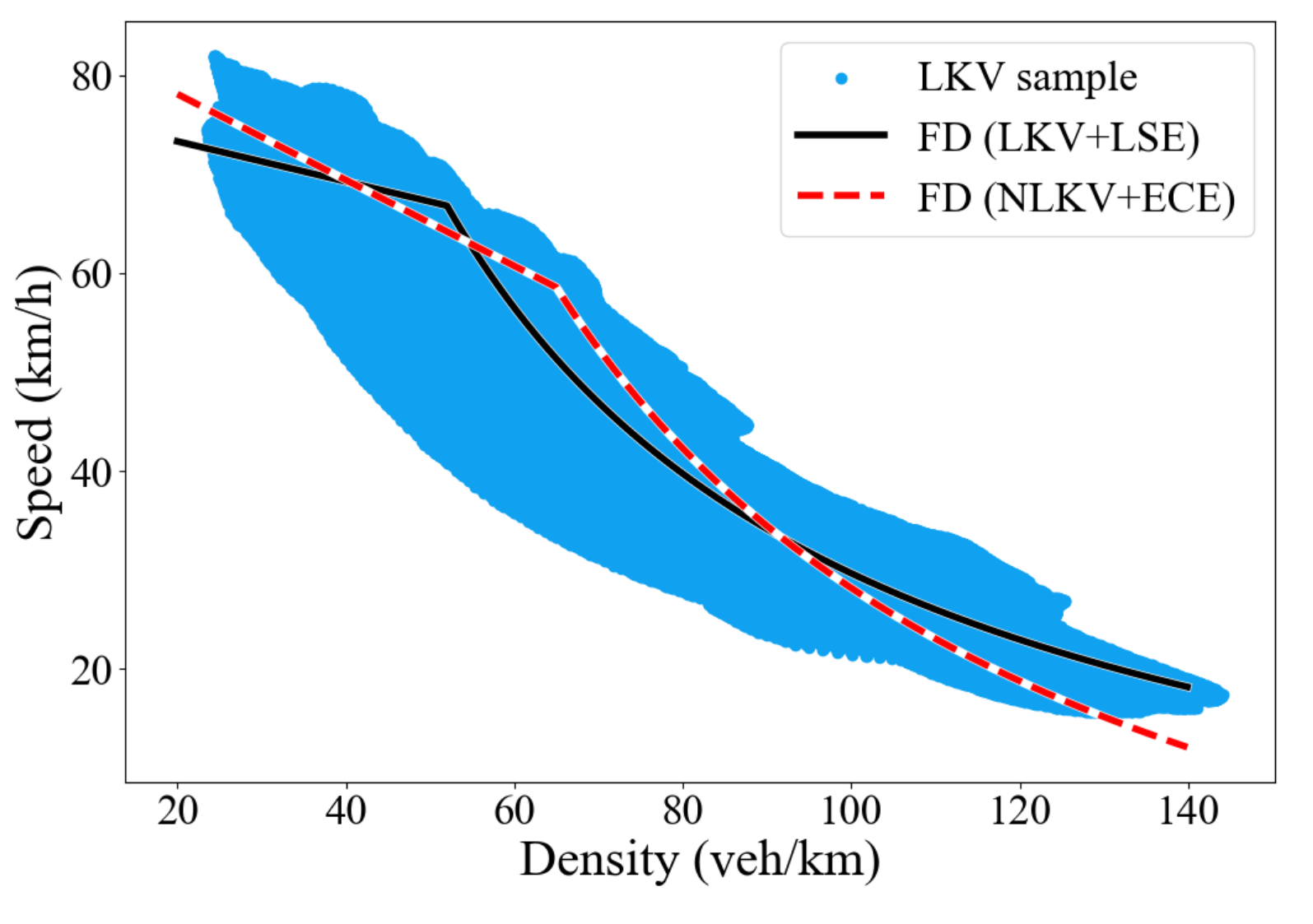}
    \label{Fig7a}
  \end{subfigure}
  \begin{subfigure}{.45\textwidth}
    \centering
    \caption{Fitted FDs with the NLKV samples}
    \includegraphics[width=.8\linewidth]{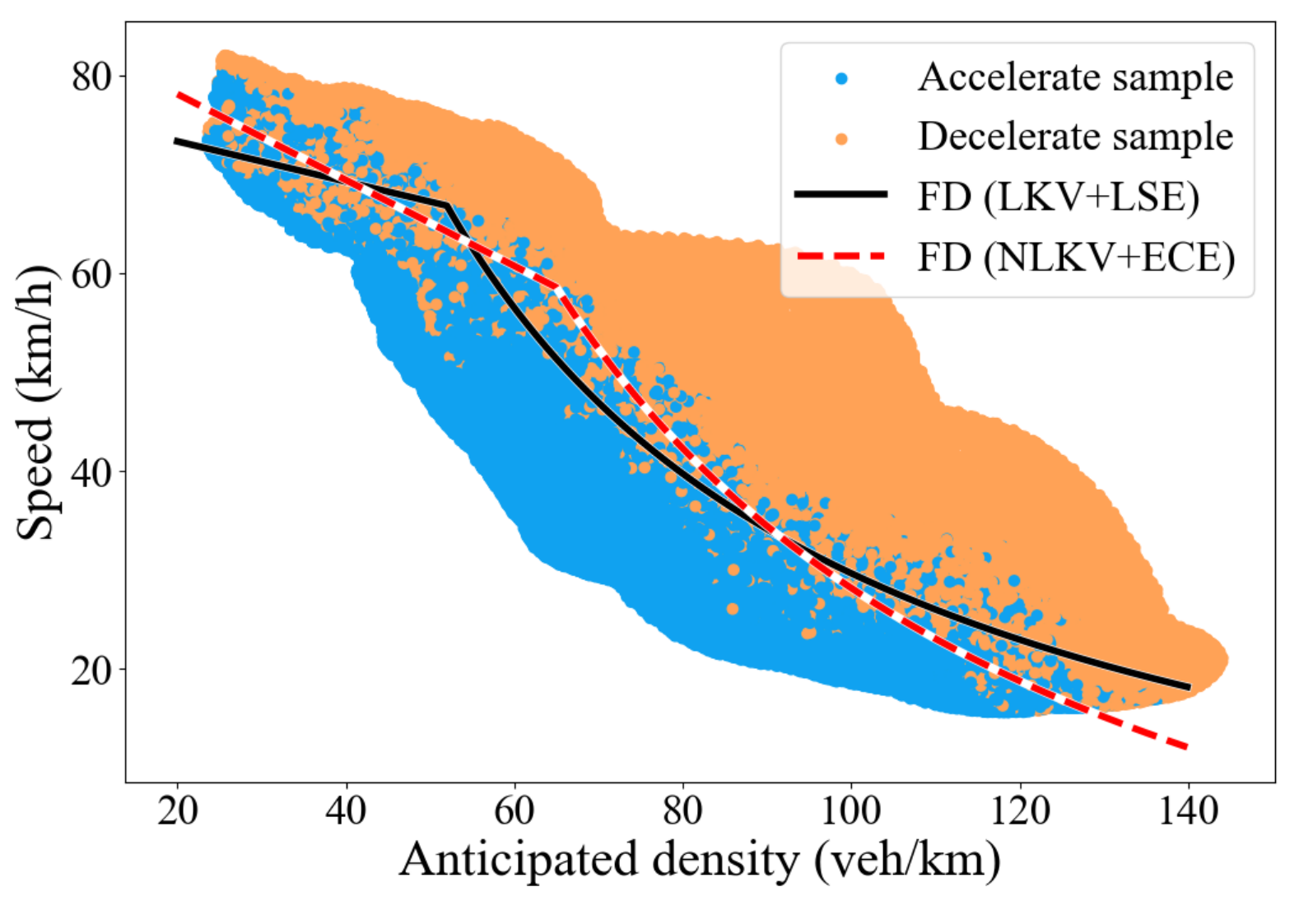}
    \label{Fig7b}
  \end{subfigure}
  \caption{Comparison of fitting results using different fitting approaches: (a) Fitted FDs with the LKV samples; (b) Fitted FDs with the NLKV samples.}
  \label{Fig7}
\end{figure}

In this subsection, we fit the parameters of Smulders’ FD model on LKV and NLKV samples from dataset 1. The fitting results of the two approaches, namely LKV+LSE and NLKV+ECE, are presented in \autoref{Fig7}. The black solid curve represents the fitting result using LKV samples, while the red dashed curve represents the fitting result using the NLKV samples. In \autoref{Fig7a}, we superimpose the two fits on the LKV samples and we superimpose the two on NLKV samples in \autoref{Fig7b}. We see a substantial difference in the estimated critical and jam densities in the two figures. The FD fitted using LKV samples underestimates the critical density and overestimates the jam density. The model fitted using LKV samples also underestimates the free flow speed. The FD fitted using NLKV samples separates acceleration and deceleration regimes better, suggesting that it captures equilibrium conditions better.
%

\subsubsection{Results of different trajectory datasets of the same road link and the same FD model}\label{Sect5.4.2}

\begin{figure}[!ht]
  \begin{subfigure}{.33\textwidth}
    \centering
    \caption{L001F002: $\mathcal{V}$}
    \includegraphics[width=1\linewidth]{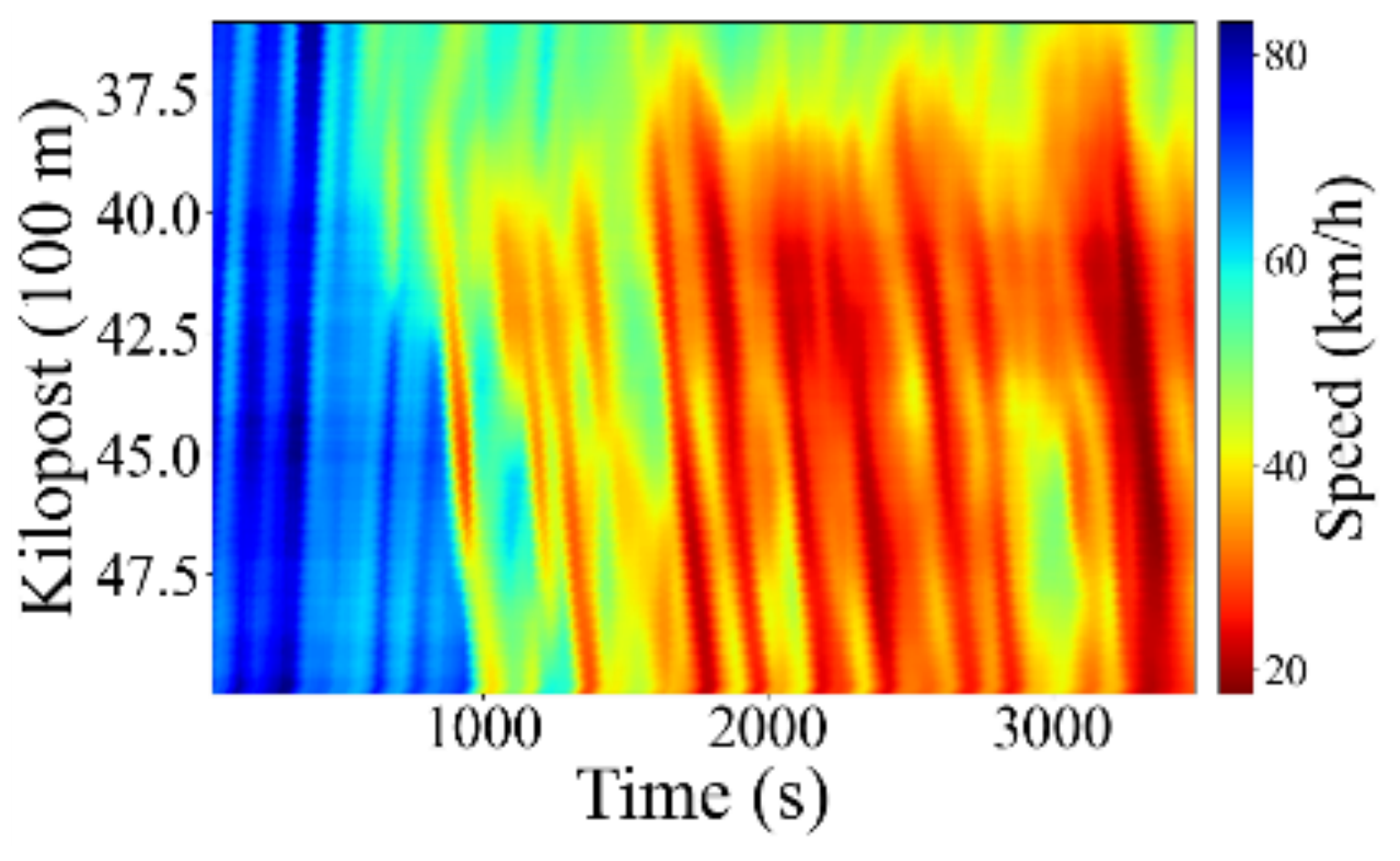}
    \label{Fig8a}
  \end{subfigure}
  \begin{subfigure}{.33\textwidth}
    \centering
    \caption{L001F002: LKV sample points}
    \includegraphics[width=1\linewidth]{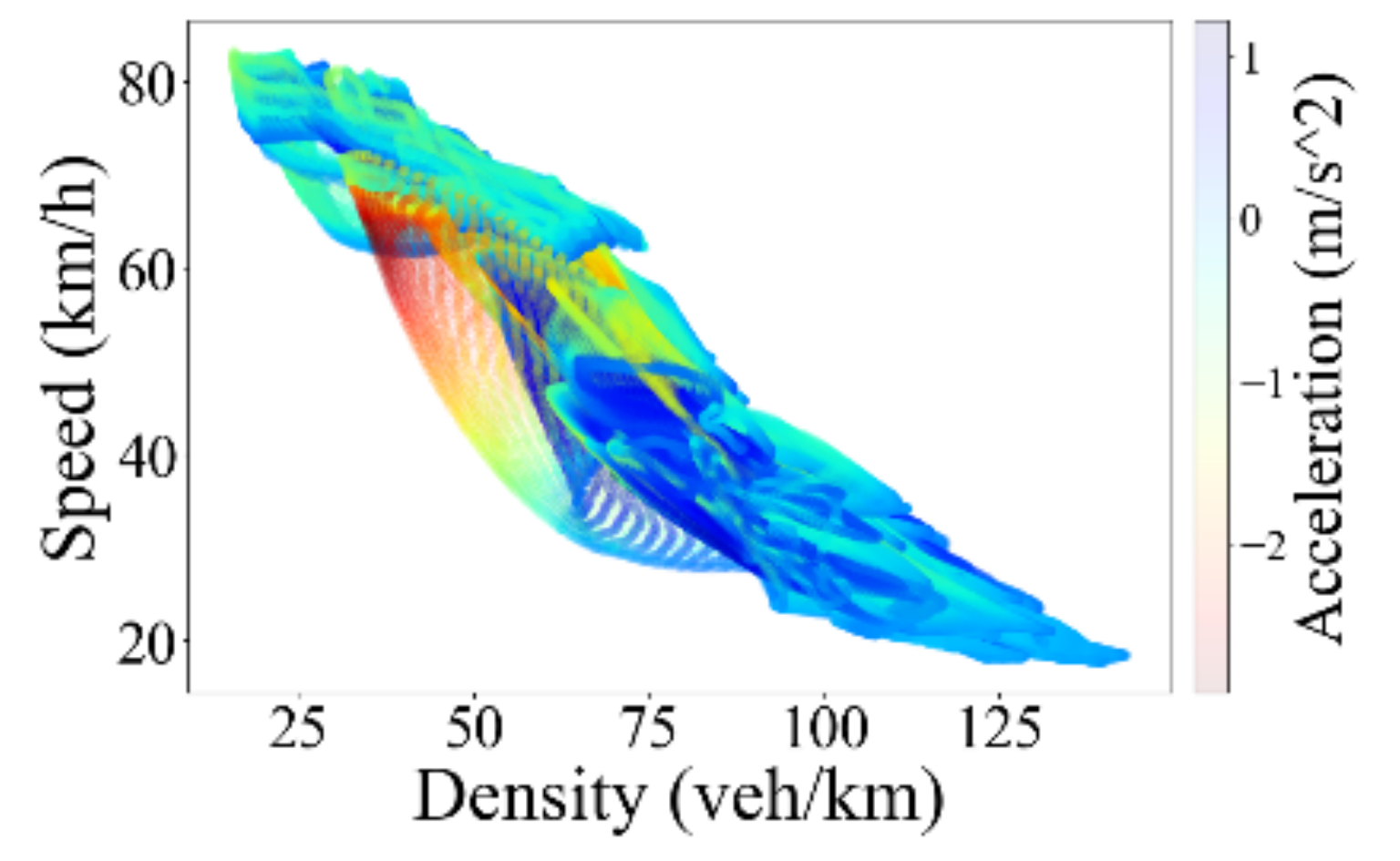}
    \label{Fig8b}
  \end{subfigure}
  \begin{subfigure}{.33\textwidth}
    \centering
    \caption{L001F002: NLKV sample points}
    \includegraphics[width=1\linewidth]{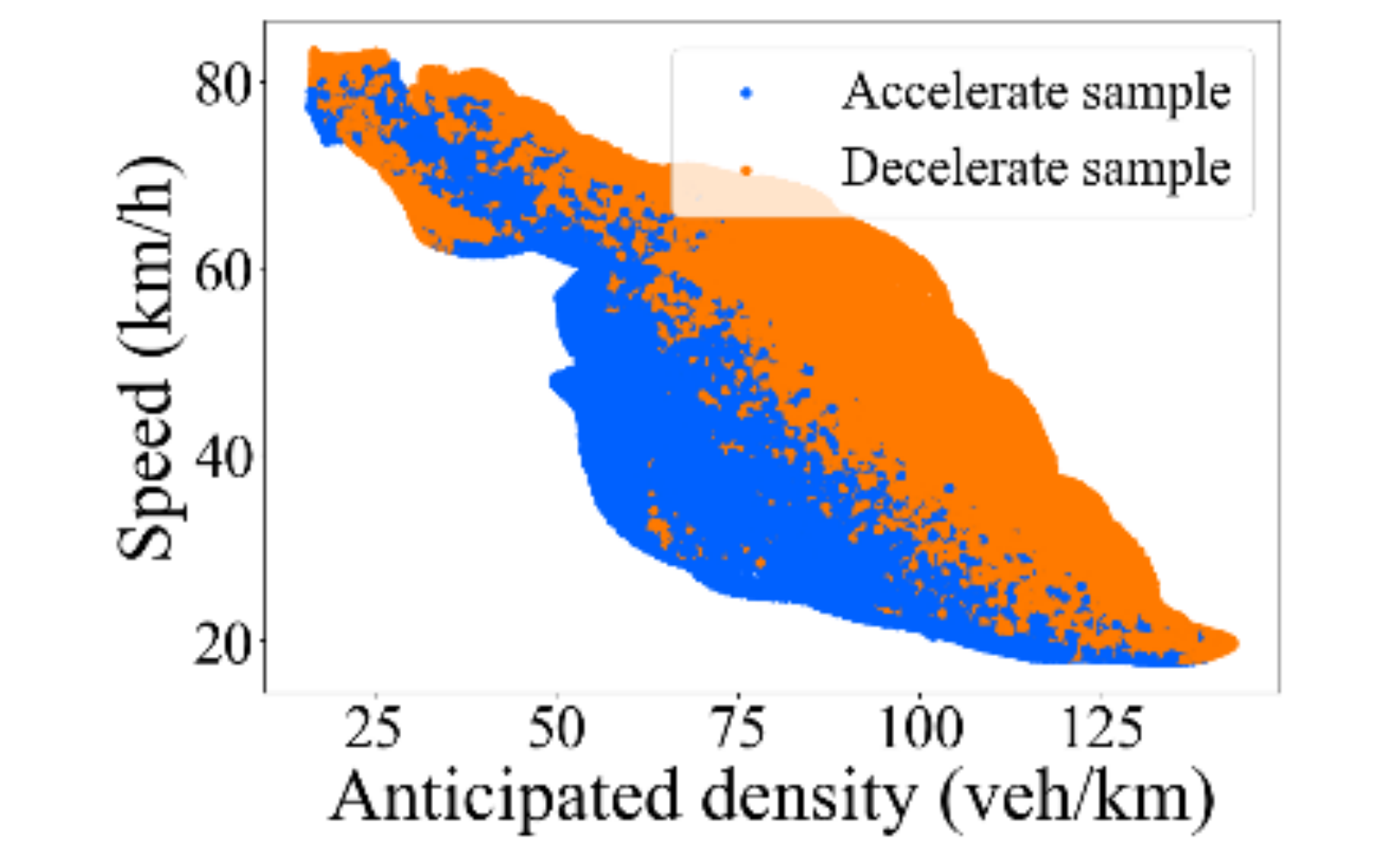}
    \label{Fig8c}
  \end{subfigure}
  \begin{subfigure}{.33\textwidth}
    \centering
    \caption{L001F003: $\mathcal{V}$}
    \includegraphics[width=1\linewidth]{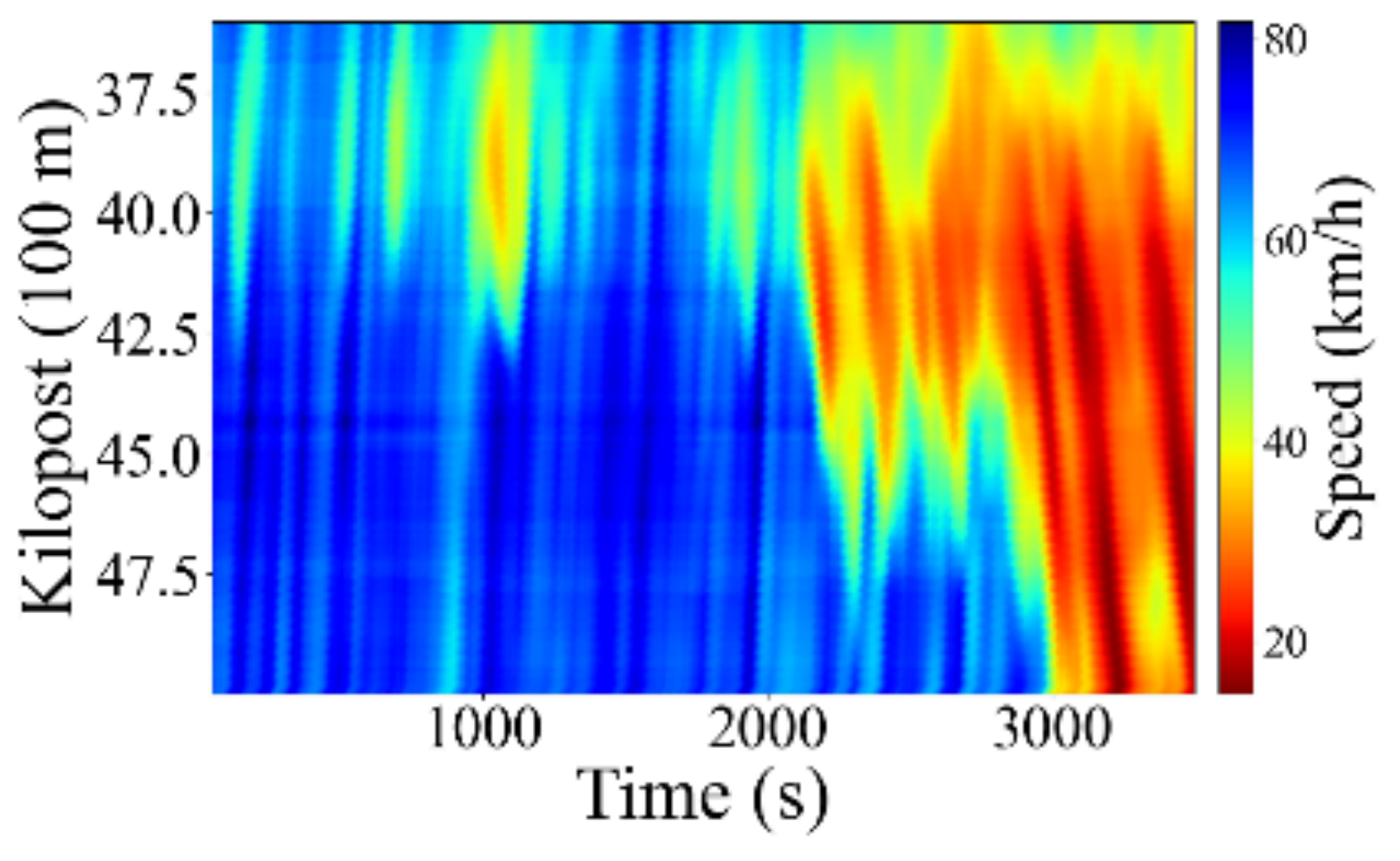}
    \label{Fig8d}
  \end{subfigure}
  \begin{subfigure}{.33\textwidth}
    \centering
    \caption{L001F003: LKV sample points}
    \includegraphics[width=1\linewidth]{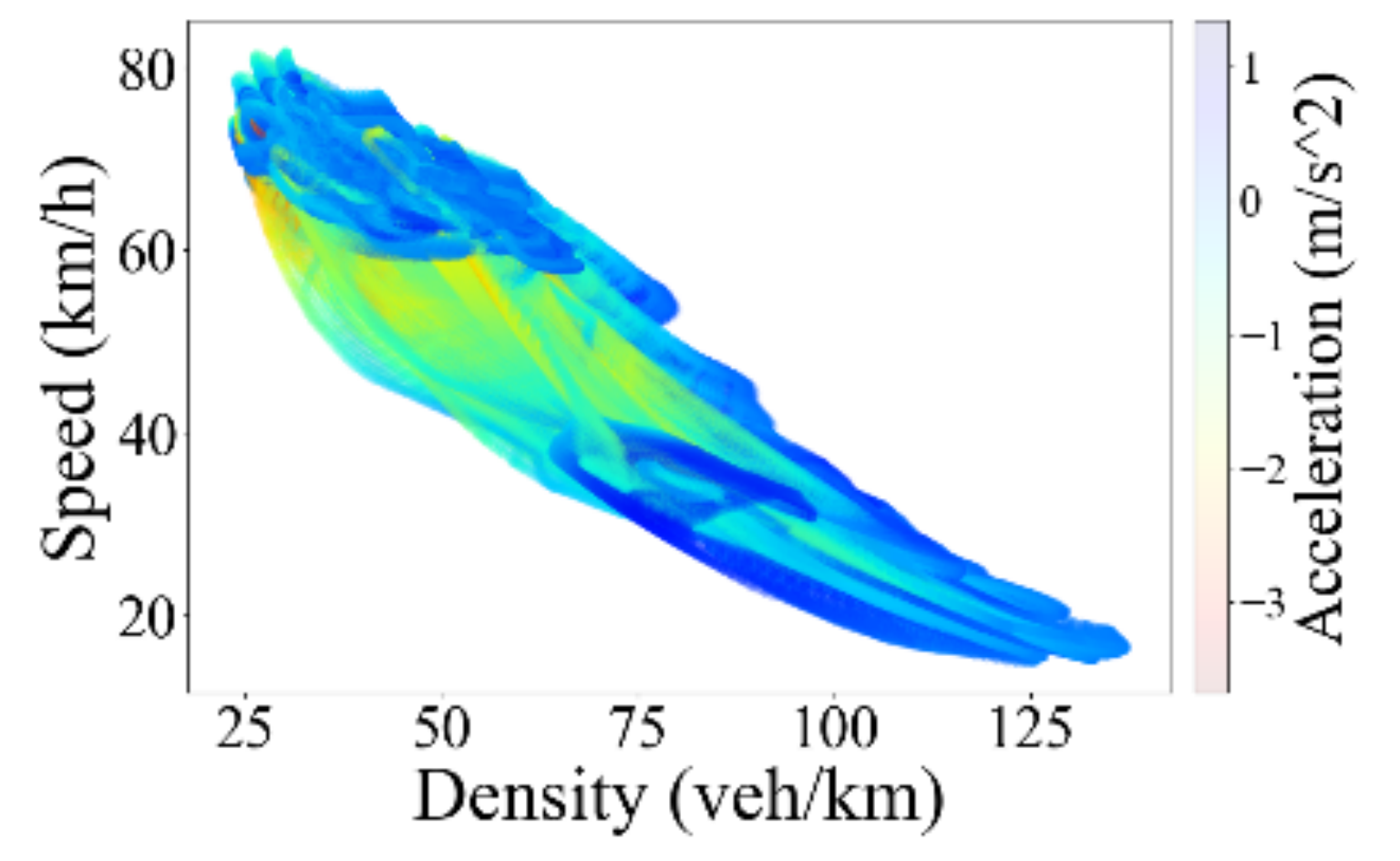}
    \label{Fig8e}
  \end{subfigure}
  \begin{subfigure}{.33\textwidth}
    \centering
    \caption{L001F003: NLKV sample points}
    \includegraphics[width=1\linewidth]{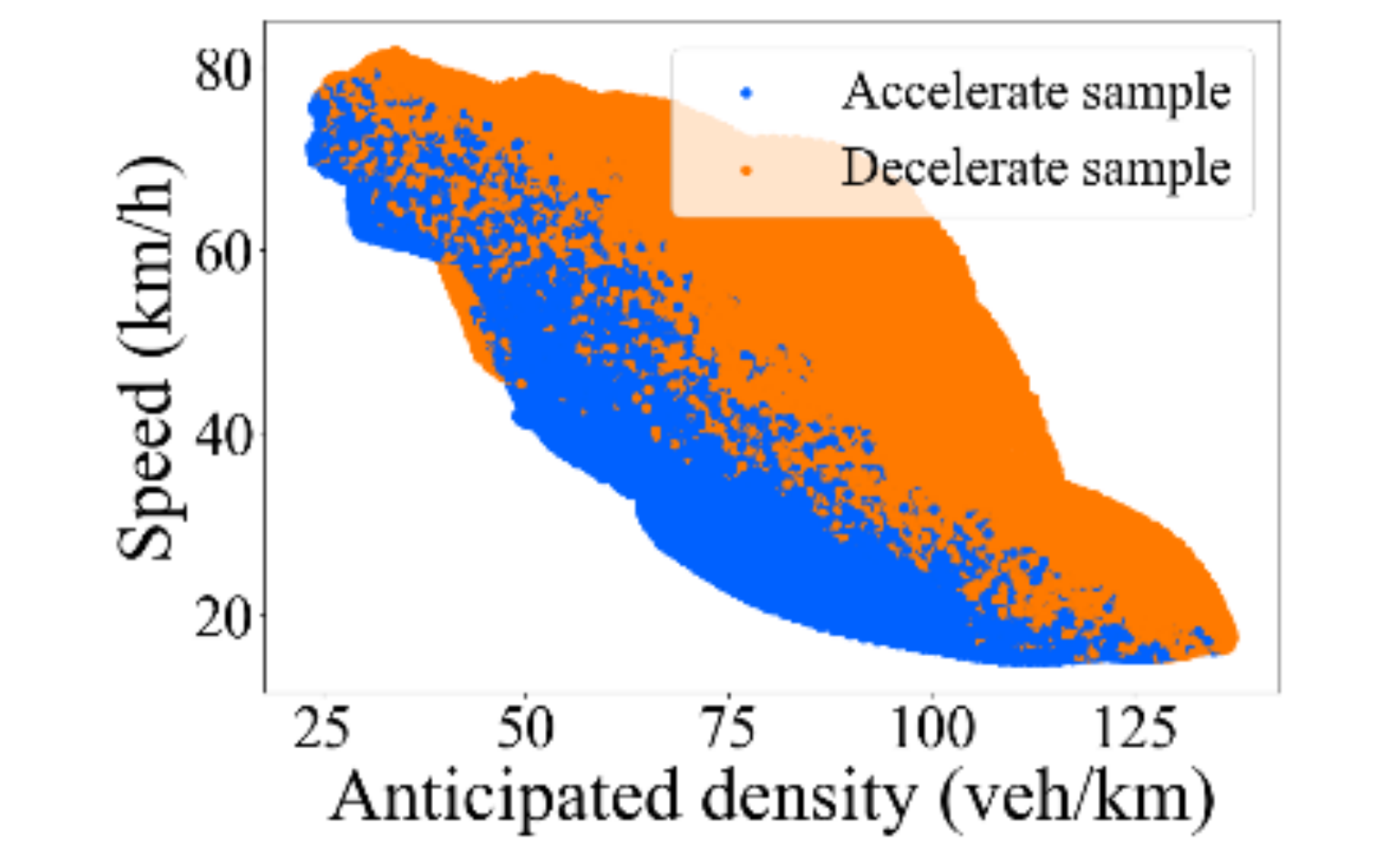}
    \label{Fig8f}
  \end{subfigure}
  \begin{subfigure}{.33\textwidth}
    \centering
    \caption{L001F004: $\mathcal{V}$}
    \includegraphics[width=1\linewidth]{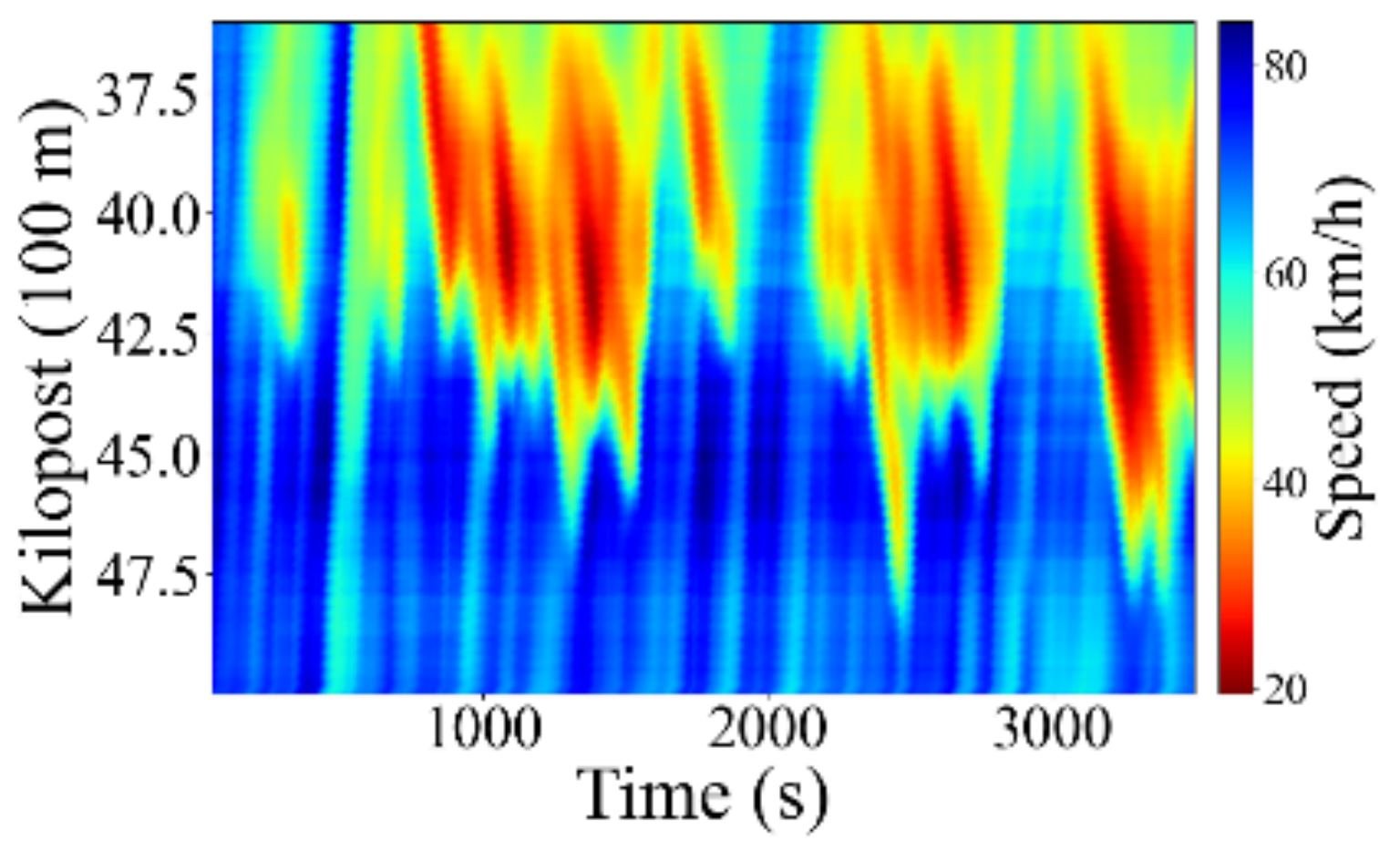}
    \label{Fig8g}
  \end{subfigure}
  \begin{subfigure}{.33\textwidth}
    \centering
    \caption{L001F004: LKV sample points}
    \includegraphics[width=1\linewidth]{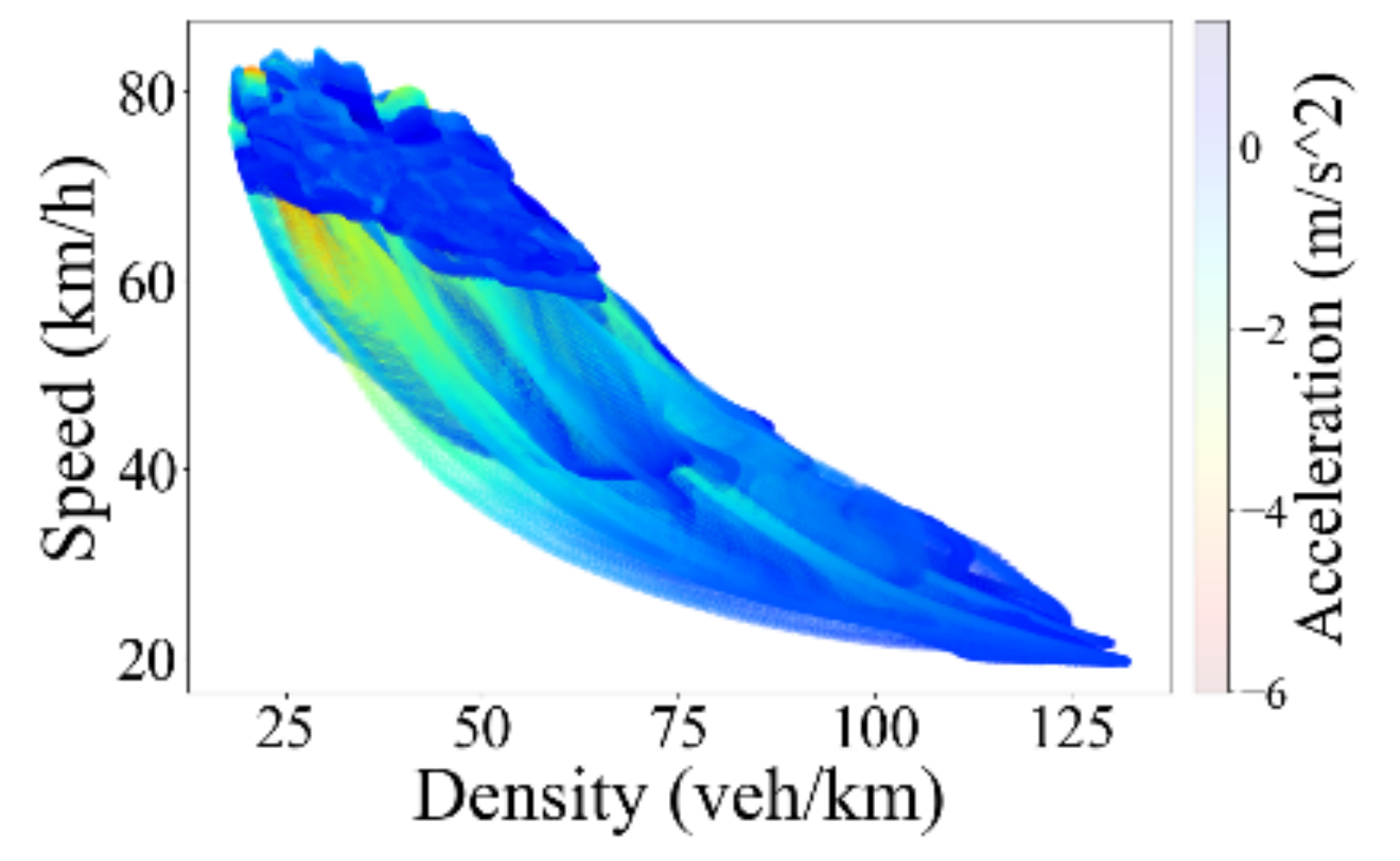}
    \label{Fig8h}
  \end{subfigure}
  \begin{subfigure}{.33\textwidth}
    \centering
    \caption{L001F004: NLKV sample points}
    \includegraphics[width=1\linewidth]{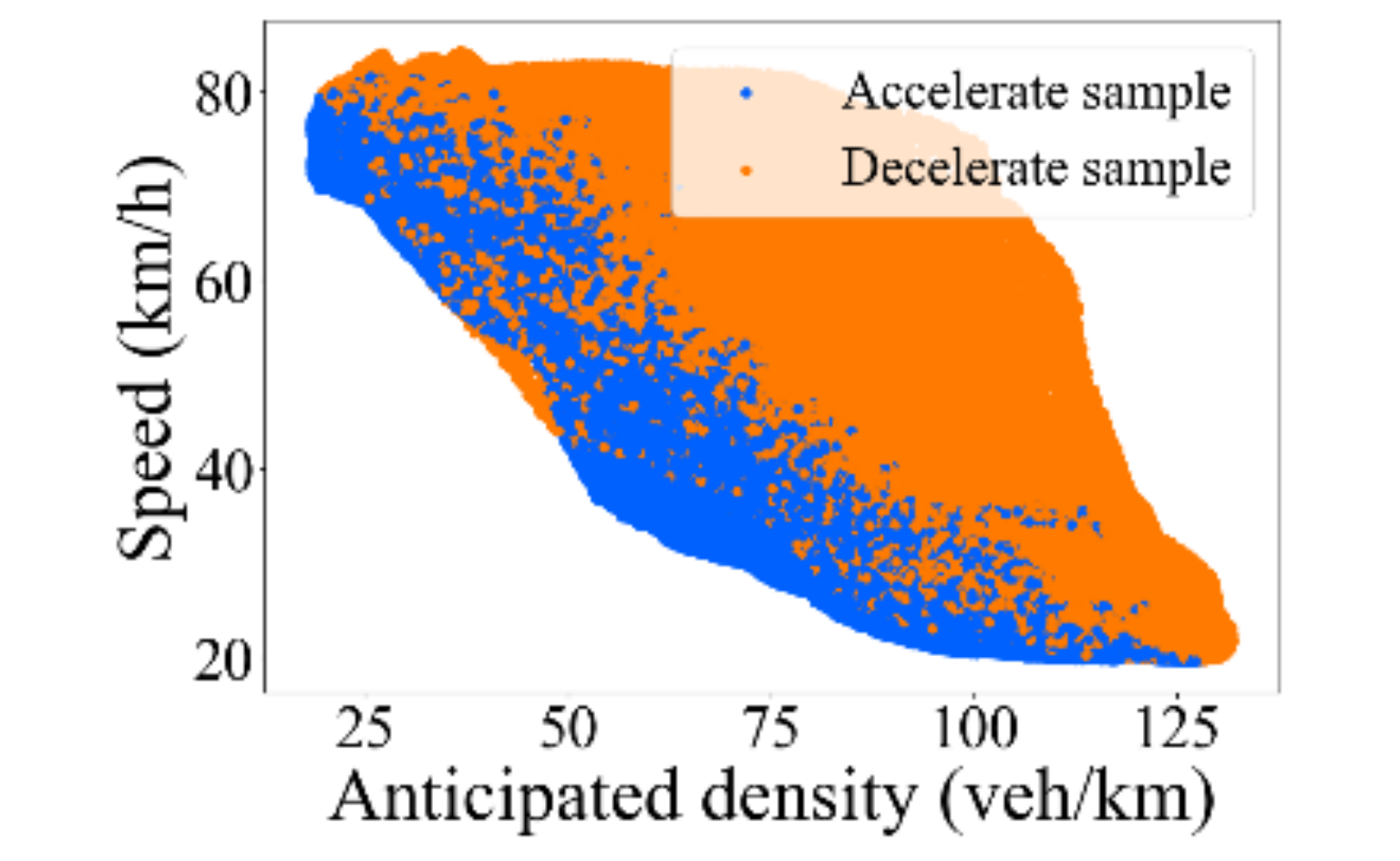}
    \label{Fig8i}
  \end{subfigure}
  \begin{subfigure}{.33\textwidth}
    \centering
    \caption{L001F005: $\mathcal{V}$}
    \includegraphics[width=1\linewidth]{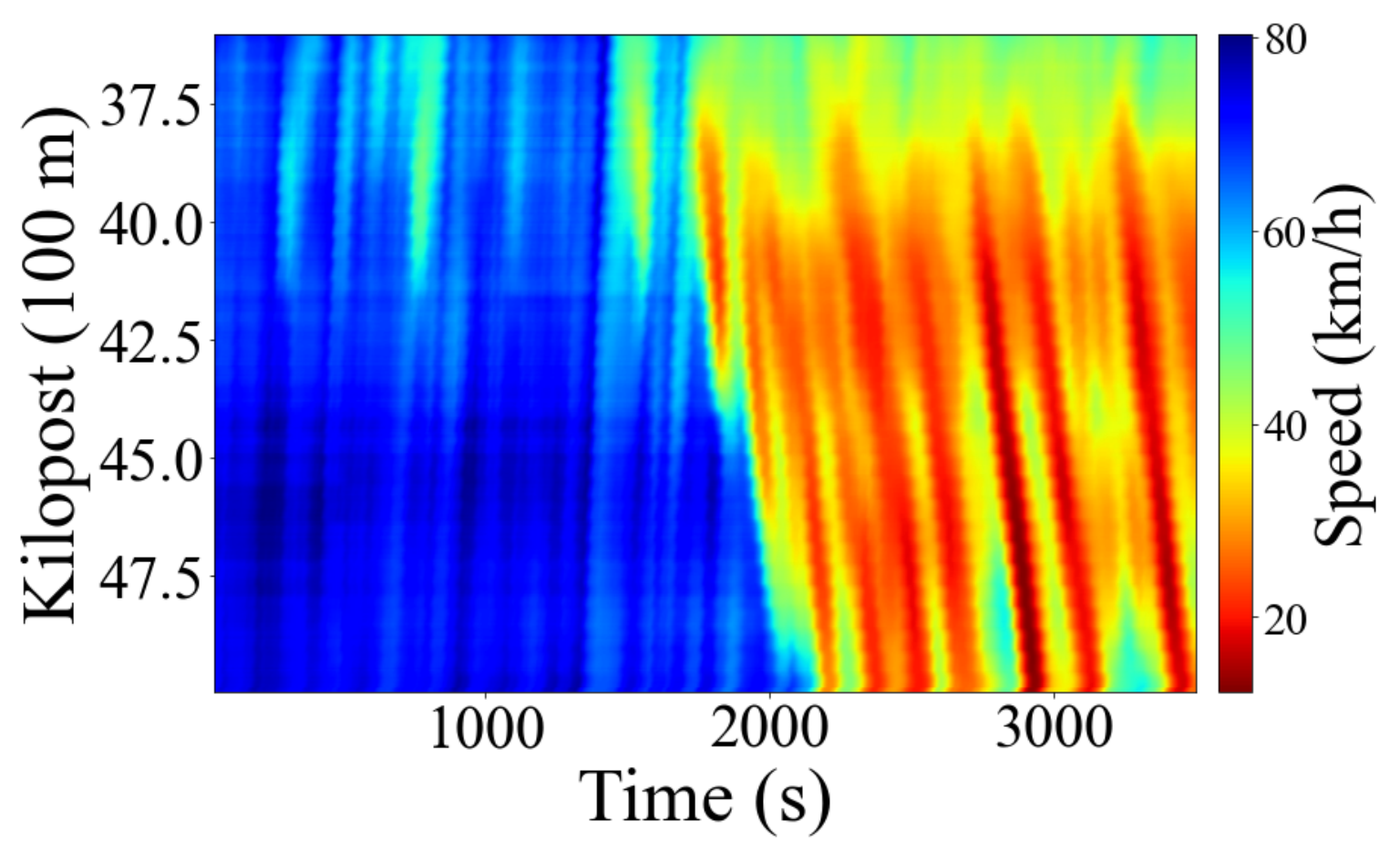}
    \label{Fig8j}
  \end{subfigure}
  \begin{subfigure}{.33\textwidth}
    \centering
    \caption{L001F005: LKV sample points}
    \includegraphics[width=1\linewidth]{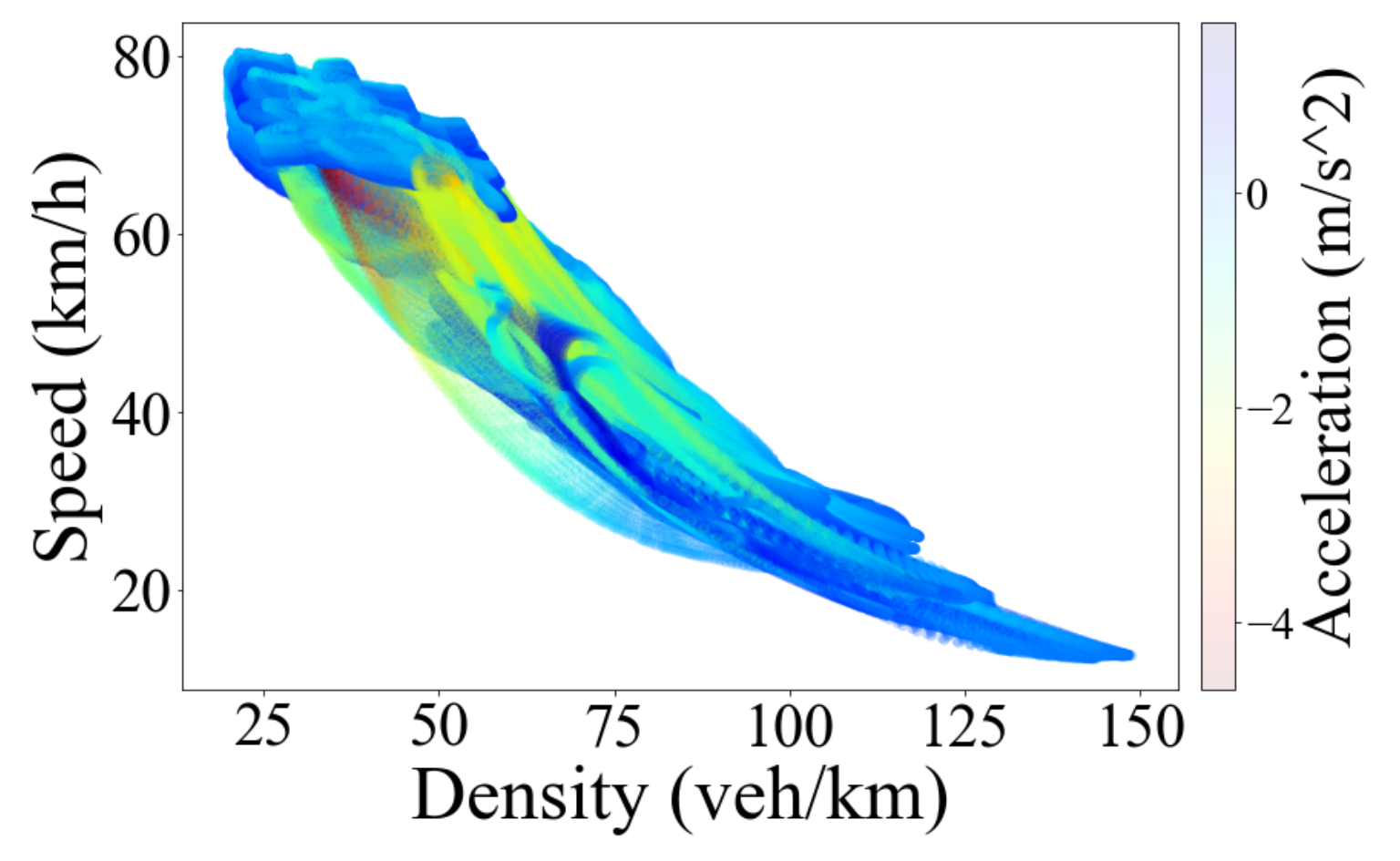}
    \label{Fig8k}
  \end{subfigure}
  \begin{subfigure}{.33\textwidth}
    \centering
    \caption{L001F005: NLKV sample points}
    \includegraphics[width=1\linewidth]{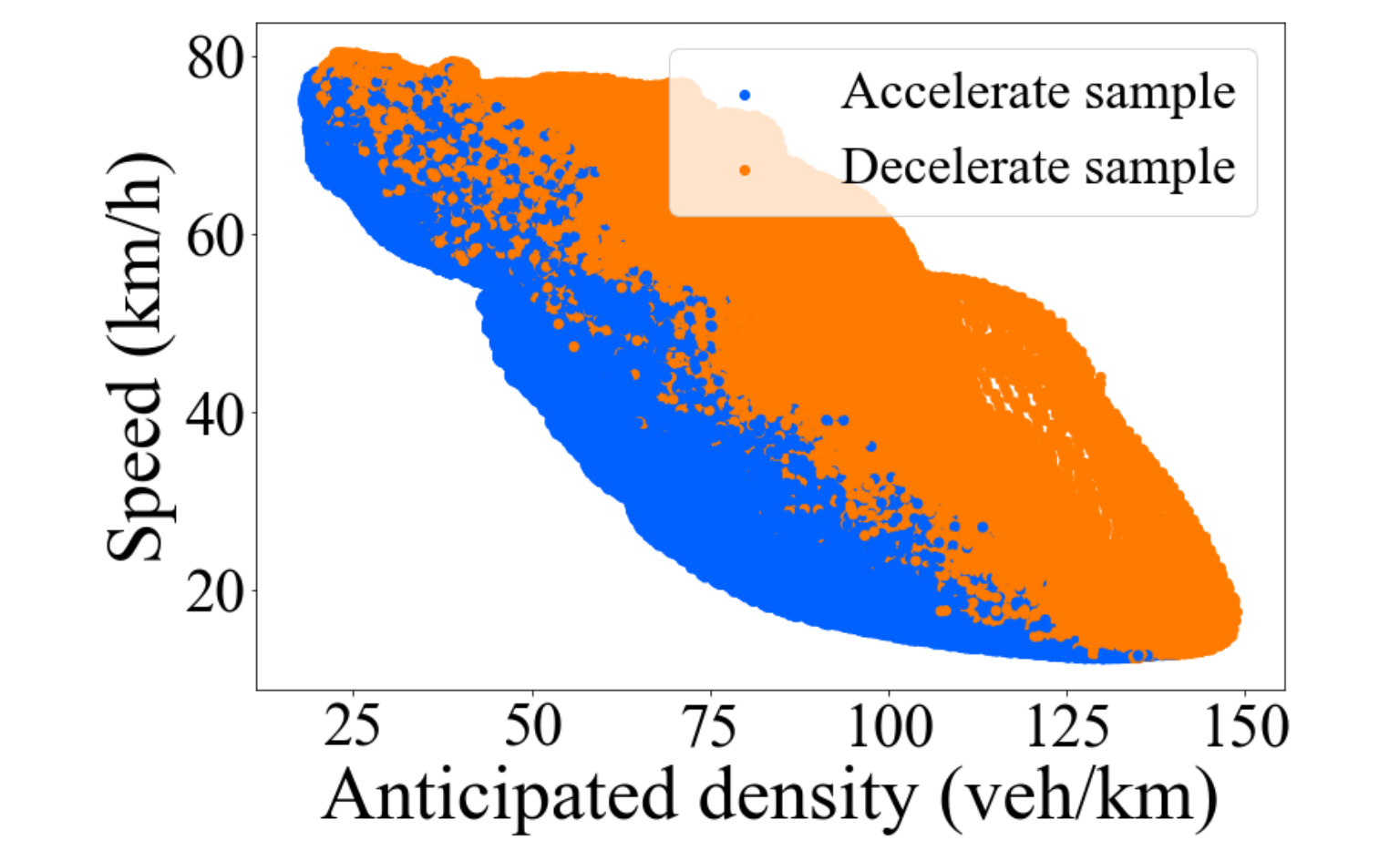}
    \label{Fig8l}
  \end{subfigure}
  \caption{The speed field, LKV and NLKV samples of L001F002-L001F005: (a)-(c) L001F002; (d)-(f) L001F003; (g)-(i) L001F004; (j)-(l) L001F005.}
  \label{Fig8}
\end{figure}

We further compare the fitting results of the two methods using a different sample of trajectories from the same location and time period, specifically, L001 of ZTD. We employ the Smulders’ and the Franklin-Newell FD models. ZTD provides five isolated detected trajectory datasets for link L001, each collected over a duration of one hour. These datasets are denoted as L001F001, L001F002, L001F003, L001F004, and L001F005. The macroscopic fields and LKV and NLKV samples of L001F001 are shown in \autoref{Fig4a}-\autoref{Fig4c} and \autoref{Fig5a} and \autoref{Fig5c}. The speed fields and samples of L001F002-L001F005 are presented in \autoref{Fig8}. It is evident from \autoref{Fig8} that all the datasets exhibit a transition from free flow to congestion, but with varying characteristics. For instance, L001F002 and L001F005 show a gradual buildup of congestion over time. L001F004 demonstrates congestion occurring in specific spatial areas, while L001F003 experiences congestion in both space and time. These variations in traffic characteristics result in distinct shapes and patterns in both the LKV and NLKV samples. The changes observed in the LKV samples have a significant impact on the distributions of traffic density and, consequently, the fitted FDs. In contrast, the NLKV samples display relatively stable separating boundaries between acceleration and deceleration regimes. 
Therefore, modeling the separating boundaries 
using NLKV samples, hence the locus of equilibrium states, is invariant to the trajectory samples compared to modeling the mean speeds using LKV samples. This is because the true FD is an emergent property of road traffic (i.e., independent of its microscopic constituents).

\begin{figure}[!ht]
  \begin{subfigure}{.49\textwidth}
    \centering
    \caption{Smulders' FD model with the LKV+LSE approach}
    \includegraphics[width=.8\linewidth]{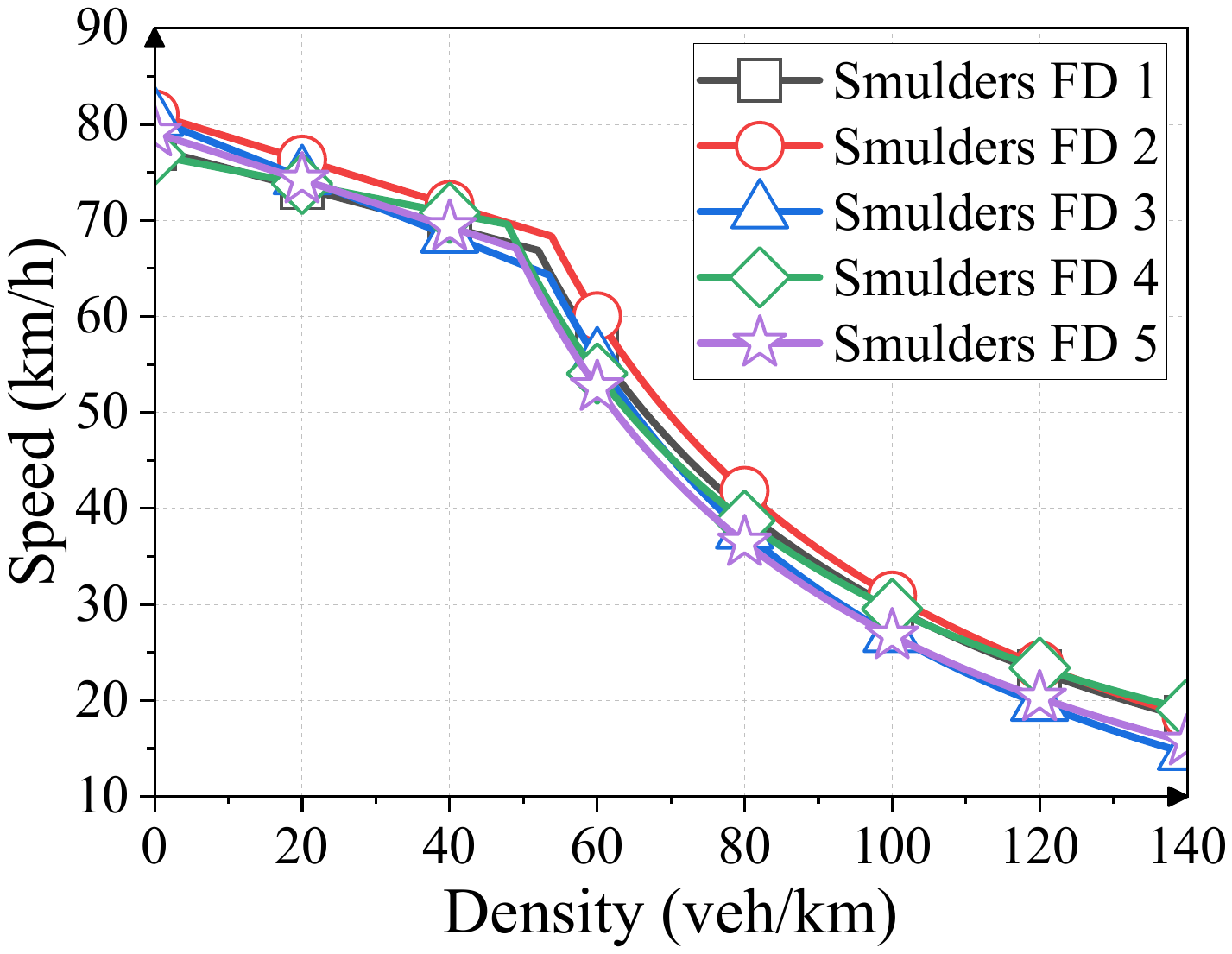}
    \label{Fig9a}
  \end{subfigure}
  \begin{subfigure}{.49\textwidth}
    \centering
    \caption{Smulders' FD model with the NLKV+ECE approach}
    \includegraphics[width=.8\linewidth]{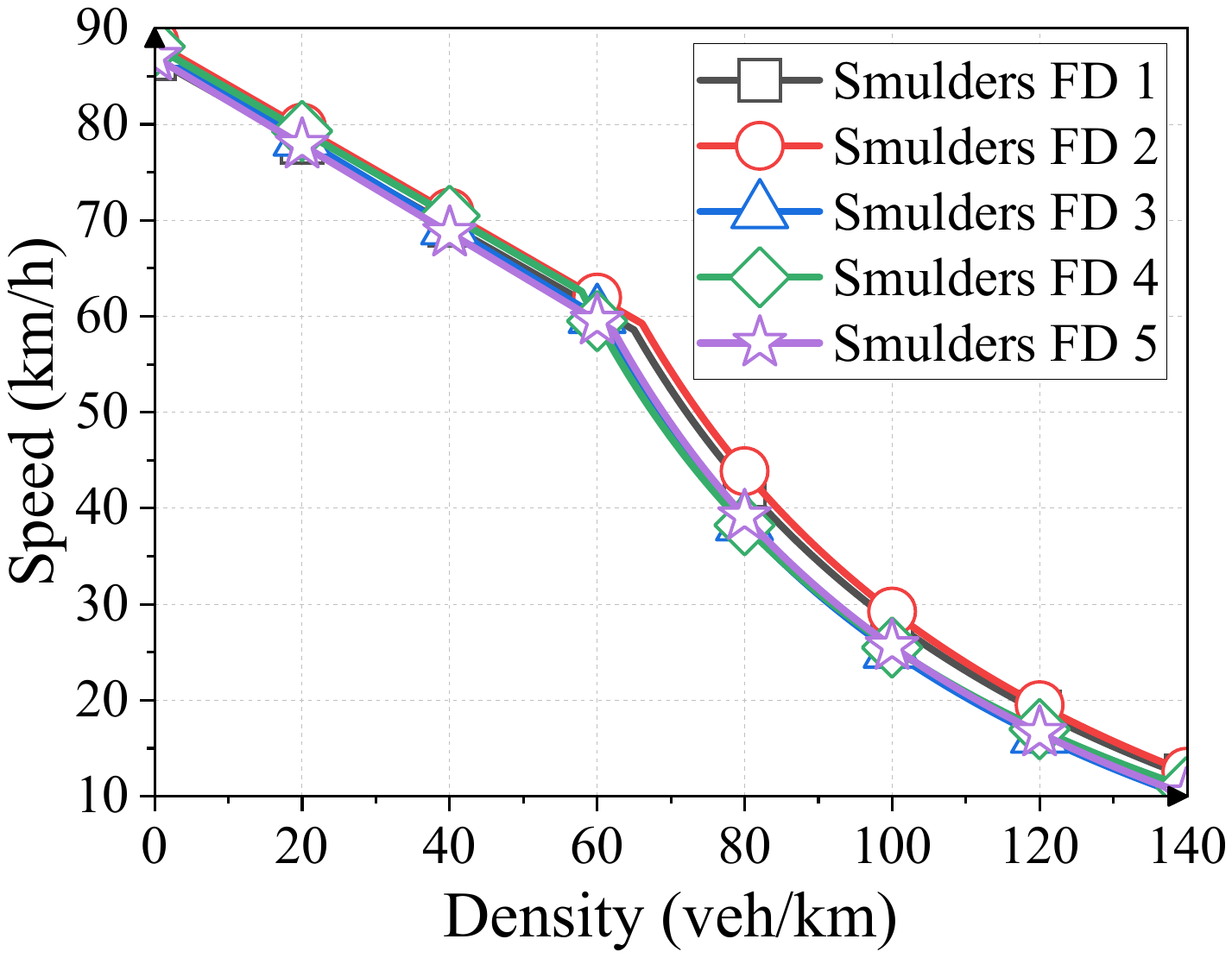}
    \label{Fig9b}
  \end{subfigure} \\
  \begin{subfigure}{.49\textwidth}
    \centering
    \caption{the Franklin-Newell FD fitting using the LKV+LSE}
    \includegraphics[width=.8\linewidth]{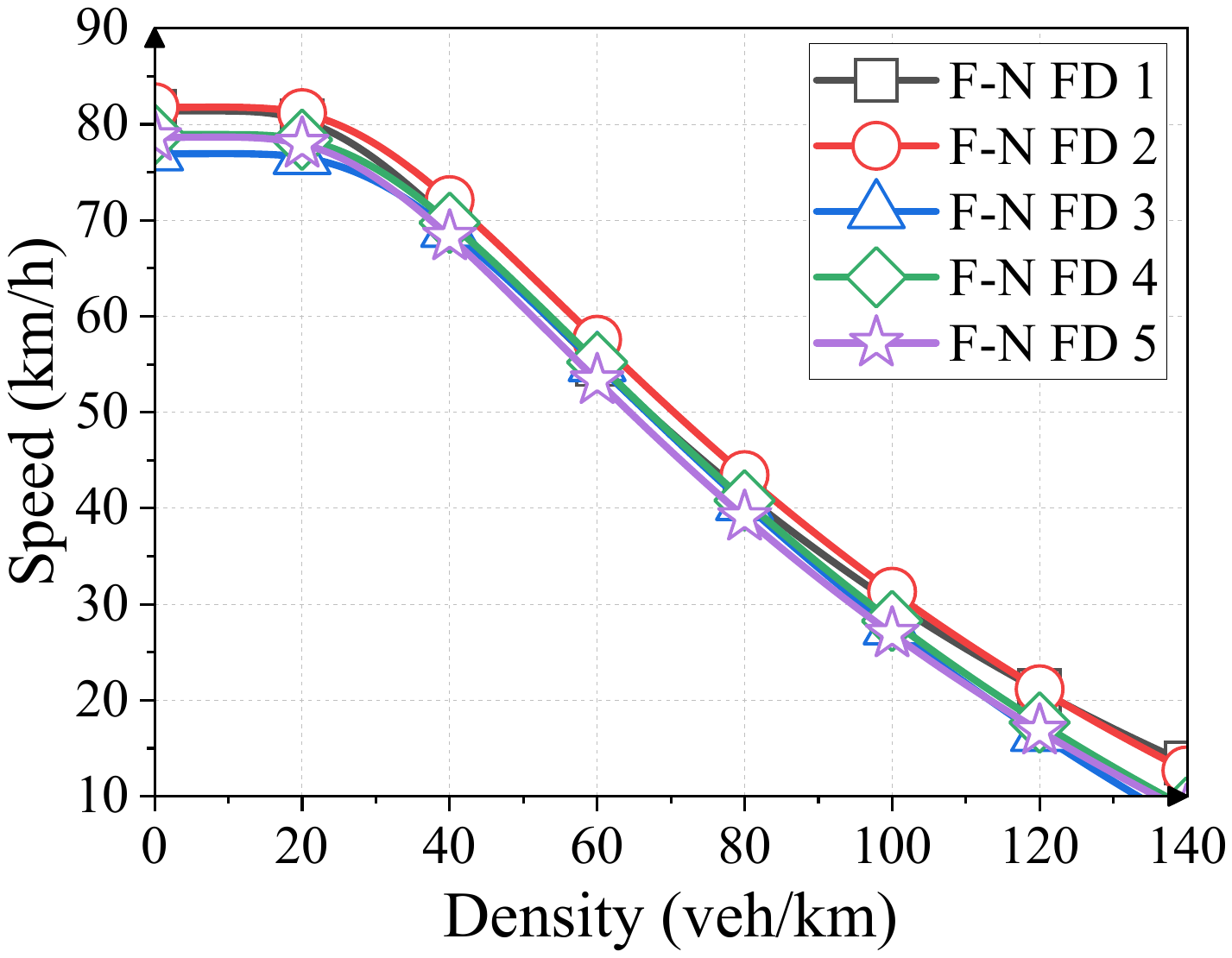}
    \label{Fig9c}
  \end{subfigure}
  \begin{subfigure}{.49\textwidth}
    \centering
    \caption{the Franklin-Newell FD fitted using NLKV+ECE minimization}
    \includegraphics[width=.8\linewidth]{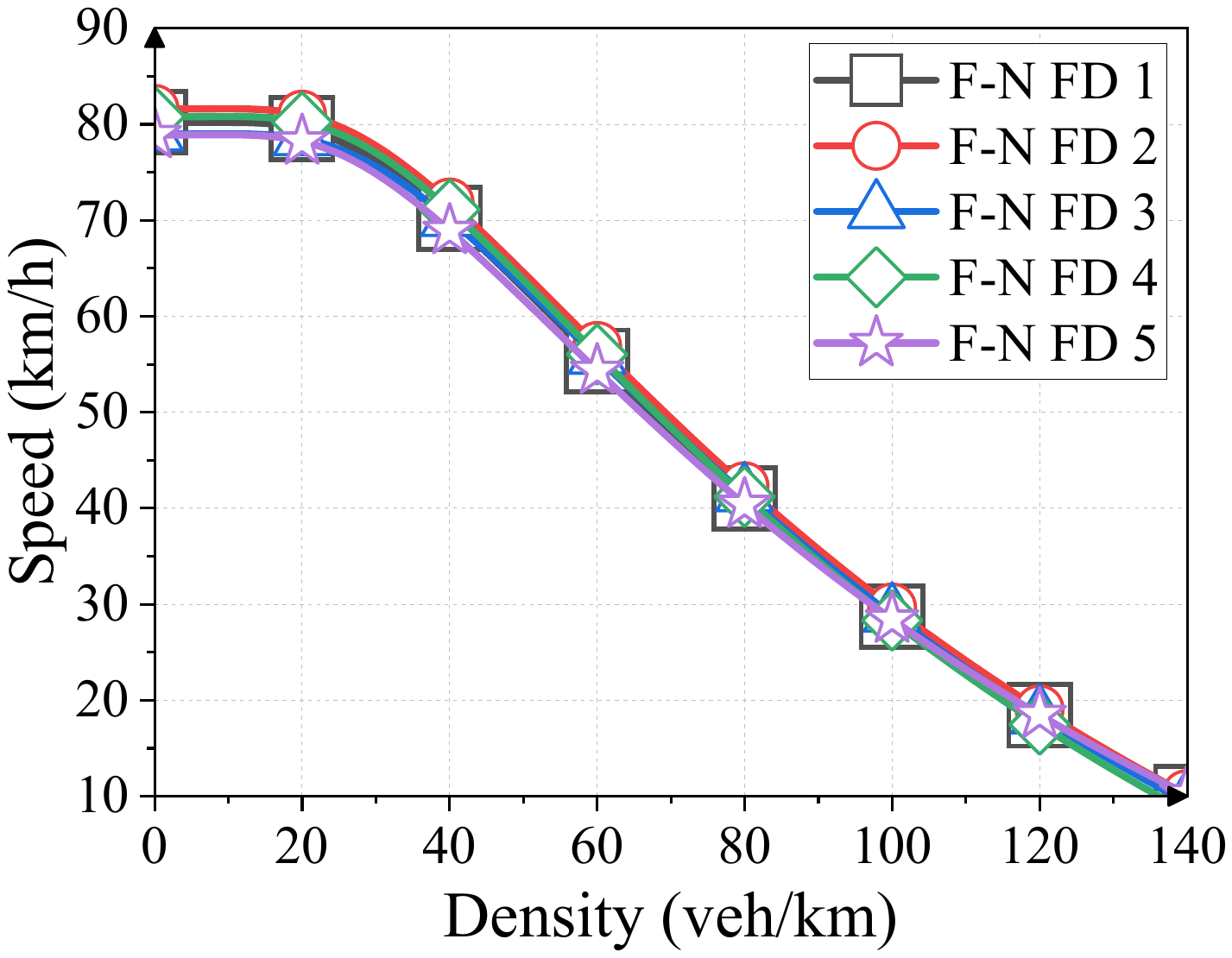}
    \caption{}
    \label{Fig9d}
  \end{subfigure}
  \caption{Comparison of FDs fitted using different datasets from L001: (a) Smulders' FD model with the LKV+LSE approach; (b) Smulders' FD model with the NLKV+ECE approach; (c) the Franklin-Newell FD fitting using the LKV+LSE; (d) the Franklin-Newell FD fitted using NLKV+ECE minimization.}
  \label{Fig9}
\end{figure}

We proceed to fit the parameters of the Smulders’ FD model and the Franklin-Newell’s FD model on these five datasets (L001F001 to L001F005) using the LKV+LSE and NLKV+ECE approaches, respectively (note that Smulders’ FD fitted to L001F001 has already been conducted in \autoref{Sect5.4.1}). The fitting results are as shown in \autoref{Fig9}, where the curves represent the fitted FDs on the different L001 datasets. \autoref{Fig9a} and \autoref{Fig9c} depict Smulders’ FDs and the Franklin-Newell FDs obtained using the LKV+LSE approach, while \autoref{Fig9b} and \autoref{Fig9d} display the Smulders’ FDs and Franklin-Newell’s FDs obtained through the NLKV+ECE approach. A careful observation of the figures that the FDs generated by the LKV+LSE approach exhibit more diversity, whereas the FDs generated by the NLKV+ECE approach demonstrate greater consistency and similarity across the different datasets.

Irrespective of environmental variations and the heterogeneity of traffic flow, the $v-k$ relationship should be determined solely by the characteristics of the road traffic and should not be influenced by the collected data. However, the fitting approach that lacks equilibrium information (i.e., LKV+LSE) heavily relies on the detected features, leading to varying fitted FDs for the same road link with different trajectories. On the other hand, the FDs fitted using NLKV samples exhibit invariance to trajectory samples from the same spatio-temporal region. 

\subsection{Comparison of different FD models using LKV+LSE and NLKV+ECE approaches}\label{Sect5.5}

\begin{figure}[!ht]
\centering
  \begin{subfigure}{.45\textwidth}
    \centering
    \caption{LKV fitting}
    \includegraphics[width=.8\linewidth]{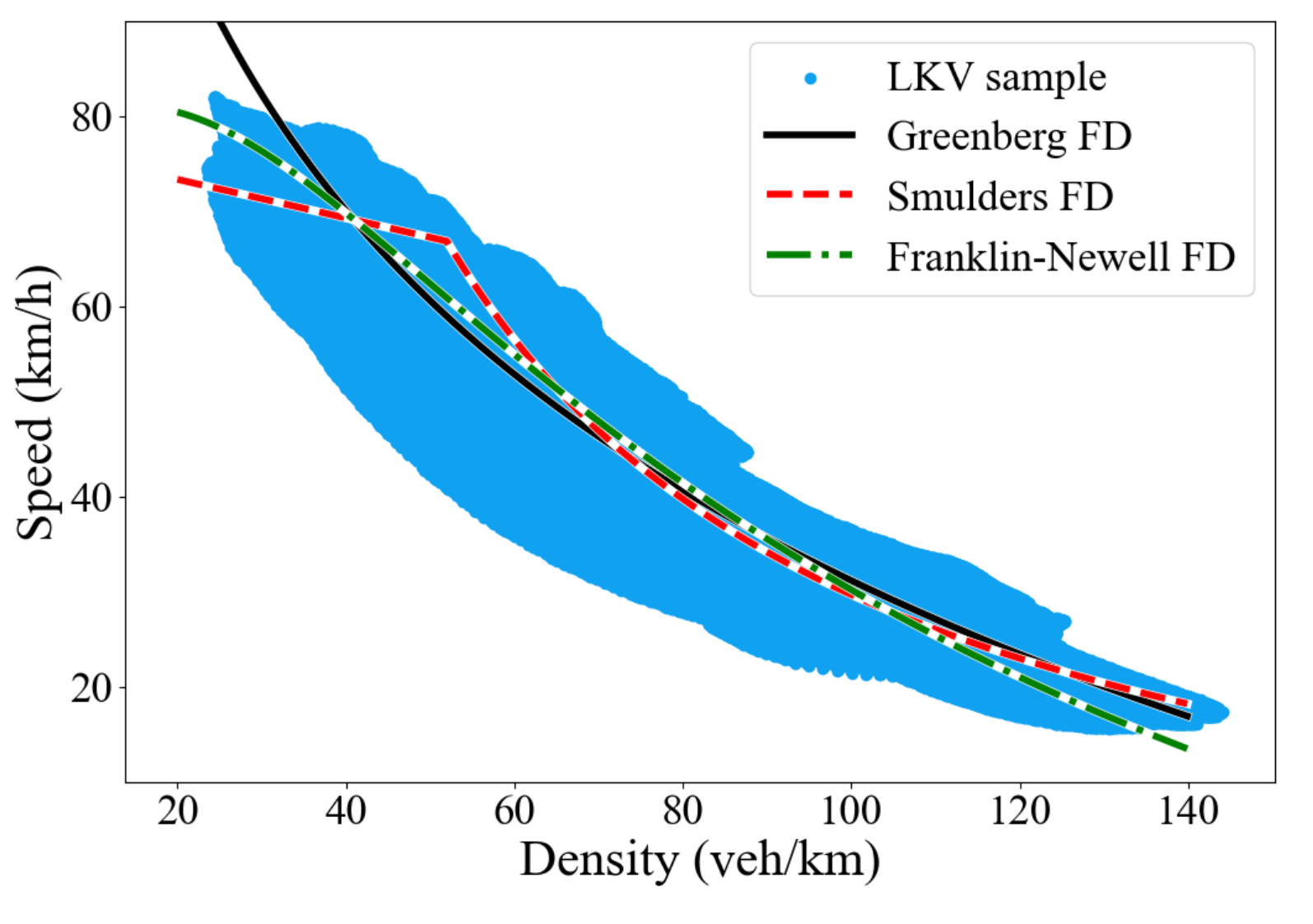}
    \label{Fig10a}
  \end{subfigure}
  \begin{subfigure}{.45\textwidth}
    \centering
    \caption{NLKV fitting}
    \includegraphics[width=.8\linewidth]{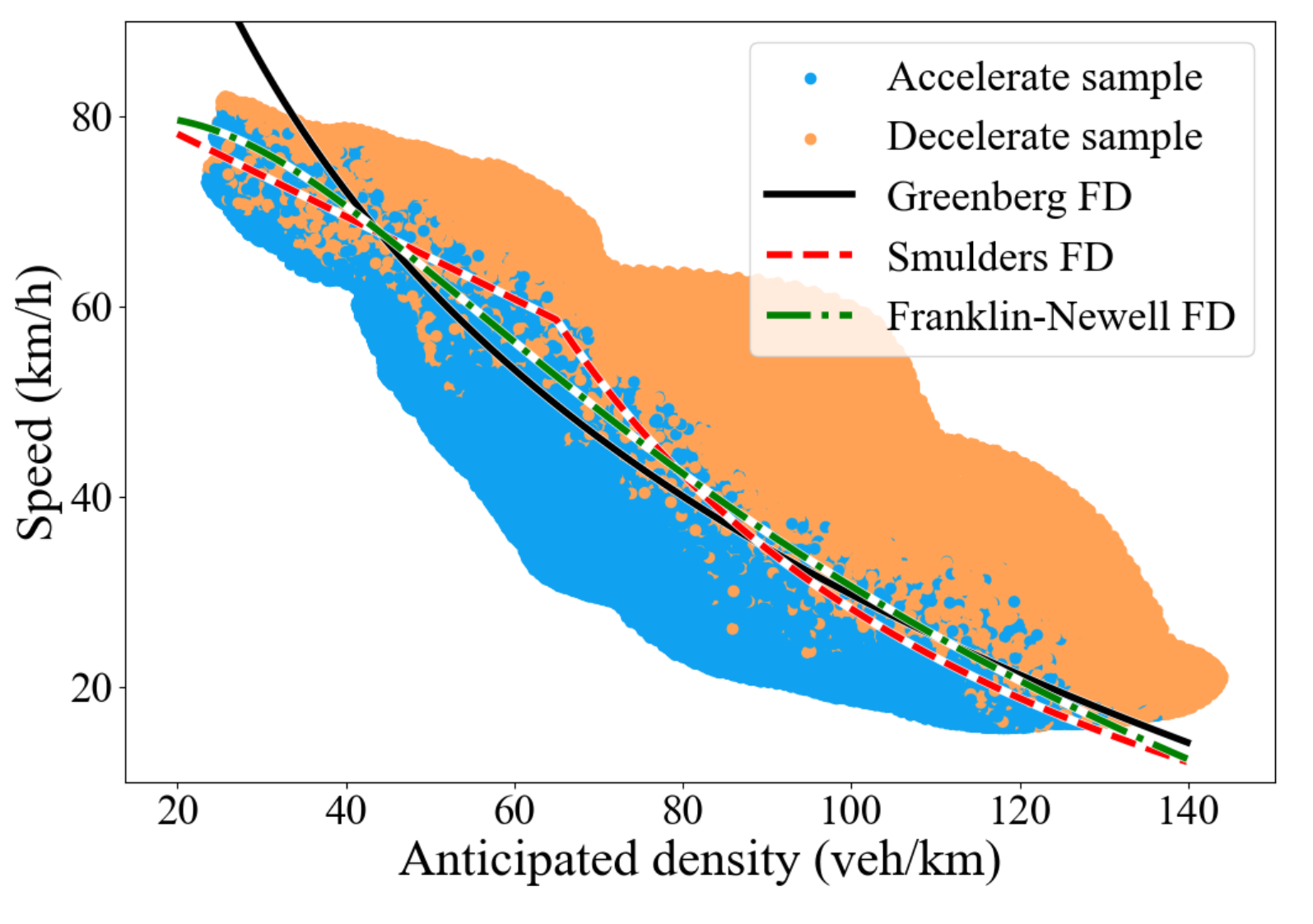}
    \label{Fig10b}
  \end{subfigure}
  \caption{The fitting results of the different FD models on the LKV and NLKV samples: (a) LKV fitting; (b) NLKV fitting.}
  \label{Fig10}
\end{figure}

In the comparison of fitting results considering FD model selection, we first fit Greenberg’s, Smulders’, and the Franklin-Newell FD models using LKV samples from dataset 1, using the LKV+LSE approach. We then select the FD with the least squared error as the optimal FD under this approach. Similarly, we fit these three models using NLKV samples and select the FD with the least ECE error as the optimal FD under the NLKV+ECE approach. The fitting results are presented in \autoref{Fig10} and \autoref{table2}.

\autoref{Fig10} illustrates the fitted FDs obtained through different approaches, capturing the characteristics of their corresponding samples. In \autoref{Fig10a}, the fitted FDs using the LKV+LSE approach exhibit a diverse free flow speed and a higher jam density compared to the FDs in \autoref{Fig10b}. \autoref{table2} provides a quantitative comparison of the fitted FDs, including the model parameters, as well as LSE loss and the ECE loss. Among the FDs fitted by LKV+LSE approach, Smulders' FD has the lowest LSE loss, indicating a closer fit to the LKV samples of dataset 1. Consequently, the Smulders model can be considered as the most suitable model among the three models for the LKV samples of dataset 1. However, in the NLKV+ECE approach, the Franklin-Newell FD exhibits the lowest ECE loss among all the fitted FDs. These results are logical since the Franklin-Newell model accounts for driver behavior.

\begin{table}[!ht]
\centering
\caption{The FD comparison of the different samples and models}\label{table2}
\begin{tabular}{ccccccccccc}
\hline
                & \multicolumn{5}{c}{LKV+LSE}                                                        & \multicolumn{5}{c}{NLKV+ECE}                                                       \\ \cline{2-11} 
                & $v_f$ & $v_0$ & $k_c$ & $k_j$ & \begin{tabular}[c]{@{}c@{}}LSE\\ loss\end{tabular} & $v_f$ & $v_0$ & $k_c$ & $k_j$ & \begin{tabular}[c]{@{}c@{}}ECE\\ loss\end{tabular} \\ \hline
Greenberg       & -     & 42.5  & -     & 208.3 & 21.627                                             & -     & 46.3  & -     & 189.9 & 2.635                                              \\
Smulders        & 77.4  & -     & 52    & 381.1 & \textbf{15.121}                                    & 86.8  & -     & 65    & 199.9 & 1.336                                              \\
Franklin-Newell & 81.4  & -     & -     & 187.9 & 16.384                                             & 80.2  & -     & -     & 168.3 & \textbf{1.206}                                     \\ \hline
\end{tabular}
\end{table}


\section{Concluding Remarks}\label{Sect6}

This paper proposes a novel way to create samples from trajectory datasets, which considers non-localities, to estimate the fundamental diagram (FD) of traffic flow. The proposed non-local density-speed samples, dubbed NLKV samples, incorporate anticipation by replacing the local density in density-speed samples, dubbed LKV, with the anticipated density. Additionally, we label the acceleration and deceleration behaviors of traffic flow. Specifically, if the current traffic flow is moving faster than the desired speed associated with its anticipated density, a negative acceleration is observed. Conversely, a positive acceleration is inferred if the current traffic flow is moving slower than the desired speed. If the traffic flow tends to maintain speed, this signifies that the current speed corresponds to the desired speed associated with the anticipated density.

By utilizing the NLKV samples, the FD modeling task can be formulated as a binary classification problem. The FD is fitted to the separating boundary between acceleration and deceleration regimes. 
We developed an enhanced cross entropy (ECE) loss as our statistical model from basic statistical assumptions that account for basic equilibrium traffic physics, which not only accounts for the cost of misclassification but also addresses sample bias. This approach stands in contrast to the conventional method employing LKV samples and least squares estimation (LSE). The NLKV samples contain equilibrium information and can be assumed to conform to the independent identically distributed hypothesis required for maximum likelihood estimation (MLE). Consequently, our proposed approach provides a more rational and methodologically sound framework for FD fitting.

Through our comparison analysis of the LKV+LSE and NLKV+ECE approaches using field trajectories, we observe distinct differences in their fitting performance. Our investigation of four different links reveals that accelerations differ across different pairs of anticipated density and current speed, exhibiting a gradient ranging from acceleration to near-equilibrium and ultimately to deceleration. Notably, the NLKV samples successfully segregate the acceleration and deceleration regions, yielding a more accurate representation of the ``true'' FD under equilibrium. In contrast, the LKV samples are heavily influenced by hysteresis effects, leading to relations that do not reflect equilibrium conditions.

Observing the same FD model fitted using these two approaches, we find that parameters can differ substantially. Specifically, the FD model fitted using LKV+LSE tends to underestimate the free flow speed and overestimate the jam density compared to the model fitted using NLKV+ECE. Additionally, when analyzing trajectories collected in the same road link but at different times, we note that the FDs fitted by NLKV+ECE demonstrate invariance to the trajectory sample 
compared to those fitted by LKV+LSE. Finally, we find that the best-fit models may vary when using these two approaches.

The use of NLKV samples open up several interesting possibilities for future research. Researchers can develop models to describe the separating boundaries of NLKV, which could provide a more direct representation of the ``true'' equilibrium FD instead of relying on a parametric models such as Greenberg's, Smulders', and the Franklin-Newell model. Additionally, since NLKV samples consider traffic dynamics, they can be coupled with continuous traffic flow modeling to enhance prediction accuracy. Furthermore, we replace the LSE loss with ECE loss, which circumvents criticisms regarding the Gaussian assumption of the noise distribution. Notably, by eradicating the hysteresis effect present in LKV samples, NLKV samples provide a promising avenue for improving the noise distribution of speed. 


\section*{Acknowledgments}

This work is supported by National Key R\&D Program of China under grant 2021YFB1600100, and Chinese Scholarship Council under grant 202207000057. The work is also supported by the NYUAD Center for Interacting Urban Networks (CITIES), funded by Tamkeen under the NYUAD Research Institute Award CG001.

\section*{Author Contribution Statement}
The authors confirm contribution to the paper as follows: study conception and design: J. Liu, F. Zheng, B. Yu, S. Jabari; data collection: J. Liu, B. Yu; analysis and interpretation of results: J. Liu, F. Zheng, S. Jabari; draft manuscript preparation: J. Liu, F. Zheng, S. Jabari. All authors reviewed the results and approved the manuscript.

\section*{Conflict of interest}
The author(s) declared no potential conflicts of interest with respect to the research, authorship, and/or publication of this article.


\section*{Appendix}
We provide Python codes to generate NLKV samples from trajectories, along with a subset of the LKV and NLKV samples of L001F001 for reference. These resources are available at \url{https://github.com/AziaJingLiu/NLKV-sample-create}.
\bibliography{references}

\end{sloppypar}

\end{document}